\documentstyle{article}

\newlength{\dinwidth}
\newlength{\dinmargin}
\makeatletter

\def\bbl{\bf}
 \def\eck{\cal}
 \def\sno{\cal}
 \def\got{\bf}
\catcode`\@=11

\newif\if@fewtab\@fewtabtrue

\global\def\draftcontrol{0}

\def\draftdate{\number\day.\number\month.\number\year\ \ \ \hourmin }
{\count255=\time\divide\count255 by 60
\xdef\hourmin{\number\count255}
\multiply\count255 by-60\advance\count255 by\time
\xdef\hourmin{\hourmin:\ifnum\count255<10 0\fi\the\count255}}
\def\ps@draft{\let\@mkboth\@gobbletwo
    \def\@oddhead{}
    \def\@oddfoot
       {\hbox to 7 cm{$\scriptstyle\bf Draft\ version:\ \draftdate$
       \hfil}\hskip -7cm\hfil\rm\thepage \hfil}
    \def\@evenhead{}\let\@evenfoot\@oddfoot}

\def\label#1{\ifnum\draftcontrol=1
 \global\def\draftnote{\scriptsize\tt #1}\fi
 \@bsphack\if@filesw {\let\thepage\relax
   \def\protect{\noexpand\noexpand\noexpand}%
\xdef\@gtempa{\write\@auxout{\string
      \newlabel{#1}{{\@currentlabel}{\thepage}}}}}\@gtempa
   \if@nobreak \ifvmode\nobreak\fi\fi\fi
  \@esphack}

\def\@eqnnum{\hbox to 3cm{\phantom{\rm(\theequation)} \draftnote
                         \hfil}\hskip -3cm {\rm(\theequation)}}

\def\eqnarray{\def\draftnote{{}}\global\@fewtabtrue
\stepcounter{equation}\let\@currentlabel=\theequation
\global\@eqnswtrue
\global\@eqcnt\z@\tabskip\@centering\let\\=\@eqncr
$$\halign to \displaywidth\bgroup\@eqnsel\hskip\@centering\@eqcnt\z@
  $\displaystyle\tabskip\z@{##}$&\global\@eqcnt\@ne
  \hskip 1\arraycolsep \hfil${##}$\hfil
  &\global\@eqcnt\tw@ \hskip 1\arraycolsep
$\displaystyle\tabskip\z@{##}$
\hfil  \tabskip\@centering&\global\@eqcnt\thr@@\llap{##}\tabskip\z@
\cr}

\def\endeqnarray{\@@eqncr\egroup
      \global\advance\c@equation\m@ne$$\global\@ignoretrue}

\def\@@eqncr{\let\@tempa\relax
    \ifcase\@eqcnt \def\@tempa{& & &}\or \def\@tempa{& &}
      \or \def\@tempa{&}
      \or\def\@tempa{}
\fi\@tempa
\if@eqnsw
\if@fewtab\@eqnnum\fi
\stepcounter{equation}\fi\global
\@eqnswtrue\global\@eqcnt\z@\global\@fewtabtrue\cr}

\@addtoreset{equation}{section}

\def\cases#1{\left\{\,\vcenter{\normalbaselines\m@th
    \ialign{$\displaystyle{##}\hfil$&\quad##\hfil\crcr#1\crcr}}\right.}

\def\ct#1{\ifnum\draftcontrol=1{\tt [#1]}\else{\cite{#1}}\fi}
\def\ctz#1#2{\ifnum\draftcontrol=1{\tt [#1,#2]}\else{\cite[#1]{#2}}\fi}
\def\draftcite#1{\ifnum\draftcontrol=1#1\else{}\fi}

\def\@lbibitem[#1]#2{\item{}\hskip -3cm \hbox to 2cm
{\hfil$\scriptstyle\draftcite{#2}$}\hskip
1cm[\@biblabel{#1}]\if@filesw
     {\def\protect##1{\string ##1\space}\immediate
      \write\@auxout{\string\bibcite{#2}{#1}}}\fi\ignorespaces}

\def\@bibitem#1{\item\hskip -3cm \hbox to 2cm
{\hfil \scriptsize\tt\draftcite{#1}}\hskip 1cm
\if@filesw \immediate\write\@auxout
       {\string\bibcite{#1}{\the\value{\@listctr}}}\fi\ignorespaces}

\def\C{{\bbl C}}
\def\Rea{{\bbl R}}

\def\Z{{\bbl Z}}
\def\Nat{{\bbl N}}

\newtheorem{theo}{Theorem}
\newtheorem{defi}{Definition}
\def\lb#1{\label{#1}}
\def\lab#1{\ifnum\draftcontrol=1{{\tt [#1]} \lb{#1}}\else{\lb{#1}}\fi}
\def\Eq#1{(\ref{#1})}
\def\theequation{{\thesection.\arabic{equation}}}
\def\[{\begin{eqnarray}}
\def\nn{\nonumber}
\def\non{\nonumber \\ }
\def\]{\end{eqnarray}}
\def\hh#1{\hspace*{#1}}

\def\een{\end{enumerate}}
\def\ben{\begin{enumerate}}
\def\IF#1{\quad\mbox{if #1}}

\def\FOR#1{\quad\mbox{for #1}}
\def\a{\alpha}

\def\d{\delta}
\def\e{\epsilon}
\def\f{\varphi}
\def\m{\mu}
\def\n{\nu}
\def\p{\psi}

\def\s{\sigma}
\def\t{\tau}
\def\th{\theta}
\def\x{\xi}
\def\y{\eta}
\def\z{\zeta}
\def\ba{\mbox{\boldmath$\alpha$}}
\def\bd{\mbox{\boldmath$\delta$}}
\def\by{\mbox{\boldmath$\eta$}}
\def\bL{\mbox{\boldmath$\Lambda$}}
\def\br{\mbox{\boldmath$\rho$}}
\def\bs{\mbox{\boldmath$\sigma$}}
\def\bw{\mbox{\boldmath$\omega$}}
\def\bx{\mbox{\boldmath$\xi$}}
\def\bz{\mbox{\boldmath$\zeta$}}

\def\F{\Phi}
\def\G{\Gamma}
\def\K{\kappa}
\def\L{\Lambda}
\def\P{\Psi}
\def\cl{{\ell}}
\def\cC{{\cal C}}

\def\cF{{\cal F}}

\def\cV{{\cal V}}
\def\cN{{\cal N}}
\def\cO{{\cal O}}
\def\cP{{\cal P}}
\def\cT{{\cal T}}

\def\Fz#1{\cF^{#1}}
\def\Pz#1{\cP^{#1}}
\def\gg{{\got g}}
\def\gh{{\got h}}

\def\sL{{\sno L}}
\def\sM{{\sno M}}
\def\sR{{\sno R}}

\def\sI{{\sno I}}
\def\ew{{\eck w}}
\def\eC{{\eck C}}
\def\eG{{\eck G}}

\def\eW{{\eck W}}

\def\va{{\ba}}
\def\vb{{\bf b}}
\def\vd{{\bd}}

\def\vk{{\bf k}}

\def\vL{{\bL}}
\def\vo{{\bf 0}}
\def\vp{{\bf p}}
\def\vq{{\bf q}}

\def\vr{{\bf r}}
\def\vs{{\bf s}}
\def\vt{{\bf t}}

\def\vv{{\bf v}}
\def\vx{{\bf x}}
\def\vP{{\bf P}}
\def\vX{{\bf X}}

\def\vxi{{\bx}}
\def\vet{{\by}}
\def\vze{{\bz}}
\def\tx{\tilde{\xi}}

\newfont\sgb{cmmib7} 

\def\sma{{\hbox{\sgb \symbol{"0B}}}}

\def\smvL{{\hbox{\sgb \symbol{"03}}}}
\def\Oint{\oint\limits}
\def\End{\mathop{\rm End}}

\def\mod{\,\mathop{\rm mod}\,}

\def\id{\mathop{\rm id}\nolimits}
\def\ad#1#2{(\mathop{\rm ad}\,#1)^{#2}}
\def\re#1{{\mathop{\rm e}\nolimits}^{#1}}
\def\frc#1#2{{\textstyle \frac{#1}{#2}}}

\def\9{{E_9}}
\def\0{{E_{10}}}
\def\exam{E_{10}^{(\smvL_7)}}
\def\dd#1#2{\frac{d#1}{d#2}}
\def\dz{\frac{d}{dz}}
\def\|{\,|\,}
\def\.{\cdot}
\def\X{\!\cdot\!}
\def\XO{\otimes}
\def\sh{^\sharp}
\def\dg{^\dagger}
\def\td{\tilde}
\def\brv{\breve}
\def\cc{\circ}
\def\res{\mathop{\rm Res}\nolimits}
\def\Res#1#2{{\mathop{\rm Res}\nolimits}_{#1}\left[#2\right]}
\def\Dm#1#2#3{#1^{-1}\d\left(\frac{#2-#3}{#1}\right)}

\def\Di#1#2#3{-#1^{-1}\d\left(\frac{-#2+#3}{#1}\right)}
\def\rL#1{L_{#1}}
\def\Ln{L_{n}}
\def\Lo{L_{0}}
\def\one{{\bf1}}
\def\Vp{\cV(\p,z)}
\def\Vf{\cV(\f,z)}
\def\VP#1{\cV(\p,{#1})}
\def\VF#1{\cV(\f,{#1})}
\def\VV#1{\cV\big(\cV(\p,#1)\f,w\big)}
\def\Vs{\cV\sh}
\def\Vd{\cV\dg}
\def\<{\langle}
\def\>{\rangle}
\def\dual#1#2{\<\!\< #1 | #2 \>\!\>}
\def\FLie{\mbox{$\cF\big/\rL{-1}\cF$\,}}
\def\PLie{\mbox{$\Pz1\big/\rL{-1}\Pz0$\,}}
\def\LR{\L_{\Rea}}
\def\Pr{\P_{\vr}}
\def\Ps{\P_{\vs}}
\def\amm{\a^\m_m}
\def\ann{\a^\n_n}
\def\:{\mbox{\bf:}}
\def\ord{\mbox{\large\bf:}}
\def\Ord{{}_\times^\times}
\def\II{I\hspace{-.2em}I_{9,1}}
\def\III{I\hspace{-.2em}I_{25,1}}
\def\ggA{\gg(A)}
\def\ggF{\gg_\cF}
\def\ggL{\gg_\L}
\def\ggI{\gg_{\II}}
\def\gfake{\gg_{\III}}
\def\ghA{\gh(A)}
\def\bra#1{\langle #1|}
\def\ket#1{|#1\rangle}
\def\bras#1#2{{}_{_{#2}}\!\langle #1|}
\def\kets#1#2{|#1\rangle_{\!_{#2}}}
\def\vev#1#2{\langle #1\| #2 \rangle}
\def\vac{\bra{\td\vo}}
\def\tr#1#2{\t_{#1#2}}
\def\ttr#1#2{\td{\t}_{#1#2}}
\def\btr#1#2{\bar{\t}_{#1#2}}
\def\vtr#1#2{\brv{\t}_{#1#2}}
\def\V#1{V^{[#1]}}
\def\VN{V^{[N]}}
\def\Vt#1#2#3{V_{#1#2#3}}
\def\tV#1{\td{V}^{[#1]}}
\def\CSV{V^{\rm CSV}}
\def\pr#1#2{\Pi_{#1#2}}
\def\R{\sR}
\def\M{\sM}
\def\Mh{\hat{\M}}
\def\N{{\sno N}}
\def\Nh{\hat{\N}}
\def\sew#1#2{\stackrel{#1\,#2}{\cup}}
\def\i{^{(i)}}
\def\j{^{(j)}}
\def\k{^{(k)}}
\def\1{^{(1)}}
\def\2{^{(2)}}
\def\3{^{(3)}}
\def\4{^{(4)}}
\def\5{^{(5)}}
\def\6{^{(6)}}
\def\RVpic#1#2#3#4{
 \put(35,15){\circle{10}}
 \put(35,15){\makebox(0,0){$#4$}}
 \put(22,15){\line(1,0){8}}
 \put(52,15){\line(1,0){8}}
 \put(35,-2){\line(0,-1){8}}
 \put(10,15){\vector(1,0){12}}
 \put(40,15){\vector(1,0){12}}
 \put(35,10){\vector(0,-1){12}}
 \put( 8,15){\makebox(0,0)[r]{$#1$}}
 \put(62,15){\makebox(0,0)[l]{$#2$}}
 \put(35,-12){\makebox(0,0)[t]{$#3$}}}
\def\Vpic#1#2#3#4{
 \put(35,15){\circle{10}}
 \put(35,15){\makebox(0,0){$#4$}}
 \put(10,15){\line(1,0){8}}
 \put(52,15){\line(1,0){8}}
 \put(35,-2){\line(0,-1){8}}
 \put(30,15){\vector(-1,0){12}}
 \put(40,15){\vector(1,0){12}}
 \put(35,10){\vector(0,-1){12}}
 \put( 8,15){\makebox(0,0)[r]{$#1$}}
 \put(62,15){\makebox(0,0)[l]{$#2$}}
 \put(35,-12){\makebox(0,0)[t]{$#3$}}}
\def\vza@[#1]{\bz^a\X\va_{#1}}
\def\vza{\@ifnextchar[{\vza@}{\bz^a}}
\def\vda#1{\vd\X\va_{#1}}
\def\vyb@[#1]{\by^b\X\va_{#1}}
\def\vyb{\@ifnextchar[{\vyb@}{\by^b}}
\def\vxc@[#1]{\bx^c\X\va_{#1}}
\def\vxc{\@ifnextchar[{\vxc@}{\bx^c}}
\def\vxd@[#1]{\bx^d\X\va_{#1}}
\def\vxd{\@ifnextchar[{\vxd@}{\bx^d}}
\def\vxe@[#1]{\bx^e\X\va_{#1}}
\def\vxe{\@ifnextchar[{\vxe@}{\bx^e}}
\def\Ar@[#1]#2#3{A^{(#1)#2}_{#3}(\vr)}
\def\Ar@@#1#2{A^{#1}_{#2}(\vr)}
\def\Ar{\@ifnextchar[{\Ar@}{\Ar@@}}
\def\As@[#1]#2#3{A^{(#1)#2}_{#3}(\vs)}
\def\As@@#1#2{A^{#1}_{#2}(\vs)}
\def\As{\@ifnextchar[{\As@}{\As@@}}
\def\Av@[#1]#2#3{A^{(#1)#2}_{#3}(\vv)}
\def\Av@@#1#2{A^{#1}_{#2}(\vv)}
\def\Av{\@ifnextchar[{\Av@}{\Av@@}}

\makeatother
\catcode`\@=12

\begin{document}


\thispagestyle{empty}
\renewcommand{\thefootnote}{\fnsymbol{footnote}}
\begin{flushright} hep-th/9505106 \\
                   DESY 95-092 \\
                   KCL-TH-95-3 \end{flushright}
\vspace*{2cm}
\begin{center}
{\LARGE \sc Multistring Vertices and \\[4mm]
            Hyperbolic Kac Moody Algebras%
            \footnote[1]{submitted to Int.\ J. Mod.\ Phys.\ A}}\\
 \vspace*{1cm}
       {\sl R. W. Gebert\footnote[2]{Supported by
        Deutsche Forschungsgemeinschaft under Contract No.\
        {\sl DFG Ni 290/3-1}.},  H. Nicolai\footnotemark[3],}\\
 \vspace*{2mm}
     IInd Institute for Theoretical Physics, University of Hamburg\\
     Luruper Chaussee 149, 22761 Hamburg, Germany\\
 \vspace*{4mm}
       {\sl and}\\
 \vspace*{4mm}
       {\sl P. C. West\footnote[3]{Supported by EU human capital and
        mobility program Contract No.\ {\sl ERBCHRXCT920069}}  }\\
 \vspace*{2mm}
     Department of Mathematics, King's College London\\
     Strand, London WC2R 2LS, Great Britain\\
 \vspace*{6mm}
\ifnum\draftcontrol=1{\LARGE \bf Draft version: \draftdate \\}
                     \else{May 17, 1995 \\}\fi

\vspace*{1cm}
\begin{minipage}{11cm}\footnotesize
  Multistring vertices and the overlap identities which they satisfy
  are exploited to understand properties of hyperbolic Kac Moody
  algebras, and $\0$ in particular. Since any such algebra can be
  embedded in the larger Lie algebra of physical states of an
  associated completely compactified subcritical bosonic string, one
  can in principle determine the root spaces by analyzing which
  (positive norm) physical states decouple from the $N$-string vertex.
  Consequently, the Lie algebra of physical states decomposes into a
  direct sum of the hyperbolic algebra and the space of decoupled
  states. Both these spaces contain transversal and longitudinal
  states. Longitudinal decoupling holds generally, and may also be
  valid for uncompactified strings, with possible consequences for
  Liouville theory; the identification of the decoupled states simply
  amounts to finding the zeroes of certain ``decoupling polynomials''.
  This is not the case for transversal decoupling, which crucially
  depends on special properties of the root lattice, as we explicitly
  demonstrate for a non-trivial root space of $\0$. Because the
  $N$-vertices of the compactified string contain the complete
  information about decoupling, all the properties of the hyperbolic
  algebra are encoded into them.  In view of the integer grading of
  hyperbolic algebras such as $\0$ by the level, these algebras can be
  interpreted as interacting strings moving on the respective group
  manifolds associated with the underlying finite-dimensional Lie
  algebras.
\end{minipage}
\end{center}
\renewcommand{\thefootnote}{\arabic{footnote}}
\setcounter{footnote}{0}
\newpage

\tableofcontents

\section{Introduction}
This paper brings together two lines of development both of which are
intimately connected to core issues of modern string theory, namely
the theory of indefinite and, more specifically, hyperbolic Kac Moody
algebras on the one hand, and (aspects of) string field theory on the
other. Little is known about the infinite-dimensional Lie algebras and
groups that have been suggested as candidates for a unified symmetry
of (super)string theory; similarly, it is far from assured that we are
already in possession of the proper physical concepts necessary to
deal with Planck scale physics and quantum gravity.  Nonetheless, the
developments of the past decade have taught us that progress in one of
these areas may well depend on advances in the other. The main purpose
of the present article is to show how methods developed in the early
days of string theory may provide a better understanding of the
mathematical structures encountered in the theory of indefinite Kac
Moody algebras.  More specifically, we will work out the connection
between hyperbolic Kac Moody algebras and multistring vertices, and in
particular make precise the sense in which the structure constants of
such algebras can be identified with the scattering amplitudes for the
physical states corresponding to the root space elements of the Lie
algebra (although this seems to be almost a folklore result in
connection with ``monstrous'' Lie algebras based on the lattice
$\III$, the point has never been made for hyperbolic Kac Moody
algebras to the best of our knowledge). In fact, we shall argue (and
offer concrete evidence) that the multistring vertices of a suitably
compactified bosonic string ``know everything'' there is to know about
certain hyperbolic Kac Moody algebras such as $\0$.

$N$-string vertices played an important role in the early history of
string theory following the discovery of the $N$-tachyon scattering
amplitude \ct{Vene68,KobNie69a}. While scattering amplitudes were
originally computed by evaluating products of suitable vertex
operators (such as $\:\exp (i\vv \X \vX) \:$ for the tachyon) between
two vacuum states in a one-string Fock space \ct{FubVen70} (for a
review see \ct{AlAmBeOl71}), it was soon realized that they could
alternatively be obtained by attaching the physical states involved in
the scattering to the legs of a multistring vertex acting on a
multistring Fock space.  The beginning of this development was the
discovery of a special three-vertex with the requisite properties in
\ct{Sciu69}; a slightly more symmetric, but physically equivalent
three-vertex was found shortly after in \ct{CaSchwVe69}. The problem
of constructing an $N$-vertex for arbitrary $N>3$ was solved in
\ct{Love70,Oliv71} by sewing $N-2$ three-vertices. Not only did these
$N$-vertices describe $N$-string scattering, but they provided
transparent proofs of various properties of the scattering amplitudes,
such as duality.  The formalism did, however, have some drawbacks: the
$N$ string vertices obtained by sewing were rather complicated and
their properties were not at all apparent. For instance, the action of
the Virasoro generators on the $N$-vertex was not properly understood;
furthermore, the investigation of loop effects in this framework led
to incorrect results as negative norm states were found to propagate
within the loops.

A new approach to multistring vertices was initiated in 1986
\ct{NevWes86b,NevWes86a,NevWes87a} and further developed in
\ct{NevWes88a,NevWes88b,West89}. It was shown there that $N$-vertices
could be directly and very explicitly determined without recourse to
sewing from a simple set of defining equations called overlap
equations. These come in two varieties, namely as unintegrated and
integrated overlaps; although the latter contain somewhat less
information than the unintegrated overlaps they are frequently more
useful for practical calculations as we will also see in this paper.
An especially appealing feature of this formalism is the elegant
geometrical interpretation of the overlap equations in terms of
coordinate patches associated with every string emitted from the
worldsheet; the $N$-vertex is, in fact, completely characterized by
the transition functions between these patches.  One finds in this way
an infinite class of vertices corresponding to different transition
functions, which are physically equivalent but differ off-shell. As an
added bonus, general properties of the multistring vertices could now
be fully elucidated. Furthermore, the multistring vertex for arbitrary
genus was found in \ct{NevWes88a} and ghosts were introduced in
\ct{NevWes86b,FreWes88}. The first of these papers on ghost vertices
enabled the Copenhagen group to refine and complete the original
sewing program \ct{DiVec-etal87a,Side87} and to derive a particularly elegant
expression for the bosonic loop measure \ct{DiVec-etal87b}. The
integrated overlap equations, which can be obtained from the overlap
equations by contour integration, were independently studied and
extensively applied in the context of the so-called Grassmannian
approach to string theory \ct{AlGoMoVa88}. For an account of the
various operator approaches and further references, see also
\ct{West89}.

Despite the well known and close links between string theory and Kac
Moody algebras \ct{BarHal71,Sega81,FreKac80,GodOli85} (see also
\ct{Kac90}, which is the standard textbook on the subject), it appears
that multistring technology has not really been exploited for its full
worth in the Lie algebra context, as much of the pertinent literature
deals solely with one-string Fock spaces and one-string vertex
operators. We will here demonstrate that multistring vertices may
serve as valuable tools for the analysis of the multiple commutators
arising in the theory of hyperbolic Kac Moody algebras, and that they
offer new and promising insights into their structure, which has so
far defied all attempts at a complete understanding. Multistring
vertices can in principle provide exhaustive information about the
root spaces of such algebras and may thus bring us closer to the
ultimate goal of finding a manageable representation of the Lie
algebra elements that would be analogous to the current algebra
representation for affine algebras, a task which is even harder than
the calculation of root multiplicities. To appreciate the difficulties
readers should recall that there is so far not a single example of a
hyperbolic Kac Moody algebra for which even the root multiplicities
are completely known. We will in this paper mainly rely on the
results of \ct{GebNic95}, where a DDF type approach adapted to the
root lattice was developed.  The crucial result invoked there is that
any hyperbolic Kac Moody algebra $\ggA$ can be embedded in a larger
Lie algebra of physical states $\ggL$ \ct{Borc86}, where $\L$ is the
root lattice associated with the Cartan matrix $A$.  Consequently, the
elements of $\ggA$ can be completely characterized as the orthogonal
complement in $\ggL$ of those states that cannot be reached by
multiple commutators of the basic Chevalley Serre generators.
Although the idea of defining the Kac Moody algebra by what is {\em
  not} in it rather than by what is in it seems completely
counterintuitive, it turns out, somewhat surprisingly, that these
``missing'' states (mostly referred to as ``decoupled states'' in this
paper) are easier to identify than the Lie algebra elements themselves
if astute use is made of multistring vertices: one must only check
that they decouple from the relevant $N$-vertex!  We emphasize that the
decoupling we are concerned with here is of a novel type, and not like
the decoupling of null states in $d=26$ string theory, in that the
decoupled states have {\em positive} norm (this casts some doubt on
the existence of a BRST-type cohomology, which would explain
decoupling).  As in \ct{GebNic95}, our principal example will be the
maximally extended hyperbolic Kac Moody algebra $\0$, even though many
of our results are valid more generally.

Our results also suggest a new interpretation of hyperbolic Kac Moody
algebras in physics: they describe a theory of interacting strings at
tree level, where the usual one-string Fock space is replaced by the
basic representation of the underlying affine Kac Moody subalgebra.
The basic representation in turn is nothing but the set of purely
transversal states built on the tachyonic groundstate $\ket{\vr_{-1}}$
and its orbit under the action of the affine Weyl group; here
$\vr_{-1}$ is the over-extended root, and the tachyonic momenta that
can be generated by the affine Weyl group are of the form $\vv =
\vr_{-1} + (\frc12 \vb^2)\vd + \vb$, where $\vd$ is the affine null
root, and $\vb$ has support on the finite subdiagram of the full
Dynkin diagram obtained by deleting the extended and overextended
nodes.  We note that our results in some sense constitute a
realization of a remark in \ct{Witt86}, where it was proposed to
interpret $\0$ as a symmetry acting on a multistring Fock space.  As
pointed out there, the group corresponding to $\0$ would admit $\9
\times U(1)$ as a subgroup, where the $U(1)$ factor is generated by
the central charge $c$ (it is hard to imagine what the group would
look like if one does not even know the Lie algebra; see, however,
\ct{Tits88} for some further information on this topic).  Accordingly,
the central charge $c$, whose integer eigenvalues are called the
``levels'' of the relevant representations of the affine subalgebra,
should be viewed as an operator counting the number of strings. If we
accept this interpretation, the following picture emerges.  As shown
in \ct{GepWit86}, the allowed one-string Fock spaces of a bosonic
string, the ``internal'' part of which propagates on a Lie group
manifold, constitute integrable highest-weight representations of the
associated affine algebra.  But since the basic representation is the
simplest of these we can think of the hyperbolic algebra as a
multistring Fock space for interacting strings moving on the group
manifold corresponding to the underlying finite Lie algebra (in which
case the target space contains no uncompactified spacetime part). The
Chevalley involution $\th$ (cf.\ Eq.\ \Eq{Chevinv}) would then play
the role of a charge conjugation operator, exchanging strings and
``antistrings'' (i.e.\ representations and contragredient
representations). Note that we employ two different string
compactifications. We start with a (subcritical) string which is
completely compactified on a Minkowski torus and into whose physical
state space the hyperbolic Kac Moody algebra is embedded as a
subalgebra; this model is merely needed in order to have an explicit
realization of the algebra and has no immediate physical
interpretation. Rather, the graded structure of the hyperbolic algebra
is tied to the interacting string picture where now the strings live
on a Lie group manifold.

For evident reasons the hyperbolic algebra $\0$ has also been advanced
as a candidate symmetry of the superstring in ten dimensions. Despite
the considerable efforts invested into the search for a proper
formulation of string field theory, however, it has not been possible
so far to exhibit a multistring Hamiltonian (or, rather, a
Wheeler-DeWitt type constraint operator) for the superstring in ten
dimensions which would admit such a symmetry and commute with it.
Besides, one would naively expect the relevant symmetry algebra to be
a superalgebra, not a bosonic Lie algebra. For a superalgebra, one
must take into account additional roots of length one on the root
lattice for the fermionic generators \ct{GodOli85}, but enlarging the
even selfdual $\0$ root lattice $\II$ in this way would put a stain on
its beauty in our opinion. To construct a superalgebra one could
conceivably compactify the string on the unique odd selfdual
Lorentzian lattice $I_{9,1}\equiv\Z^{9,1}$; see \ct{Borc92} for some
comments and speculations. But since $I_{9,1}$ does not contain $\II$
as a sublattice, $\0$ could certainly not emerge as a bosonic
subalgebra in such a construction. A related question is whether there
exists a (different?) superalgebra which can be embedded into the
space of bosonic and fermionic physical states of the fully
compactified superstring in the same way as $\0$ is embedded into the
space of physical states of a bosonic string. Curiously, such an
algebra would have no real roots at all due to the absence of
tachyonic states in the superstring. Hence, neither superalgebra would
deserve to be called ``super-$\0$'' because neither would contain $\0$
as a bosonic subalgebra\footnote{It is known that the {\em
    finite}-dimensional exceptional Lie algebras of $E$-type possess
  no superextensions \ct{Kac77}.}! Therefore, a more attractive
possibility from our point of view is that there is actually no need
for a superalgebra because $\0$ may already be secretly ``aware of''
supersymmetry even though it is a bosonic symmetry by all appearances.
That this is not an entirely far-fetched idea was explicitly shown in
the context of two-dimensional Poincar\'e supergravities, whose local
supersymmetry transformations can be ``bosonized'' into certain Kac
Moody gauge transformations in the associated linear systems
\ct{NicWar89,Nico94}. It would, however, mean that the theory
appropriate for $\0$ is not the superstring, but some other theory
(perhaps involving supergravity in eleven dimensions \ct{CrJuSch78}).

The implications of our results for string field theory and the
possible relation to quantum Liouville theory remain fascinating
topics for future research. Namely, the decoupling of certain positive
norm physical states holds not only on the lattice, but also in the
continuum. This raises the question of whether systematically
discarding these states can lead to a consistent quantum theory in the
``forbidden zone'' $1<d<25$. Furthermore, we suspect that radiative
(string loop) corrections will also have a role to play in the context
of hyperbolic Kac Moody algebras, although at our present level of
understanding it is not at all clear where they might enter. In
particular, it would be necessary to find out whether these
corrections are compatible with our decoupling mechanism; in this
case, the vertices with loops will almost certainly contain
information about the $\0$ states at arbitrary level, as the latter
would propagate inside loops.  The multistring description also
suggests a subtle link between the worldsheet topology (i.e.\ its
genus and punctures, see \ct{Mand73,GidWol87} for a nice description)
and the level which grades the hyperbolic algebra, indicating that
hyperbolic symmetries might naturally involve all orders of string
perturbation theory.

Let us now briefly summarize the content of this paper.

Section \ref{VERTEXOP} is devoted to an exposition of the one-string
model. We have decided to work within the modern framework of vertex
algebras since it makes the Lie algebra structure of the space of
physical string states especially transparent. After a short review of
vertex algebras we shall present in detail the string model relevant
for the paper, i.e.\ the closed bosonic in $d<26$ dimensions with {\it
  all} target space coordinates toroidally compactified. In
particular, the relation of the Lie algebra of physical states and the
embedded Kac Moody algebra will be worked out. The difference between
the string scalar product and the invariant bilinear form for the Lie
algebra will be explained.

In Sect.\ \ref{VERTICES} we will deal with the multistring formalism.
Starting from the operatorial definition of $N$-string vertices via
overlap identities it will be shown how to obtain their explicit form
in the usual oscillator representation. Then the conformal
transformation properties of the vertices will be derived.

The content of the first two sections will be brought together in
Sect.\ \ref{THREEVERTEX} where set up a correpondence between
one-string vertex operators and three-vertices taking into account
some additional physical assumptions. Then the machinery of sewing
three-vertices is presented. With this method at hand we subsequently
discuss the notions of locality and duality for the four-vertex and
their generalizations to $N$-vertices. Finally two specific
three-vertices are explicitly constructed, one of which leading to a
vertex operator that satisfies all the axioms of a vertex algebra. In
particular, we find a rather simple, diagrammatical proof of the
Jacobi identity. Moreover, we establish the equivalence of the overlap
identities for this three-vertex and the so-called Jacobi identity for
intertwining operators.

Section \ref{E10} contains the core results of this paper. Making
extensive use of the overlap equations, we discuss the decoupling of
longitudinal and transversal states in detail, and show how root
spaces can be determined without explicit computation of commutators.
On the way, structure constants for $N$-fold commutators will be
introduced and it will be explained in which sense the structure
constants of hyperbolic algebras can be identified with the scattering
amplitudes for the underlying compactified string. These issues will
be addressed in Sect.\ \ref{E10-1}, where we describe in detail how to
rewrite (multiple) commutators in terms of $N$-string vertices.
Section \ref{E10-2} summarizes the essential results of \ct{GebNic95}
needed here and describes how to recover them in the multistring
formalism.  Section \ref{E10-3} is devoted to an analysis of the
longitudinal decoupling; there we will introduce the ``decoupling
polynomials'' from which one can directly read off which longitudinal
states couple and which decouple, and compute these polynomials up to
degree seven. These results go considerably further than the ones
previously obtained. In Sect.\ \ref{E10-4} we deal with the decoupling
of transversal states. As this is the least understood part of our
construction we will make no attempt at a systematic treatment but
rather illustrate the phenomenon in terms of a concrete (and
non-trivial) example.

\section{The vertex algebra associated with the compactified string}
\lab{VERTEXOP}
We shall study a chiral sector of a closed bosonic string living on a
Minkowski torus as spacetime, i.e.\ we compactify all target space
coordinates. Consequently, the center of mass momenta of the string
form a lattice with Minkowski signature. It turns out that the states
and vertex operators realize a mathematical structure called vertex
algebra \ct{Borc86}. The latter framework will be convenient to
establish the connection between string vertices and multiple
commutators. In our exposition we will closely follow the review
\ct{Gebe93}.
\subsection{Vertex algebras} \lab{VERTEXOP-1}
In the vertex algebra formalism one works with {\it formal} variables
$z$, $w$, $y$, $\ldots$ and {\it formal} Laurent series. For a
vector space $S$, say, we set
$S[\![z,z^{-1}]\!]=\left\{\sum_{n\in\Z}s_nz^{-n-1}|s_n\in S\right\}$.
For a formal series we use the following residue notation:
\[ \res_{z}\bigg[\sum_{n\in\Z}s_nz^{-n-1}\bigg]=s_0, \lb{Res} \]
so that we may think of $\Res{z}{\ldots}$ as the formal expression for
the operation $\oint_0\frac{dz}{2\pi i}[\ldots]$ in complex analysis.
We shall also need the analogue of the $\d$ function which is
defined as $\d(z):=\sum_{n\in\Z}z^n$. Formally, this is the Laurent
expansion of the classical $\d$ function at $z=1$ and indeed it enjoys
similar properties. Formal calculus is presented in great detail in
\ct{FLM88}.
\begin{defi} \lab{def1} \hh{1em} \\[1.5ex]
A {\bf vertex algebra} is a $\Z$-graded vector space
\[ \cF=\bigoplus_{n\in\Z}\Fz{n}, \]
equipped with a linear map $\cV:\cF\to(\End\cF)[\![z,z^{-1}]\!]$, which
assigns to each state $\p\in\cF$ a {\bf vertex operator} $\Vp$, and
the vertex operators satisfy the following axioms:
\ben
\item {\bf (Regularity)} If $\p,\f\in\cF$ then
\[ \Res{z}{z^n\Vp\f}=0 \FOR{ n sufficiently large} \lb{Reg} \]
and n depending on $\p$ and $\f$.
\item {\bf (Vacuum)} There is a preferred state $\one\in\cF$, called the
vacuum, satisfying
\[ \cV(\one,z)=\id_\cF. \lb{Vac} \]
\item {\bf (Injectivity)} There is a one-to-one correspondence between
states and vertex operators:
\[ \Vp=0\quad\iff\quad\p=0. \lb{Inj} \]
\item {\bf (Conformal vector)}
There is a preferred state $\bw\in\cF$, called the
conformal vector, such that its vertex operator,
\[ \cT(z)\equiv\cV(\bw,z)=\sum_{n\in\Z}\Ln z^{-n-2}, \lb{Con} \]  \ben
\item gives the {\bf Virasoro algebra} with some central charge
$c\in{\bbl R}$,
\[ [\rL{m},\Ln]=(m-n)\rL{m+n}+\frac c{12}(m^3-m)\d_{m+n,0}; \lb{Vir} \]
\item provides a {\bf translation generator}, $\rL{-1}$,
\[ \cV(\rL{-1}\p,z)=\dz\Vp \FOR{ every }\p\in\cF; \lb{Tra} \]
\item gives the grading of $\cF$ via the eigenvalues of $\Lo$,
\[ \Lo\p=n\p\equiv h_\p\p\FOR{ every }
          \p\in\Fz{n},n\in\Z; \lb{Wei} \]
the eigenvalue $ h_\p$ is called the {\bf (conformal) weight} of
$\p$.
\een
\item {\bf (Jacobi identity)} For every $\p,\f\in\cF$,
\[ \lefteqn{\Dm yzw \VP{z}\VF{w} \Di ywz \VF{w}\VP{z}} \hspace{8mm} \non
   & & =\Dm wzy \VV{y}, \hspace*{40mm} \lb{Jac} \]
where binomial expressions have to be expanded in nonnegative
integral powers of the second variable.
\een \end{defi}
Since vertex operators are operator-valued formal Laurent series, we
can give an alternative formulation (see e.g.\ \ct{Borc86}) of the
axioms of a vertex algebra using the mode expansion
\[ \Vp=\sum_{n\in\Z}\p_nz^{-n-1}. \lb{Modeex} \]
{}From the axioms one can derive some important properties of vertex
algebras. For example, the Jacobi identity implies the relations
\[ [\rL{-1},\Vp] &=& \dz\Vp, \lb{Trans-com} \\ {}
   [\Lo,\Vp] &=& \left(z\dz+h_\p\right)\Vp
                  \IF{}\p\in\Fz{h_\p}, \lb{Scale-com} \]
which give, respectively,
\\[1ex]
{\bf (Translation property)}
\[ \re{x\rL{-1}} \Vp \re{-x\rL{-1}} = \cV(\p,z+x), \lb{Trans-prop} \]
{\bf (Scaling property)}
\[ \re{x\Lo} \Vp \re{-x\Lo} = \re{xh_\p}\cV(\p,\re{x}z)
   \IF{}\p\in\Fz{h_\p}, \lb{Scale-prop} \]
for every $x\in w\C[\![w]\!]$.
Thus $\rL{-1}$ and $\Lo$ generate translations and scale
transformations, respectively.

Combining injectivity with the Jacobi identity one arrives at
\\[1ex]
{\bf (Skew symmetry)}
\[ \Vp\f=\re{z\rL{-1}}\cV(\f,-z)\p, \lb{Skew-prop} \]
which shows that the vertex operator $\Vp$, when applied to the
vacuum, ``creates'' the state $\p\in\cF$\ :
\[ \Vp\one=\re{z\rL{-1}}\p. \lb{Create-prop} \]
We shall denote by $\Pz{h}\subset\Fz{h}$ the space of {\bf (conformal)
highest weight vectors {\rm or} primary states} satisfying
\[ \Lo\p &=& h\p, \lb{Phys-cond1} \\
   \Ln\p &=& 0\qquad\forall n>0. \lb{Phys-cond2} \]
For example, in any vertex algebra the vacuum is a primary state of
weight 0 and therefore $\Fz0$ is always at least one-dimensional.
We can deduce from the Jacobi identity that, for $\p\in\Pz{h}$,
\[ [\Ln,\Vp] = z^n\left\{z\dz+(n+1)h\right\}\Vp
               \qquad\forall n\in\Z, \lb{Primary} \]
or, in terms of mode operators,
\[ [\Ln,\p_m] = \{(h-1)(n+1)-m\}\p_{m+n}, \lb{Primary-mode} \]
i.e.\ $\Vp$ is a so-called {\bf(conformal) primary field} of weight
$h$. From now on we shall use the notation $\F(z)$ for a primary
field. Upon exponentiation we find that
\\[1ex]
{\bf (Projective change)}
\[ \re{x\Ln} \F(z) \re{-x\Ln}
   =\left(\frac{\partial p(z)}{\partial z}\right)^h \F\big(p(z)\big)
    \qquad\forall n\neq0, \lb{Proj-prop} \]
for every $x\in w\C[\![w]\!]$ where $p(z)=z(1-nxz^n)^{-1/n}$.

Since the operators $\{\rL{-1},\Lo,\rL1\}$ satisfy the
${\got su}(1,1)$ Lie algebra
\[ [\Lo,\rL1]=-\rL1,\qquad[\Lo,\rL{-1}]=\rL{-1},\qquad[\rL1,\rL{-1}]=2\Lo, \]
we have the following M\"obius transformation properties of the
vertex operators (see also \ct{Godd89b}):
If $\p\in\cF$ is a {\bf quasiprimary state} of weight $h$, i.e.\
$\p$ satisfies $\Lo\p=h\p$ and $\rL1\p= 0$, then
\[ \Mh\Vp\Mh^{-1}
   =\left[\dd{\M(z)}{z}\right]^h\cV(\p,\M(z)), \lb{Moeb-trans} \]
where
\[ \M(z):=\frac{az+b}{cz+d},\qquad
   \Mh:=\re{\frac bd\rL{-1}}\left(\frac{\sqrt{ad-bc}}d\right)^{2\Lo}
            \re{-\frac cd\rL1}, \lb{Moeb-form} \]
for $a,b,c,d\in w\C[\![w]\!]$.

For our purposes it will be very useful to have a relation between the
Jacobi identity for vertex algebras and the notions of locality and
duality from conformal field theory. This can be achieved by
considering matrix elements of products of vertex operators w.r.t.\
some nondegenerate bilinear form $\vev{\_}{\_}$.
It follows from the axioms that any matrix element of
a vertex operator is a Laurent {\it polynomial} in $z$,
\[ \vev{\chi}{\Vp\f}\in\C[z,z^{-1}]
   \qquad\forall\ \chi,\p,\f\in\cF, \lb{3-point} \]
so that these three-point correlation functions may be regarded as
meromorphic functions of $z$. Of course, we formally identify
$\chi$ with an ``out-state'' inserted at $z=\infty$ and
$\f$ with an ``in-state'' inserted at $z=0$.

We recall that in the Jacobi identity axiom \Eq{Jac} we used
the terminology `expansion in nonnegative integral powers of the second
variable'. We have to make this convention mathematically more precise.
Let $\C[z,w]_S$ denote the subring of the field of rational functions
which can be written as
\[ f(z,w) = \frac{g(z,w)}{w^s  \prod_{l=1}^r (a_l z + b_l w )} \lb{fzw} \]
where $g(z,w)$ is any polynomial in the formal variables $z$ and $w$;
$r,s \in \Nat , a_l \neq 0$ for $l=1,...,r$. We can define two
embeddings of such rational functions into the space
$W [\![ z,z^{-1},w,w^{-1}]\!]$ of formal Laurent series in $z$ and $w$.
The first of these is designated as
$\iota_{zw}$ and obtained by expanding the product factors in \Eq{fzw}
in non-negative integral powers of the variable $w$.
The second, denoted by $\iota_{wz}$, is obtained by pulling out
the $z$ factors in the denominator of \Eq{fzw} instead
and expanding in non-negative powers
of $z$. We note that even for the same
rational function $f(z,w)$ the formal power series $\iota_{zw} f$
and $\iota_{wz}f$ will differ as elements of
$W[\![ z,z^{-1},w,w^{-1}]\!]$ unless $f$ is a Laurent polynomial.

We have the following important theorem due to \ct{FLM88}.

\begin{theo} \lab{thm1} \hh{1em} \\[-2.5ex]  \ben
\item {\bf (Locality $\equiv$
          rationality of products $+$ commutativity)} \\
For $\chi,\p,\f,\xi\in\cF$, the formal series
$\vev{\chi}{\VP{z}\VF{w}\xi}$ which involves only finitely many negative
powers of $w$ and only finitely many positive powers of $z$, lies
in the image of the map $\iota_{zw}$:
\[ \vev{\chi}{\VP{z}\VF{w}\xi}=\iota_{zw}f(z,w), \lb{Locality-1} \]
where the (uniquely determined) element $f\in\C[z,w]_S$
is of the form
\[ f(z,w)=\frac{g(z,w)}{z^rw^s(z-w)^t} \]
for some polynomial $g(z,w)\in\C[z,w]$ and $r,s,t\in\Nat$.
We also have
\[ \vev{\chi}{\VF{w}\VP{z}\xi}=\iota_{wz}f(z,w), \lb{Locality-2} \]
i.e.\ $\VP{z}\VF{w}$ agrees with $\VF{w}\VP{z}$ as operator-valued
rational functions.
\item {\bf (Duality $\equiv$
          rationality of iterates $+$ associativity)} \\
For $\chi,\p,\f,\xi\in\cF$, the formal series
$\vev{\chi}{\VV{y}\xi}$ which involves only finitely many negative powers
of $y$ and only finitely many positive powers of $w$, lies in the
image of the map $\iota_{wy}$:
\[ \vev{\chi}{\VV{y}\xi}=\iota_{wy}f(y+w,w), \lb{Duality-1} \]
with the same $f$ as above, and
\[ \vev{\chi}{\cV(\p,y+w)\VF{w}\xi}
    =\iota_{yw}f(y+w,w), \lb{Duality-2} \]
i.e.\ $\VP{z}\VF{w}$ and $\VV{z-w}$ agree
as operator-valued rational functions, where the right hand expression
is to be expanded as a Laurent series in $z-w$.
\een \end{theo}
For a proof see \ct{FLM88} or \ct{FrHuLe93}.

The first part of the theorem in particular states that these
matrix elements may be viewed as meromorphic functions of the
formal variables. In other words, for \Eq{Locality-1} and \Eq{Locality-2}
there exist meromorphic functions
of $z$ and $w$ which upon expansion in $z$ or $w$ agree with the formal
power series on the left hand side in their respective domains of
convergence. Thus vertex algebras can be seen as a rigorous
formulation of meromorphic conformal field theories.
Note that the second part of the theorem should be interpreted as
crossing symmetry of the four-point correlation function on
the Riemann sphere. It establishes a precise formulation of an
operator product expansion in two-dimensional conformal field theory
(see e.g.\ \ct{LueMac76}, \ct{BePoZa84}, \ct{Gins89}, \ct{Godd89b})
in the sense that $\VP{z}\VF{w}$ agrees with
$\sum_{n\in\Z}(z-w)^{-n-1}\cV(\p_n\f,w)$ as an
operator-valued rational function. The theorem then also ensures
that this operator product expansion involves only finitely many
singular (at ``$z=w$'') terms.

For establishing the Jacobi identity in concrete models the following
theorem is particularly useful. (For a proof see \ct{FrHuLe93})

\begin{theo} \lab{thm2} \hh{1em} \\[-2.5ex]  \ben
\item The Jacobi identity follows from locality and duality.
\item In the definition of a vertex algebra the Jacobi identity
may be replaced by the principle of locality, Eq.\ \Eq{Trans-com}
and Eq.\ \Eq{Scale-com}.
\een \end{theo}

If we regard our formal variables as {\it complex} variables, then the
formal expansions of rational functions that we have been discussing
converge in suitable domains. The matrix elements in Eq.\
\Eq{Locality-1} and Eq.\ \Eq{Locality-2} converge to a common rational
function in the disjoint domains $|z|>|w|>0$ and $|w|>|z|>0$,
respectively. The matrix elements in Eq.\ \Eq{Duality-1} and Eq.\
\Eq{Duality-2} for $y=z-w$ converge to a common rational
function in the domains $|w|>|z-w|>0$ and$|z|>|w|>0$,
respectively, and in the common domain $|z|>|w|>|z-w|>0$
these two series converge to the common function.

We will now briefly discuss the issue of invariant bilinear form and
adjoint vertex operator for a vertex algebra.
Define the {\bf restricted dual} of $\cF$,
\[ \cF^\prime\equiv\bigoplus_{n\in\Z}(\Fz{n})^*, \]
the direct sum of the dual spaces of the homogeneous subspaces $\Fz{n}$,
i.e.\ the space of linear functionals on the vertex algebra $\cF$
vanishing on all but finitely many $\Fz{n}$. Suppose we have a
grading-preserving linear isomorphism
$\R:\cF\to\cF^\prime$, $\p\mapsto \R(\p)\equiv\p^*$, satisfying
$\R\cc\p_n=\p^*_n\cc\R$, then this
amounts to choosing a nondegenerate bilinear form $(\_|\_)$ on $\cF$
as $(\chi|\f):= \dual{\R(\chi)}{\f}$ where
$\dual{\_}{\_}$ denotes the natural pairing between $\cF$ and $\cF^\prime$.
We may define the {\bf adjoint vertex
operator} w.r.t.\ the bilinear form by stipulating
$(\Vp\chi|\f)=(\chi|\Vs(\p,z)\f)$ and putting (cf.\ \ct{FrHuLe93})
\[ \Vs(\p,z)
   &=&\sum_{n\in\Z}\p_n\sh z^{-n-1} \non
   &:=&\cV\left(\re{z\rL1}(-z^{-2})^{\Lo}\p,z^{-1}\right)
       \in(\End\cF)[\![z,z^{-1}]\!]. \lb{vertop-sh} \]
We observe that the expression on the right hand side
in general is not summable unless $(\rL1)^n\p=0$ for $n$ large
enough, which is assured if the spectrum of $\Lo$ is bounded from below. For
the string moving on a Minkowski torus the spectrum of $\Lo$ is
unbounded. But for our purposes it is sufficient to know the adjoint
vertex operator associated with quasiprimary states, and for these
states the above definition indeed makes sense:
\[ \Vs(\p,z)=(-1)^{h_\p} z^{-2h_\p}\cV(\p,z^{-1})
       \qquad\mbox{for $\p$ quasiprimary}, \]
or, in terms of the mode expansion \Eq{Modeex},
\[ \p_n\sh=(-1)^{h_\p}\p_{-n+2h_\p-2}. \lb{Modesh} \]
The Virasoro generators satisfy the relation $\Ln\sh=\rL{-n}$ (note
the special mode expansion \Eq{Con} for the Virasoro generators!) or,
in terms of the ``stress--energy tensor'',
$\cT\sh(z)=\frac1{z^4}\cT(z^{-1})$.  From this it is easy to see that
the homogeneous subspaces $\Fz{n}$, $n\in\Z$, are orthogonal to each
other with respect to this bilinear form,
\[ (\Fz{m}|\Fz{n})=0\IF{}m\neq n. \]
One can also show that the bilinear form is symmetric and that the
conjugation operation $\sharp$ satisfies $\cV^{\sharp\sharp}=\cV$
(for proofs see \ct{FrHuLe93}).

We shall provide a certain subspace of the Fock space $\cF$ with the
structure of a Lie algebra (cf.\ \ct{Borc86}, \ct{FLM88}).
We define a bilinear product on $\cF$ by
\[ [\p,\f]:=\Res{z}{\Vp\f}\equiv\p_0\f, \lb{Lie-com} \]
which is antisymmetric on the quotient space \FLie due to the
skew symmetry property \Eq{Skew-prop}. By considering the mode
expansion of the Jacobi identity we get
$\p_0(\f_0\x)-\f_0(\p_0\x)=(\p_0\f)_0\x$.
But this equation translates precisely into the classical Jacobi
identity for Lie algebras,
\[ [[\p,\f],\x]+[[\f,\x],\p]+[[\x,\p],\f]=0, \lb{Lie-jac} \]
on \FLie. Hence, \Eq{Lie-com} defines a Lie bracket on this quotient space.
At the level of vertex operators, dividing out the subspace
$\rL{-1}\cF$ reflects the fact that the zero mode $\p_0$ of a vertex
operator $\Vp$ remains unchanged when a total derivative is added to
$\Vp$; for if \( \p= \rL{-1}\f\in \rL{-1}\cF \) for some
$\f\in\cF$, then \( \p_0= \Res{z}{\dz\Vf}= 0 \) by \Eq{Tra}
and \Eq{Res}.
If one prefers a representation of a Lie algebra in terms of linear
operators we point out that the operator $\p_0=\Res{z}{\Vp}$ is just
the adjoint action of $\p$ on \FLie, viz.
\[ {\rm ad}_\p(\f)=[\p,\f]=\p_0(\f). \]
In other words the Lie algebra of zero mode operators,
$\{\p_0\|\p\in\cF\}$, is the adjoint representation of \FLie.
Consequently, the above Jacobi identity establishes the homomorphism
property of the map $\FLie\to\End\cF$, $\p\mapsto\p_0$:
\[ {\rm ad}_{[\p,\f]}=([\p,\f])_0
                     =[\p_0,\f_0]
                     =[{\rm ad}_\p,{\rm ad}_\f]. \]

The Lie algebra \FLie will be too large for our purposes.
In string theory, a distinguished role
is played by the primary states of weight $h=1$, which we shall call
{\bf physical states} from now on. In fact, we learn from Eq.\
\Eq{Primary-mode} that for a physical state $\p$ the corresponding zero
mode operator $\p_0$ commutes with the Virasoro algebra thereby
preserving all subspaces $\Pz{n}$ of primary states of weight $n$.
In particular, it maps physical states into physical states, i.e.\
$[\Pz1,\Pz1]\subset\Pz1\mod\rL{-1}\Pz0$.
Hence we will be mainly interested in the {\bf Lie algebra of
physical states},
\[ \ggF:=\PLie, \lb{Lie-def} \]
where we used the fact that
\[ \rL{-1}\Fz0\cap\Pz1=\rL{-1}\Pz0 \]
in any vertex algebra (see \ct{Gebe94} for the proof).

For the benefit of readers not familiar with the abstract formalism
of vertex algebras we pause to briefly explain the connection between
the formula \Eq{Lie-com} and the definition used in
\ct{GodOli85}. There the commutator of two integrated vertex operators
is defined by means of the following contour integrals
\[ [\p_0 , \f_0 ] :=
   \Oint_{|z|>|w|} \frac{dz}{2\pi i} \VP{z}
   \Oint_{0} \frac{dw}{2\pi i} \VF{w} -
   \Oint_{|w|>|z|} \frac{dw}{2\pi i} \VF{w}
   \Oint_{0} \frac{dz}{2\pi i} \VP{z}.  \lb{GO-com1}     \]
The arrangement of contours is important here because
the operator products must be expanded in their respective
domains of convergence for the above expressions to make sense;
the two expansions correspond precisely to the operations
$\iota_{zw}$ and $\iota_{wz}$ in the calculus of formal power series
introduced above. To evaluate \Eq{GO-com1} we recall the locality
property of vertex operators, i.e.\ $\VP{z}\VF{w}$ and $\VF{w}\VP{z}$
are the same as operator-valued rational functions. Hence we may
deform the contour such that the two contributions can be combined
into a single one to obtain \ct{GodOli85}
\[ [\p_0 , \f_0 ] =
  \Oint_{0} \frac{dw}{2\pi i} \Oint_{w} \frac{dz}{2\pi i}
    \VP{z} \VF{w}, \lb{GO-com2} \]
where the integral over $z$ is to be performed with a small contour
encircling the point $w$. By duality (i.e.\ Part 2 of Theorem
\ref{thm1}), this expression can be rewritten as
\[ [\p_0 , \f_0 ]
    &=& \Oint_{0} \frac{dw}{2\pi i} \Oint_{w} \frac{dz}{2\pi i}
        \VV{z-w} \non
    &=& \Oint_{0} \frac{dw}{2\pi i} \Oint_{0} \frac{dz}{2\pi i}
        \VV{z}.    \lb{GO-com3}     \]
Hence the resulting vertex operator coincides
with the operator $(\p_0 \f )_0$ corresponding to the state \Eq{Lie-com},
and therefore the definition \Eq{Lie-com} and the commutator
prescription are entirely equivalent.

The problem of finding an invariant bilinear form for the Lie algebra
of physical states is resolved quite elegantly; for it turns out that
the bilinear form $(\_|\_)$ defined above gives rise to an invariant
bilinear form on $\ggF$. We first have to convince ourselves that the
form on $\cF$ projects down to a well-defined form on $\ggF$. Indeed,
since $\rL{-1}\sh= \rL1$ we have $(\rL{-1}\chi|\f)= 0$ for any
quasiprimary state $\f$, i.e.\ the space $\rL{-1}\cF$ is orthogonal
to all quasiprimary states and thus deserves to be called null space.
In particular, $\rL{-1}\Pz0$ consists of {\bf null physical states},
physical states orthogonal to all physical states including
themselves. Hence the Lie algebra $\ggF$ is obtained from $\Pz1$ by
dividing out (unwanted) null physical states. Recall that when
defining the Lie algebra \FLie we had to divide out the space
$\rL{-1}\cF$ for mathematical reasons. But with the bilinear form at
hand we are now led to a physical interpretation of that maneuver.

It is well known that there are additional null physical states in
$\Pz1$ if and only if the central charge takes the critical value
$c=26$, namely the space $(\rL{-2}+\frac32\rL{-1}^2)\Pz{-1}$ (see
\ct{GSW88} for the calculations). The existence of these additional
null physical states is used in the proof of the no-ghost theorem
\ct{GodTho72} which we shall refer to in the last chapter when we
exploit the DDF construction.

As already mentioned, the adjoint vertex operator in
Eq.\ \Eq{vertop-sh} is summable for any quasiprimary state irrespective of
the spectrum of $\Lo$. Together with Eq.\ \Eq{Modesh}, this implies
that the form always satisfies
$(\chi|\p_n\f)=-(\p_{-n}\chi|\f)$ for physical states $\chi$, $\p$,
$\f$. Specializing to $n=0$ we indeed recover the invariance property:
\[ ([\chi,\p]|\f)=(\chi|[\p,\f])\qquad\forall\chi,\p,\f\in\ggF.
   \lb{invform} \]
If we put $n=1$ and $\chi=\one$, then we obtain
\[ (\p|\f)=(\one|-\p_1\f)\qquad\forall\p,\f\in\ggF, \]
so that we can think of the projection of $-\p_1\f$ on the vacuum
as the bilinear form.

\subsection{Toroidal compactification of the bosonic string} \lab{VERTEXOP-2}
It is by no means obvious that nontrivial examples of vertex
algebras exist. However, an important class of vertex algebras is
provided by the following result, which was announced in \ct{Borc86}
and was proved in \ct{FLM88}.

\begin{theo} \lab{thm3} \hh{1em} \\[1ex]
Associated with each nondegenerate even lattice $\L$
there is a vertex algebra.
\end{theo} \medskip

In fact, the above examples of vertex algebras gave rise to the very
notion and the abstract definition of vertex algebras. As we shall see
below, the physics described by these vertex algebras is the chiral
sector of a first quantized closed bosonic string moving on a
spacetime torus.  This section will be concerned with the explicit
construction of the vertex algebra stated above and the discussion of
the Lie algebra of physical states with the invariant and covariant
bilinear forms.  For further details, the reader may also wish to
consult the articles \ct{GodOli85}, \ct{Godd86} and \ct{GodOli86} or
the comprehensive review \ct{LeScheWa89}.

Let $\L$ be an even lattice of rank $d<\infty$ with a symmetric
nondegenerate $\Z$-valued $\Z$-bilinear form $\_\X\_$ and
corresponding metric tensor $\y^{\m\n}$, $1\le\m,\n\le d$ ($\L$ even
means that $\vr^2 \in2\Z$ for all $\vr\in\L$). The vertex algebra
which we shall construct can be thought of as a chiral sector of a
first quantized closed bosonic string theory with $d$ spacetime
dimensions compactified on a torus. Thus $\L$ represents the allowed
momentum vectors of the theory.

Introduce {\bf oscillators} $\amm$, $m\in\Z$, $1\le\m\le d$, satisfying
the commutation relations of a $d$-fold {\bf Heisenberg algebra},
\[ [\amm,\ann]=m\y^{\m\n}\delta_{m+n,0}, \]
and {\bf zero mode states} $\Pr$, $\vr\in\L$. We want the latter to
carry momentum $\vr$ and to be annihilated by the positive oscillator
modes, i.e.
\[ \amm\Pr&=&0\IF{}m>0, \\ p^\m\Pr&=&r^\m\Pr, \]
where $p^\m\equiv\a^\m_0$ denotes the center of mass momentum operator
for the string and $r^\m$ are the components of $\vr\in\L$. While the
operators $\amm$ for $m>0$ by definition act as annihilation
operators, the operators $\amm$ for $m<0$ will be called creation
operators, since they generate an irreducible Heisenberg module
$\cF^{(\vr)}$ with highest weight $\vr\in\L$ from any ground state
$\Pr$.

We take $\LR:=\Rea\XO_{\Z}\L$ to be the real vector space in which $\L$
is embedded, and for notational convenience we define
\[ \vr(m):=\sum_{\m=1}^dr_\m\amm\equiv\vr\X\va_m \]
for $\vr\in\LR$, $m\in\Z$, such that
\[ [\vr(m),\vs(n)]=m(\vr\X\vs)\delta_{m+n,0}, \lb{osccom} \]
with the $\Z$-bilinear form on $\L$ to be extended to an $\Rea$-bilinear
form on $\LR$. We denote the $d$-fold Heisenberg algebra spanned by
the oscillators by
\[ \hat{\bf h}:=\{\vr(m)\|\vr\in\LR,m\in\Z\}\oplus\Rea\X1, \]
and for the vector space of finite products of creation operators
($\equiv$ algebra of polynomials on the negative oscillator modes) we
write
\[ S(\hat{\bf h}^-):=
   \bigoplus_{N\in\Nat}\bigg\{\prod_{i=1}^N\vr_i(-m_i)\|
                    \vr_i\in\LR,m_i>0\ {\rm for}\ 1\le i\le N\bigg\}, \]
where ``$S$'' stands for ``symmetric'' because of the fact that the
creation operators commute with each other. Hence the Heisenberg module
built on some ground state $\Pr$ is given by $\cF^{(\vr)}=
S(\hat{\bf h}^-)\Pr$.

If we formally introduce center of mass position operators $q^\m$,
$1\le\m\le d$, commuting with $\amm$ for $m\neq0$ and satisfying
\[ [q^\n,p^\m]=i\y^{\m\n}, \]
then we find that
\[ \re{i\vr\.\vq}\Ps=\P_{\vr+\vs}, \]
i.e.\ the zero mode states can be generated from the vacuum $\P_\vo$:
\[ \Pr=\re{i\vr\.\vq}\P_\vo. \]
Thus the operators $\re{i\vr\.\vq}$, $\vr\in\L$, may be identified
with the zero mode states and form an Abelian group which is called
the {\bf group algebra} of the lattice $\L$ and is denoted by $\Rea[\L]$.
Collecting all the Heisenberg modules $\cF^{(\vr)}$ one might expect
the full Fock space $\cF$ of the vertex algebra to be
$S(\hat{\bf h}^-)\XO\Rea[\L]$. However, it is well-known that we
need to replace the group algebra $\Rea[\L]$ by something more delicate
in order to adjust the signs in the Jacobi identity for the vertex
algebra. We will multiply $\re{i\vr\.\vq}$ by a so-called {\bf cocycle
factor} $c_\vr$ which is a function of momentum $\vp$. This means that
it commutes with all oscillators $\amm$ and satisfies the eigenvalue
equations
\[ c_\vr\Ps=\e(\vr,\vs)\Ps. \]
This can be implemented by imposing the conditions
\[ \e(\vr,\vs)\e(\vr+\vs,\vt)&=&
      \e(\vr,\vs+\vt)\e(\vs,\vt), \lb{ecoc1} \\
      \e(\vr,\vs)&=&(-1)^{\vr\.\vs}\e(\vs,\vr), \lb{ecoc2} \\
      \e(\vr,-\vr)&=&(-1)^{\frac12{\vr}^2}, \lb{ecoc3} \\
      \e(\vo,\vo)&=&1. \lb{ecoc4} \]
Note that the cocycle condition \Eq{ecoc1} implies
$\e(\vo,\vo)=\e(\vo,\vr)=\e(\vr,\vo)\ \forall\vr$.
Without loss of generality we can assume that the function $\e$ is
bimultiplicative, i.e.\ $\e(\vr+\vs,\vt)=\e(\vr,\vt)\e(\vs,\vt)$ and
$\e(\vr,\vs+\vt)=\e(\vr,\vs)\e(\vr,\vt)\ \forall\vr,\vs,\vt$.
Together with \Eq{ecoc3} and the normalization condition \Eq{ecoc4},
this then implies that
$\e(m\vr,n\vr)=[\e(\vr,\vr)]^{mn}=(-1)^{\frac{1}{2} mn{\vr}^2}\
\forall\vr,\ m,n\in\Z$.
We will take the {\bf twisted group algebra} $\Rea\{\L\}$ consisting of
the operators
\[ e_{\vr}:=\re{i\vr\.\vq}c_\vr, \lb{evr} \]
for  $\vr\in\L$, instead of $\Rea[\L]$. This means just that we
are working with a certain section in the double cover
$\hat{\L}$ of the lattice $\L$. Due to the identification of the
operators $e_\vr$ with zero mode states it is suggestive to adopt the
ket notation $e_\vr\equiv\ket\vr$ which we will use where appropriate.

In summary, the Fock space associated with the lattice $\L$ is defined
to be
\[ \cF:=S(\hat{\bf h}^-)\XO\Rea\{\L\}. \]
Note that the oscillators $\vr(m)$, $m\ne0$, act only on the first
tensor factor, namely, creation operators as multiplication operators
and annihilation operators via the adjoint representation, i.e.\ by
\Eq{osccom}. The zero mode operators $\a^\m_0$, however, are only
sensible for the twisted group algebra, viz.
\[ \vr(0) \ket\vs=(\vr\X\va_0) \ket\vs=(\vr\X\vs) \ket\vs
     \qquad\forall\vr\in\LR,\vs\in\L,  \lb{zeroact} \]
while the action of $ e_\vr$ on $\Rea\{\L\}$ is given by
\[ e_\vr e_\vs = \e(\vr,\vs) e_{\vr+\vs}. \]

We shall define next the {\bf vertex operators} $\Vp$ for $\p\in
\cF$. We introduce the {\bf Fubini--Veneziano coordinate field},
\[ X^\m(z)
   := q^\m-ip^\m\ln z+i\sum_{m\in\Z}\frac1m\amm z^{-m},
   \lb{FubVen-coo} \]
which really only has a meaning when exponentiated, and the {\bf
  Fubini--Veneziano momentum field},
\[ P^\m(z):= i\dz X^\m(z) = \sum_{m\in\Z}\amm z^{-m-1}.
   \lb{FubVen-mom} \]
For $\vr\in\LR$ we define the formal sum
\[ \vr(z) \equiv \vr\X\vP(z) =\sum_{m\in\Z}\vr(m)z^{-m-1}, \lb{current} \]
which is an element of $\hat{\bf h}[\![z,z^{-1}]\!]$ and may be regarded
as a generating function for the operators $\vr(m)$, $m\in\Z$, or as a
``current'' in contrast to the ``states'' in $\cF$.
It is convenient to split $X^\m(z)$ into three parts:
\[ X^\m(z)=X^\m_<(z)+ (q^\m-ip^\mu\ln z) +X^\m_>(z), \] where
\[ X^\m_<(z):=-i\sum_{m>0}\frac1m\a^\m_{-m}z^m \,, \qquad
   X^\m_>(z):=i\sum_{m>0}\frac1m\amm z^{-m}. \lb{FubVen-split} \]
We will employ the usual normal-ordering procedure, i.e.\ colons
indicate that in the enclosed expressions, $q^\n$ is written to the
left of $p^\m$, as well as the creation operators are to be placed to
the left of the annihilation operators:
\[ \:\vr(m)\vs(n)\:&=&\cases{ \vr(m)\vs(n) & if $m\le n$, \cr
                              \vs(n)\vr(m) & if $m>n$,} \non[.5ex]
   \:q^\n p^\m\:=\:p^\m q^\n\:&=&q^\n p^\m. \]
For $\ket\vr\in\Rea\{\L\}$, the associated vertex operator then takes the
familiar form
\[ \cV(\ket\vr,z) &:=& \re{i\vr\.\vX_<(z)} e_\vr z^{\vr(0)}
                       \re{i\vr\.\vX_>(z)} \non
                  &=& \:\re{i\vr\.\vX(z)}\:c_\vr , \lb{vertexop1} \]
Note that the cocycle factors of the vertex operators are hidden in the
elements of the twisted group algebra, $\Rea\{\L\}$.

Let $\p=\big[\prod_{j=1}^N\vs_j(-n_j)\big]\ket\vr$ be
a typical homogeneous element of $\cF$ and define
\[ \Vp&:=&\ord\;\cV(\ket\vr,z)\prod_{j=1}^N\frac1{(n_j-1)!}
       \left(\dz\right)^{n_j-1}\big(\vs_j\X\vP(z)\big)\;\ord \non
    &\equiv&i\;\ord\;\re{i\vr\.\vX(z)}\prod_{j=1}^N
         \frac1{(n_j-1)!)}\left(\dz\right)^{n_j}\big(\vs_j\X\vX(z)\big)
         \;\ord\;c_\vr. \lb{vertexop2} \]
Extending this definition by linearity we finally obtain a
well-defined map
\[ \cV:\cF&\to&(\End\cF)[\![z,z^{-1}]\!], \non
      \p&\mapsto&\Vp=\sum_{n\in\Z}\p_nz^{-n-1}. \]
We choose the vacuum to be the zero mode state with no momentum and
without any creation operators, i.e.\
\[ \one:=\ket\vo\equiv 1\XO e_{\vo}. \]
The state
\[ \bw:=\frc12\sum_{\m=1}^d\y_{\m\n}\a^\m_{-1}\a^\n_{-1}\ket\vo
      \equiv \frc12\va_{-1}\X\va_{-1}\ket\vo \]
provides a conformal vector of dimension $d$. Indeed, by \Eq{vertexop2} and
\Eq{current}, we have
\[ \cT(z)\equiv\cV(\bw,z)
          =\frc12\sum_{\m=1}^d\:\vP(z)\X\vP(z)\: \nn \]
so that
\[ \Ln\equiv\bw_{n+1}=\frc12\sum_{m\in\Z}\:\va_m\X\va_{n-m}\:\ ,
                                                  \lb{Virosc} \]
in agreement with the well-known expression from string theory. Using
the oscillator commutation relations one finds that the $\Ln$'s
obey \Eq{Vir} with central charge $c=d$ (see e.g.\ \ct{GSW88} for the
calculation).

Finally, let $\p=\big[\prod_{j=1}^N\vs_j(-n_j)\big]\ket\vr$ be
a typical homogeneous element of $\cF$. Then
\[ \Lo\p&=&\bigg(\frc12\vr^2+\sum_{j=1}^Nn_j\bigg)\p \lb{numberop} \]
yields the desired grading of $\cF$. We observe that if $\L$ is
Lorentzian then $\vr^2$ can be arbitrarily negative so that the
spectrum of $\Lo$ is unbounded from above as well as from below.

With the above definitions it is straightforward to verify the first
four axioms for a vertex algebra, but to prove the
Jacobi identity, into which most of the information about the
vertex algebra is encoded, is much harder. A proof based on the
old normal-ordering string techniques can be found in \ct{FLM88}.
In Sect.\ \ref{SEWING} we will present an alternative much more
elegant proof using overlap identities for the three-vertex.

We turn now to the analysis of the Lie algebra of physical states,
$\ggL$. The simplest physical states are the {\bf tachyonic states}
$ \ggL^{[0]}:= \{\ket\vr\| \vr\in\L, \vr^2=2\} $
and the {\bf photonic states:}
$ \ggL^{[1]}:=\{\vs(-1)\ket\vr\|
                 \vr\X\vs=0, \vs\in\LR, \vr\in\L, \vr^2=0\} $,
where the superscript of $\ggL$ counts the oscillator excitations.
We want to stress again that the physical states in $\ggL$ are
only defined modulo $\rL{-1}\Pz0$, which means for example that
$\vr(-1)\ket\vr= \rL{-1}\ket\vr\equiv0$
in $\ggL$ for a vector $\vr\in\L$ with $\vr^2=0$.

The commutators between the first nontrivial physical states become
\[ [\br(-1)\ket\vo,\bs(-1)\ket\vo]&=&0, \lb{Comm1} \\ {}
   [\br(-1)\ket\vo,\ket\vr]&=&(\br\X\vr)\ket\vr, \\ {}
   [\ket\vr,\ket\vs]
     &=&\cases{ 0 & if $\vr\X\vs\ge0$, \cr
                \e(\vr,\vs)\ket{\vr+\vs} & if $\vr\X\vs=-1$, \cr
                -\vr(-1)\ket\vo & if $\vr\X\vs=-2$, \cr
                \ldots & if $\vr\X\vs\le-3$, } \lb{Comm3} \]
for $\br,\bs\in\LR$ and $\vr,\vs\in\L$, $\vr^2=\vs^2=2$. Note that
for an Euclidean lattice the Schwarz inequality would imply
$|\vr\X\vs|\le2$ leading to a finite-dimensional Lie algebra of
physical states; but here we are interested in the much more
complicated case of a Lorentzian lattice.

We have seen that a special role is played by the norm 2 vectors of
$\L$ which we call {\bf real roots} of the lattice. The {\bf
reflection} $\ew_\vr$ associated with a real root $\vr$ is defined
as $\ew_\vr(\vx)=\vx-(\vx\X\vr)\vr$ for $\vx\in\LR$. It is easy to see
that a reflection in a real root is an automorphism of the lattice.
The hyperplanes perpendicular to these real roots divide the
vector space $\LR$ into regions called {\bf Weyl chambers}. The
reflections in the real roots of $\L$ generate a group called the
{\bf Weyl group} $\eW$ of $\L$, which acts simply transitively on the
Weyl chambers of $\L$. This means that if we fix one Weyl chamber
$\eC$ once and for all, then any real root from the interior of
another Weyl chamber can be transported via Weyl reflection to a
unique real root in $\eC$. The real roots $\vr_i$ that are
perpendicular to the faces of $\eC$ and have inner product at most 0
with the elements of $\eC$ are called the {\bf simple roots} of
$\eC$. The {\bf (Coxeter--)Dynkin diagram} $\eG$ of $\eC$ is the set of
simple roots of $\eC$, drawn as a graph with one vertex for each
simple root of $\eC$ and two vertices corresponding to the distinct
roots $\vr_i, \vr_j$ are joined by $-\vr_i\X\vr_j$ lines.

Returning to the vertex algebra associated with the even
Lorentzian lattice $\L$, it is clear that, for any simple root
$\vr_i$, the elements $\ket{\vr_i}$, $\ket{-\vr_i}$, and
$\vr_i(-1)\ket\vo$ describe physical states, i.e.\ they lie in
$\Pz1$. Define generators for a Lie algebra $\ggA$ by
\[ e_i&\mapsto&\ket{\vr_i}, \non
   f_i&\mapsto&-\ket{-\vr_i}, \non
   h_i&\mapsto&\vr_i(-1)\ket\vo. \lb{Chevalley}  \]
Then, by \Eq{Comm1} -- \Eq{Comm3}, we find the following relations to
hold:
\[ [h_i,h_j]&=&0, \\ {}
   [h_i,e_j]&=&a_{ij}e_j,\quad [h_i,f_j]=-a_{ij}f_j, \\  {}
   [e_i,f_j]&=&\d_{ij}h_i, \]
where we defined the {\bf Cartan matrix} $A=(a_{ij})$ associated with
$\eC$ by $a_{ij}:=\vr_i\X\vr_j$. The elements $h_i$ obviously
form a basis for an abelian subalgebra of $\ggA$ called the {\bf
Cartan subalgebra} $\ghA$. In technical terms, from the above
commutators we learn that the elements $\{e_i,f_i,h_i\|i\}$ generate
the so-called free Lie algebra associated with $A$. But even more is
true; for we can show that the {\bf Serre relations}
\[ \ad{e_i}{1-a_{ij}}e_j=0, \quad
   \ad{f_i}{1-a_{ij}}f_j=0, \lb{Serre} \]
are also fulfilled for all $i\ne j$. They follow from the physical
state condition \Eq{Phys-cond1} by combining Eq.\ \Eq{numberop} with
the natural $\L$-gradation which the Lie algebra of physical states
inherits from $\cF$,
\[ \ggL^{(\vx)}:=\ggL\cap S(\hat{\bf h}^-)\ket\vx, \]
for $\vx\in\L$. Of course, some of the subspaces $\ggL^{(\vx)}$
may be empty, e.g.\ for $\vx^2>2$; but if $\ggL^{(\vx)}$
is nonempty we shall refer to $\vx\in\L$ as a {\bf root} of $\ggL$
with {\bf root space} $\ggL^{(\vx)}$ and {\bf multiplicity}
$\dim\ggL^{(\vx)}$. Hence the number of linearly independent
polarization vectors for a physical state with certain momentum $\vx$
accounts for the multiplicity of $\vx$ as a root for the Lie algebra
of physical states.
Having established the Serre relations, the Gabber--Kac theorem
\ctz{Theorem 9.11}{Kac90} tells us that the Lie algebra $\ggA$ generated
by the elements $\{e_i,f_i,h_i\|i\}$ is just the {\bf Kac Moody
algebra} associated with the Cartan matrix $A$. Namely, the latter is
defined as the above free Lie algebra divided by the maximal ideal
intersecting $\ghA$ trivially, and the theorem states that this
maximal ideal is generated by the elements
$\{\ad{e_i}{1-a_{ij}}e_j,\ \ad{f_i}{1-a_{ij}}f_j\|i\ne j\}$.

We emphasize the remarkable fact that the physical state condition
$\Lo\p=\p$ accounts for all Serre relations whereas these are usually
very difficult to deal with in the theory of Kac Moody algebras; or,
in string theory language, the absence of particles with squared mass
below the tachyon reflects the validity of the Serre relations for
the Lie algebra $\ggA$.

To summarize (cf.\ \ct{Borc86}): The physical states
$\{\ket{\vr_i},\ket{-\vr_i},\vr_i(-1)\ket\vo\| i\}$ generate via
multiple commutators the Kac Moody algebra $\ggA$ associated with the
Cartan matrix $A=(\vr_i\X\vr_j)$ which is a subalgebra of the Lie
algebra of physical states, $\ggL$.
Only in the Euclidean case these two Lie algebras coincide.
In general, we have a {\it proper} inclusion
\[ \ggA\hookrightarrow\ggL, \lb{inclus} \]
and the characterization of the elements of $\ggL$ not contained in
the Lie algebra $\ggA$ is the key problem for the vertex operator
construction of hyperbolic Kac Moody algebras. In the following we
shall refer to such physical states that cannot be obtained as linear
combinations of multiple commutators of the generators of $\ggA$ as
{\bf missing} or {\bf decoupled (physical) states}.  For it will turn
out that they are characterized by the property that they decouple
from the $S$-matrix for $\ggA$, i.e.\ putting a missing state on one
leg of the $N$-string vertex and saturating the other ones with
elements of $\ggA$ will give zero for the matrix element.

The special feature of
\Eq{inclus} is that the root system of the Kac Moody algebra $\ggA$
is well understood though its root multiplicities are not completely
known for a single example; whereas the root system of $\ggL$ is
a priori not related to that of a Kac Moody algebra although
the root multiplicities are always known. Thus a complete
understanding of \Eq{inclus} requires a ``mechanism'' which tells us
how $\ggA$ has to be filled up with physical states to reach the
complete Lie algebra of physical states.

Finally we turn to the construction of a nondegenerate bilinear form
$(\_|\_)$ on $\ggL$ satisfying the condition $(\Vp\chi|\f)=
(\chi|\Vs(\p,z)\f)$ with the adjoint vertex operator as defined
in \Eq{vertop-sh}. In fact, this ``invariance'' condition
is strong enough to determine the bilinear form uniquely up to a
normalization. First, we deduce that
\[ \vr(-m)\sh=-\vr(m)\qquad\mbox{or}\qquad
    (\amm)\sh=-\a^\m_{-m} \lb{oscadj} \]
for $\vr\in\LR$, $m\in\Z$. Thus the oscillator part of the bilinear
form is uniquely fixed and it remains to calculate the zero modes.
To do so, we evaluate $(\vr(0)e_\vr|e_\vs)$ and
$(e_\vr|\vs(0)e_\vs)$ in both ways to find that $(e_\vr|e_\vs)$ vanishes
for $\vr,\vs\in\L$ unless $\vr=-\vs$. On the other hand, an explicit
computation using the vertex operators shows that
$(e_\vr|e_{-\vr})= (\one|\one)$ for all $\vr\in\L$.
Thus we have
\[ (e_\vr|e_\vs)=\d_{\vr+\vs,\vo}(\one|\one), \]
which, together with \Eq{oscadj}, indeed uniquely fixes the bilinear
form up to the normalization of $(\one|\one)$. For reasons which will become
clear in a moment, we shall choose the normalization
\[ (\one|\one):=-1 \]
so that
\[ (e_\vr|-e_{-\vs})&=&\d_{\vr,\vs}, \non[.5ex]
   \Big(\vr(-1)\ket\vo\,\Big|\,\vs(-1)\ket\vo\Big)&=&\vr\X\vs \]
for $\vr,\vs\in\L$. When we go over to the induced form on $\ggL$ these
relations respectively give
\[ (e_i|f_j)&=&\d_{ij}, \non[.5ex]
   (h_i|h_j)&=&a_{ij}, \]
for the generators of the Kac Moody algebra $\ggA$. Together with
Eq.\ \Eq{invform} this shows that $(\_|\_)$ induces on $\ggA$ a
{\bf standard invariant bilinear form} \ct{Kac90}. This justifies our
choice of normalization.

Whereas the bilinear form defined above yields a
non-degenerate pairing of the states $\ket\vr$
and $\ket{-\vr}$, in physics we prefer a bilinear form which
pairs $\ket\vr$ with itself, because ultimately we
are looking for a symmetric scalar product\footnote{If we worked over
the complex numbers, as one usually does in a Hilbert space, we would
want an Hermitian scalar product.} which does not lead to physical
states of negative norm. For this purpose we need the
{\bf Chevalley involution} $\th$ which is given by
\[ \th(\ket\vr)&:=&\ket{-\vr}, \non[.5ex]
   \th\cc\vr(-m)\cc\th^{-1}&:=&-\vr(-m)\qquad\mbox{or}\qquad
   \th\cc\a^\m_{-m}\cc\th^{-1}:=-\a^\m_{-m}. \lb{Chevinv} \]
Note that this definition of $\th$ is consistent with the
Chevalley involution on the Kac Moody algebra $\ggA$, viz.
\[ \th(e_i)&=&-f_i, \non[.5ex]
   \th(f_i)&=&-e_i, \non[.5ex]
   \th(h_i)&=&-h_i. \]
Since the Virasoro generators are bilinear in the oscillators it is
clear that $\th$ commutes with the $L_n$'s.

We introduce a {\bf contravariant bilinear form} by defining
\[ \vev{\p}{\f}:=-(\th(\p)|\f) \lb{contraform} \]
for all $\p,\f\in\cF$.
We immediately see that, with respect to $\vev{\_}{\_}$, the zero
mode states are orthonormal to each other and $\vr(n)$ is the
adjoint of $\vr(-n)$,
\[ \vev{\vr}{\vs}&=&\d_{\vr,\vs}, \\
   \vr(m)\dg&=&\vr(-m)\qquad\mbox{or}\qquad
   (\amm)\dg=\a^\m_{-m}. \]
Hence the contravariant bilinear form $\vev{\_}{\_}$ is nothing but
the familiar {\bf string scalar product} to which the no-ghost theorem
applies. In general we have
\[ \Vd(\p,z)=\th\cc\Vs(\p,z)\cc\th^{-1}=\Vs(\th\p,z), \]
due to the fact that $\th$ is an automorphism of the vertex algebra.
For example,
\[    -[X^\m(z)]\dg =[X^\m(z)]\sh   &=&X^\m(z^{-1}), \lb{X-sharp} \\ {}
   -[P^\m(z)dz]\dg =[P^\m(z)dz]\sh &=&P^\m(z^{-1})d(z^{-1}), \\ {}
    [\cT(z)(dz)^2]\dg =[\cT(z)(dz)^2]\sh &=&\cT(z^{-1})[d(z^{-1})]^2, \]
so that $\Ln\dg=\Ln\sh=\rL{-n}$ for all $n$. These identities suggest
that, from the viewpoint of string theory, the $\sharp$-conjugation is more
natural than the (usually preferred) $\dagger$-conjugation. Namely,
whereas the latter comes from hermitian conjugation of the
oscillators the former is nothing but the PT conjugation for it
implements the transformation $z\to z^{-1}$ of the worldsheet
coordinate $z=\re{t+ix}$.

\section{General properties of multistring vertices} \lab{VERTICES}
$N$-string vertices were first introduced as an efficient way of
describing string scattering and computing amplitudes. In particular,
their sewing properties ensured that factorization, and more
specifically duality, were satisfied by the resulting string
scattering amplitudes. However, their form and derivation were rather
complicated and many of their properties unknown. One of the most
important subsequent steps in the development of $N$-string vertices
was the realization that they can be entirely characterized by a
simple set of equations called overlap identities which they satisfy
\ct{NevWes86b}. Whereas the computations required to determine the
$N$-vertices by sewing used to be long and tedious in the early days
of string theory (see e.g.\ \ct{AlAmBeOl71}), the use of overlap
identities simplifies the task enormously while at the same time
leading to a beautiful geometrical picture. In addition they allowed
for a simple derivation of all the properties of the $N$-string
vertices. We can here provide only a brief summary; we shall, however,
make an effort to present the pertinent results in a self-contained
way and to explain the basic ideas as clearly as possible. In
particular, we will explicitly show how to determine the vertices
easily and efficiently by means of the overlap equations. In Sect.\
\ref{E10} we will see that, remarkably, the multistring formalism also
furnishes the requisite tools for the analysis of hyperbolic Kac Moody
algebras and their root spaces.

\subsection{Basic definitions} \lab{VERTICES-1}
It is well known that the amplitude for the scattering of $N$ physical
string states $\p_1,\ldots, \p_N\in\Pz1$ at tree level is obtained by
evaluating the multistring vertex $\VN$ on these states and
integrating the result with a suitable measure over the Koba Nielsen
parameters\footnote{These variables which carry indices should not be
  confused with the formal variables $z,w,y$.} $z_1,\ldots,z_N$. For
open and closed string tree level amplitudes, the integrals are to be
carried out over real line and the whole complex plane (the Riemann
sphere) for each $z_i$, respectively (see e.g.\ \ct{GSW88,LueThe89}
for further explanations). However, the calculational rules
appropriate for our needs, namely the computation of Lie algebra
multiple commutators, by necessity differ from the conventional ones.
We are essentially dealing with the chiral half of a fully
compactified closed string, and we accordingly replace the
two-dimensional integrals by one-dimensional contour integrals. The
relevant integrand can be regarded as the ``holomorphic square root''
of the corresponding formula for the closed string. We have
\[ \lefteqn{ W(\p_1,\ldots,\p_N) } \hh{8mm} \non
    &=& \Oint\frac{dz_1}{2\pi i} \ldots \Oint\frac{dz_N}{2\pi i}
        \frac{(z_1 - z_2)(z_2 - z_3)(z_1 - z_3)} {\big(z_1 - z_1^{(0)}\big)
        \big(z_2 - z_2^{(0)}\big) \big(z_3 - z_3^{(0)}\big)} \,
        \m(z_1,\ldots,z_N) \, \VN(z_1,\ldots,z_N)
        \kets{\p_1}1\ldots\kets{\p_N}N. \hh{5mm} \lb{Amplitude} \]
By Cauchy's theorem, the pole terms $\big(z_i-z_i^{(0)}\big)^{-1}$
effectively act as $\d$-functions fixing three Koba Nielsen variables
$(z_1,z_2,z_3)$ to three arbitrarily given points
$(z_1^{(0)},z_2^{(0)},z_3^{(0)})$ in the complex plane (often taken to
be $\infty$, $0$ and $1$.) To maintain the M\"obius invariance of the
full amplitude the gauge fixing function must be accompanied by a
Faddeev Popov determinant.  For the chiral string this is the
holomorphic square root of the corresponding expression for the closed
string (see e.g.\ \ct{GSW88}), which accounts for the factor
$\Delta_{\rm FP} = (z_1-z_2)(z_2-z_3)(z_1-z_3)$.  For the above
formula to be unambiguously defined we must in addition specify the
integration contours over which the integrals are to be performed;
this will be done in due course.  Because the vertex $\VN$ is not
unique, an extra measure factor $\m$ is needed to compensate for the
conformal transformations relating different $N$-vertices to one
another, so as to obtain a unique and well-defined scattering
amplitude after integration over the Koba Nielsen variables. By
construction, this factor is not affected by the global M\"obius
transformations which act in the same way on all $z_i$; it will be
explicitly given in the next section.

According to the approach developed in
\ct{NevWes86b,NevWes86a,NevWes87a,NevWes88a,NevWes88b},
multistring vertices can be characterized in more mathematical
terms as follows.

\begin{defi} \lab{def2} \hh{1em} \\[1.5ex]
An {\bf N-string vertex (at tree level)} is a multilinear map
$\VN:\cF_1\XO\ldots\XO\cF_N\to\C$, where each $\cF_i$ is isomorphic to
the Fock space $\cF$ of the free string, depending on $N$ complex
parameters $z_1,\ldots,z_N$, which satisfies the (unintegrated)
{\bf overlap identities}
\[ \VN\F\i(\x_i)
    =\VN\F\j(\x_j)\left(\dd{\x_j}{\x_i}\right)^h
     \qquad\forall i,j=1,\ldots,n. \lb{overlap0} \]
relating the action of a conformal operator $\F(\x)$ of weight $h$ on one
external line (i.e.\ tensor factor) of the vertex to its action on
another external line.
Here $\x_i$ denotes a coordinate patch on the Riemann sphere around
the Koba Nielsen point $z_i$, i.e.\ $\x_i(z_i)=0$.
\end{defi}

This definition determines the $N$-vertex only up to multiplication
by an arbitrary function $f=f(z_1,\ldots,z_N)$, which could be
fixed for instance by demanding $\VN \kets{\vo}1\ldots\kets{\vo}N =1$.
A more physical way of normalizing $\VN$ is to require the string
scattering amplitudes to be unitary and thereby also to fix the
measure \ct{NevWes86a,NevWes87a}.

\begin{defi} \lab{def3} \hh{1em} \\[1.5ex]
A {\bf physical string scattering amplitude} is a multilinear map
$W:\cF_1\XO\ldots\XO\cF_N\to\C$ constructed by means of \Eq{Amplitude}
from an $N$-vertex satisfying {\bf unitarity}, i.e.\ null physical
states should decouple from the $S$-matrix.
Thus for physical states $\p_1,\ldots,\p_N$, we require
\[ \exists i:\ \vev{\p_i}{\f}=0\ \forall\f\in\Pz1
    \qquad\Longrightarrow\qquad W(\p_1,\ldots,\p_N)=0. \]
\end{defi}

These definitions, and especially the overlap identities with their
implications will be discussed in the following section.  Here we only
make some preliminary remarks. Since $\VN$ is only determined up to a
normalization, we are free to absorb $\m$ and the Faddev Popov
determinant into it and will frequently do so. It should be clear that
the overlap equations are defined only for overlapping charts, but
since the Riemann surface appropriate for tree level scattering of
strings is the Riemann sphere and this surface only needs two
coordinate patches to cover it, the coordinates $\x_i$ will be defined
everywhere except for one point. However, Eq.\ \Eq{overlap0} is only
valid if the action of $\F\i(\xi_i)$ on the vertex converges, this
will not be the case as $\xi_i$ approaches any of the Koba Nielsen
points $z_j, j\neq i$. Rigorously speaking, the overlap identities
should be understood in the sense of analytic continuation of matrix
elements. Interpreted as formal series, they look somewhat different.
In Sect.\ \ref{THREEVERTEX} we will establish for the three-vertex the
precise relation between the overlaps and the Jacobi identity for
intertwining operators. In the appendix we will demonstrate how the
unitarity condition leads to set of first order differential equations
which determine the measure $\m$.

Let us now explain our notation. Where appropriate we will suppress
the dependence of the vertex on the Koba Nielsen points $z_i$. For the
$N$-vertex $\VN$ we will also sometimes write $V_{k_1\ldots k_N}$,
explicitly exhibiting the numbering of the external legs. The ket
notation $\kets{\p_i}k$ denotes the state $\p_i\in\cF$ as an element
of the $N$-string Fock space, i.e.\ as living in the $k$-th tensor
factor, $\cF_k$. Upper indices $(k)$ on an operator $\cO$ indicate the
tensor factor, $\cF_k$, on which the operator acts, i.e.\ $\cO\k =
{\bf 1}\otimes\ldots \otimes {\bf 1} \otimes \cO \otimes {\bf 1}
\otimes \ldots \otimes {\bf 1}$ with the operator $\cO$ in the $k$-th
place.

Starting from the above intrinsic definition of a multistring vertex
we shall see that in the usual oscillator representation the vertex
will be of the form
\[ \VN=\vac\cO(\{\a^{(i)\m}_m\}), \]
with
\[ \vac:=\sum_{\vr_i\in\L\ \forall i \atop \vr_1+\ldots+\vr_N=\vo}
         \bras{\vr_1}1\XO\ldots\XO\bras{\vr_N}N\ , \lb{bra-vactilde} \]
for some operator $\cO$. Note that $\vr_1+\ldots+\vr_N=\vo$ is just
momentum conservation and that the state $\vac$ is not normalizable
(even for discrete momenta). $\cO$ will turn out to be an exponential
of bilinears in the oscillators. Due to the Fourier transform of the
bra vacuum, $\vac$, in front of $\cO$ it is clear that no creation
operators, $\amm$ for $m<0$, can occur and thus no normal-ordering is
required. It will also be important later that our definition of the
bra-states in \Eq{bra-vactilde} includes the cocycle factors
$c_{\vr_i}$ (cf.\ Eq.\ \Eq{evr} and the discussion following it), and
this prescription will automatically take care of unwanted factors of
$(-1)$ in the commutator of multistring vertices.

We will also need the mathematical prescription for turning around a
leg of the vertex. This is essentially provided by the isomorphism
$\R$ of Sect.\ \ref{VERTEXOP-1} which identifies the one-string Fock
space $\cF$ with its restricted dual space, $\cF^\prime$. Recall that
$\R$ is defined as $\R: \ket\vr\mapsto -\bra{-\vr}\equiv (\vr|\_),\
\a^\m_{-m}\mapsto-\amm$, so that $\dual{\R(\p)}{\f}= (\p|\f)=
-\vev{\th(\p)}{\f}$ by \Eq{oscadj} and \Eq{contraform}. Then the {\bf
  reversing operator}, $\R_i$, which turns leg $i$ around acts, in the
oscillator representation, on the $N$-vertex as follows: it leaves the
tensor factors $\cF_k$ for $k\ne i$ invariant and maps $\cF_i$ to
$\R_i(\cF_i)=\cF_i^\prime$, i.e.\
\[ \R_i\ :\ \a^{(k)\m}_m &\longmapsto& \a^{(k)\m}_m\FOR{}k\ne i, \non
          \a^{(i)\n}_n &\longmapsto& -\a^{(i)\n}_{-n}, \non
          \vac\cO   &\longmapsto&
           -\sum_{\vr_j\in\L\ \forall j \atop \vr_1+\ldots+\vr_N=\vo}
           \bras{\vr_1}1\ldots\bras{\vr_{i-1}}{i-1}
           \bras{\vr_{i+1}}{i+1}\ldots\bras{\vr_N}N
           \ \R_i\cO\ \kets{-\vr_i}i . \]
The above notational conventions can be most conveniently summarized
by the diagrams below:
\[ \VN_{1\ldots N}(z_1,\ldots,z_N) \qquad&\longleftrightarrow&\qquad
  \unitlength.5mm
  \begin{picture}(70,24)
   \put(35,15){\circle{10}}
   \put(35,15){\makebox(0,0){$z_i$}}
   \put(51.13,9.62){\line(3,-1){8}}
   \put(52,15){\line(1,0){8}}
   \put(35,-2){\line(0,-1){8}}
   \put(39.74,13.42){\vector(3,-1){12}}
   \put(40,15){\vector(1,0){12}}
   \put(35,10){\vector(0,-1){12}}
   \put(60.61,6.46){\makebox(0,0)[l]{$2$}}
   \put(62,15){\makebox(0,0)[l]{$1$}}
   \put(35,-12){\makebox(0,0)[t]{$N$}}
   \put(43,0){\circle*{1.5}}
   \put(46,2){\circle*{1.5}}
   \put(48.5,4.67){\circle*{1.5}}
  \end{picture} \nn \\[5ex]
  \R_1\VN_{1\ldots N}(z_1,\ldots,z_N) \qquad&\longleftrightarrow&\qquad
  \unitlength.5mm
  \begin{picture}(70,24)
   \put(35,15){\circle{10}}
   \put(35,15){\makebox(0,0){$z_i$}}
   \put(22,15){\line(1,0){8}}
   \put(51.13,9.62){\line(3,-1){8}}
   \put(52,15){\line(1,0){8}}
   \put(35,-2){\line(0,-1){8}}
   \put(39.74,13.42){\vector(3,-1){12}}
   \put(10,15){\vector(1,0){12}}
   \put(40,15){\vector(1,0){12}}
   \put(35,10){\vector(0,-1){12}}
   \put(60.61,6.46){\makebox(0,0)[l]{$3$}}
   \put( 8,15){\makebox(0,0)[r]{$1$}}
   \put(62,15){\makebox(0,0)[l]{$2$}}
   \put(35,-12){\makebox(0,0)[t]{$N$}}
   \put(43,0){\circle*{1.5}}
   \put(46,2){\circle*{1.5}}
   \put(48.5,4.67){\circle*{1.5}}
  \end{picture} \nn \] \vspace*{10mm} \\
where the arrows on the legs of the vertex $\VN$ by definition point
outwards and only those point inwards which are turned around.

\subsection{Overlap identities} \lab{OVERLAPS}
As we will not consider loop
amplitudes in this paper, the worldsheet of the string is
always taken to be the complex plane (alias the Riemann sphere),
which we parametrize by the complex variable $\z$ in the region of
its south pole and $\td\z=\z^{-1}$ in the region of its north
pole. The $N$-vertex $\VN$ quite generally describes
the scattering of $N$ string states, which are emitted or absorbed at
the Koba-Nielsen points $z_i\in\C$ ($i=1,...,N$). Hence
the vertex will depend on the variables $z_i$. As already pointed out,
these must be integrated
over before one obtains the final string scattering amplitude.
Each $z_i$ belongs to a coordinate neighbourhood
parametrized by a complex analytic coordinate $\x_i = \x_i (\z )$,
meaning an analytic function of $\z$ in the neighbourhood of $z_i$,
which vanishes at the corresponding Koba Nielsen point, i.e.\
$\xi_i (z_i ) =0$. Apart from this restriction, the functions
$\x_i (\z )$ can be chosen arbitrarily; this freedom accounts for
the multitude of $N$-string vertices. We will often identify the
variable $\z$ with one of the coordinates, viz.\ $\x_2 \equiv \z$.
In principle, all of the coordinate patches can be
extended to cover the whole complex plane, with the exception
of one point, and we will implicitly make use of this fact below
in assuming that any two given patches can be made to overlap.
The point here is that when the string, i.e.\ $\F\i(\xi_i)$, acts on
the vertex, the result can be expanded in a Laurent series around
$\x_i =0$; when $\x_i$ approaches any other Koba Nielsen point, this
series will diverge in general. So one should think of the
neighbourhoods surrounding the Koba Nielsen points as
defining the domains of convergence of the relevant series rather than
coordinate patches in the sense of differentiable manifolds.

On the domain where they overlap, the local coordinates $\x_i$ and
$\x_j$ are related by transition functions $\tr ij$ such that
\[ \x_i = \x_i (\x_j) = \tr ij (\x_j). \lb{xivonxj}  \]
If, for example, $\z \equiv \x_2$ we have $\x_i = \tr i2 (\z)$, and
then the Koba Nielsen variables are given by $z_i = \tr 2i (0)$. The
transition functions obey the self-evident relations
\[ \tr ii &=& \id ,  \non
   \tr ij \tr ji &=& \id   ,   \non
   \tr ij \tr jk &=& \tr ik    \lb{tauij}  \]
(the second identity obviously follows from the other two).
Let $\F\i (\x_i )$ be a conformal field of weight $h$ associated with the
$i$-th Fock space $\cF_i$ and the Koba Nielsen point $z_i$.
The fundamental (``unintegrated'') overlap identity \Eq{overlap0}
defining the vertex can
be rewritten in terms of differentials as
\[ \VN \F\i \big( \x_i (\z )\big) \big[ d\x_i (\z )\big]^h  =
   \VN \F\j \big( \x_j (\z )\big) \big[ d\x_j (\z )\big]^h  .
   \lb{overlap1}    \]
At a superficial glance this might just be regarded as expressing
the transformation properties of a differential of weight $h$ on the
complex plane under analytic coordinate transformations.  However, the
content of the above equation is far more subtle. All the difference
is made by the superscripts on the conformal fields indicating to
which string Fock space they belong. The overlap identities thus
relate {\em different} one-string Fock spaces in a highly non-trivial
manner.

Although \Eq{overlap1} is the most general form of the overlap
identities, the so-called integrated overlap identities are sometimes
even more useful in practical calculations. To derive them we choose
an arbitrary function $f(\x_i)$, which is analytic except at $\x_i=0$
where it may have a pole, and integration contours $\cC_j$ surrounding
the Koba Nielsen points $\z = z_j$, i.e.\ $\x_j =0$. Deforming the
contour $\cC_i$ and keeping in mind that the poles at $\z = z_j$ for
$j\neq i$ in general prevent us from pulling it over the other Koba
Nielsen points away to infinity, we get
\[  \VN \Oint_{\cC_i} d\x_i (\z ) \F\i (\x_i (\z)) f(\x_i(\z)) =
  - \sum_{j\neq
 i} \VN \Oint_{\cC_j} d\x_i (\z ) \F\i (\x_i (\z ) ) f(\x_i (\z) ). \]
The non-trivial step is now to apply the unintegrated overlap equation
\Eq{overlap1} to this expression to obtain the integrated overlap
equations. This step is required in order to maintain convergence as
discussed above. Now taking the coordinates $\x_j$ as integration
variables we find \ct{NevWes86b,NevWes86a,NevWes87a}
\[ \VN \sum_{j=1}^N \Oint_{\x_j = 0} d\x_j \F\j (\x_j )
   \bigg[ \frac{d\x_j (\x_i )}{d\x_i} \bigg|_{\x_i = \x_i (\x_j)} \bigg]^{h-1}
         f(\x_i (\x_j )) = 0  .   \lb{overlap2} \]
Of course, this identity is valid for all $i$.

For each set of transitions functions $\big\{ \tr ij \big\} $
and Koba Nielsen variables $\big\{ z_i \big\}$ one can
explicitly construct a unique $N$-vertex associated with them.
It is most compactly represented in the form
\[ \VN \big( \{z_i\}, \{\tr ij \} \big)  =
      \vac \exp \bigg\{ -\frc i2{\sum_{i,j}}' \Oint_{\x_i =0} d\x_i \,
   \vP\i (\x_i ) \cdot \vX\j_> (\x_j (\x_i ))  \bigg\}
  \cN \big( \{ \va_0\j \} \big) \lb{Nvertex1}  \]
where the prime on the sum indicates that we sum over all $i,j$ with
$i\neq j$ and $\vP \cdot \vX \equiv P^\m X_\m$ with
the Fubini--Veneziano fields defined as in Eqs.\ \Eq{FubVen-mom} and
\Eq{FubVen-split}.
The zero mode part in \Eq{Nvertex1} is given by (cf.\ \ct{NevWes88a})
\[ \cN \big( \{ \va_0\j \} \big) :=
\prod_{i < j} \bigg[ \frac{d}{d\x} \big( \G \circ \tr ij \big)
        (\x ) \Big|_{\x =0} \bigg]^{- \frac12 \vp\i\.\vp\j }.
  \lb{Nzeromode}  \]

To prove the general formula \Eq{Nvertex1} one could use the
unintegrated overlap formula \Eq{overlap0} for the conformal field
$ \F = \vX$; however, it is simpler to first check its  non-zero mode
part by means of the integrated overlap conditions \Eq{overlap2} and
then the zero mode part by \Eq{overlap1}. For this purpose
we need to expand powers of the transition functions as follows
(these expansions were already introduced in \ct{AlAmBeOl71})
\[ \frac{1}{\sqrt{m}}\Big\{ \big[ \t(\x)\big]^m -\big[\t (0)\big]^m \Big\}
   = \sum_{n\geq 1} C_{mn} (\t) \frac{\x^n}{\sqrt{n}} ,   \lb{Cmntau}  \]
where $\t$ stands for any $\tr ij$. Substituting $\F\i (\x_i ) = \vP\i
(\x_i )$ (which is of weight one, i.e.\ $h=1$) into the integrated
overlap \Eq{overlap2} with the complete set of allowable functions
$f (\x_i)=(\x_i )^{-m}$ and making use of the above expansions, we
immediately deduce the overlap equations for the oscillators
\[ \VN \bigg\{ \a^{(i)\m}_{-m} + \sqrt{m} \sum_{j\neq i} \sum_{n \geq 1}
   C_{mn}^{ij} \frac{\a^{(j)\m}_n}{\sqrt{n}} +
   \sum_{j\neq i} \big[ (\G \cc \tr ij ) (0) \big]^m
              \a^{(j)\m}_0 \bigg\} = 0   ,
   \lb{oscoverlap} \]
where we have defined
\[  C_{mn}^{ij} := C_{mn} \big(\G \cc \tr ij \big)  ,   \lb{Cmnij} \]
with
\[ \G (\x) := \frac{1}{\x} .   \]
Note that $C_{mn}^{ij}=C_{nm}^{ji}$, i.e.\ $(C^{ij})^{\rm T}=C^{ji}$
for the matrices\footnote{ If $\tau$ is a M\"obius function, we have
  $C(\tau)^{\rm T} = C(\G\cc\tau^{-1}\cc\G )$.}.
It is then not difficult to check that the following expression
in terms of oscillators satisfies Eq.\ \Eq{oscoverlap}:
\[ \VN = \vac \exp \bigg\{-\frc12{\sum_{i,j}}'
                \sum_{m,n \geq 1} \frac{\a^{(i)\m}_m}{\sqrt{m}}
     C_{mn}^{ij} \frac{\a\j_{n\m}}{\sqrt{n}}
     - {\sum_{i,j}}' \sum_{m\geq 1} \frac{\a^{(i)\m}_m}{m}
      \big[(\G \cc \tr ij ) (0) \big]^m \a\j_{0\m} +
       \ln\cN(\va\i_0)
       \bigg\}    \lb{Nvertex2}     \]
with an as yet undetermined function $\cN$ of the zero mode oscillators.
To fix the latter terms we have to make use of the unintegrated
overlap with $\F\i (\x_i ) = \vX\i (\x_i )$. By expanding the
Fubini--Veneziano fields in \Eq{Nvertex1} in terms of oscillators
it is now straightforward to verify that the non-zero mode part of
\Eq{Nvertex1} indeed coincides with the above formula \Eq{Nvertex2}
for the $N$ vertex. We stress once more that \Eq{Nvertex2} can still
be multiplied by an arbitrary function without affecting the overlap
equations.

Finally, we have to spell out the measure $\m$ occurring in our general
formula \Eq{Amplitude}. Relegating the details of the derivation
to Appendix \ref{MEASURE}, where also some examples are worked out, we
just quote the result:
\[ \m (z_1,\ldots,z_N)
    = \prod_{j=1}^N \frac{\partial \x_j (\z)}{\partial \z}
                    \bigg|_{\z =z_j}. \lb{Measure} \]
In the ``gauge'' $\z = \x_1$, this expression becomes
\[ \m (z_1,\ldots,z_N )
    = \prod_{j=2}^N \frac{\partial \tr j1 (\x_j)}{\partial \x_j}
                    \bigg|_{\x_j =0}. \lb{Measure1} \]
{}From \Eq{Measure1} one infers that the measure $\m (z_1,\ldots,z_N)$
is insensitive to those variable transformations which leave the
transition functions (and hence the vertex) invariant. We will see
at the end of this section (cf.\ \Eq{inertrafo} below) that
there is one such conformal transformation which acts in the same
manner on all $z_i$.

To demonstrate the power of the integrated overlap equations let us
consider the example of primary fields of weight
$h=1$, i.e.\ vertex operators associated with physical states. The
choice $f\equiv1$ in Eq.\ \Eq{overlap2} immediately gives
\[ \VN \sum_{j=1}^N \Oint_{\x_j = 0} d\x_j \F\j (\x_j)=0.
   \lb{tDDFoverlap} \]
An important special case of this formula, which we will
return to in Sect.\ \ref{E10}, is obtained by taking
the weight one primary fields to be transversal and longitudinal
DDF operators; the identity tells us that these can just be moved
through the vertex without any change or extra contributions.
Note that the form of the overlap equations for weight one
primary fields is universally valid for arbitrary $N$-vertices
whereas the overlap equations for other conformal weights
will explicitly depend on the choice of transition functions.

The overlap identities also make sense for objects
which transform in an anomalous but still well defined way
under conformal transformations. An example is the
energy momentum tensor $\cT$ for which
$\td\cT(\tx) = \cT(\tx)\left(\dd{\tx}{\x}\right)^2
+ \frac{c}{12}(S\tx)(\x)$, where
$(Sf)(\x) = \frac{f'''(\x)}{f'(\x)}
           -\frc32\left(\frac{f''(\x)}{f'(\x)}\right)^2$ is the
Schwarzian derivative and $c=26-d$ the conformal anomaly.
Hence apart from the last term,
the energy momentum tensor is of weight two.
The associated (unintegrated) overlap identity is
\[ \VN\cT\i(\x_i)
    =\VN\cT\j(\x_j)\left(\dd{\x_j}{\x_i}\right)^2
     + \frac{c}{12} \VN\,(S\x_j)(\x_i)
     \qquad\forall i,j=1,\ldots,n.\]
Just as before we can derive a corresponding
integrated overlap equation for the Virasoro generators from the
the above unintegrated equation. In this process one would seem to acquire
additional contributions from the Schwarzian. However, for tree level
vertices these contributions vanish. This can be shown either directly
by taking the integrated overlaps and saturating all legs with the
true vacuum, using the explicit form of the vertex given above, or by
explicitly performing the integrations over the $\xi_j$'s which
vanish as the Schwarzian term contains no relevant poles. This is what
we should also expect on physical grounds without any calculation
because the central term
in the Virasoro algebra giving rise to the Schwarzian is a quantum
effect (carrying a factor of $\hbar$) and thus not visible
at string tree level. As a result we find
\[ \VN \sum_{j=1}^N  \Oint_{\x_j =0} d\x_j \cT\j (\x_j)
   \frac{d \tr ji (\x_i )}{d\x_i} \bigg|_{\x_i = \tr ij (\x_j )}
   f(\tr ij (\x_j )) = 0 .  \lb{Loverlap1} \]
Of course, in a string loop expansion one does encounter additional
contributions to the integrated overlap equation for the energy
momentum tensor at higher orders in $\hbar$ (unless $d=26$), and these
play a very important role in the determination of string loop
corrections to $\VN$.

Next we study the conformal properties of the $N$-vertex; the freedom
of performing such transformations is considerably greater than for
one-string vertex operators and constitutes one of the main advantages
of the multistring formalism. If $\M$ is a conformal transformation
$\xi \rightarrow \M (\x )$, we denote by $\Mh$ the realization of the
{\it same} conformal transformation as an operator on Fock space. In
general this conformal transformation need not be analytic at
$\xi=0$. As before, a superscript indicates on which Fock space the
transformation acts. For instance, $\Mh\j_i$ corresponds to the
realization of the conformal transformation $\M_i$ on the
$j$-th Fock space $\cF_j$. Explicitly,
\[ \Mh\j_i = \exp \bigg\{ \sum_{n=-\infty}^\infty c_n^i L\j_n
                          \bigg\}, \lb{conftrafo} \]
where the $c_n^i$ are arbitrary parameters. Of course, this expression
has to be taken with a grain of salt as the Virasoro algebra cannot be
globally exponentiated to a group (see e.g.\ \ct{AlGoMoVa88} for a
discussion of this point). We can then define a new $N$-vertex $\tV N$
from $\VN$ by
\[ \tV N \equiv \VN (\{\td z_i\},\{\ttr ij\}) :=
   \VN (\{ z_i\},\{\tr ij\}) \prod_{j=1}^N \Mh\j_j , \]
where we can choose the transformations $\M_i$ independently on each
leg of the vertex. If the conformal transformations $M_i(\x)$ are
analytic at $\xi=0$ then one finds that one obtains the same
scattering amplitude and due to this freedom there is an enormous
variety of $N$-string vertices which has no counterpart in terms of
one-string vertices. One way to see this result is that if the
transformation is analytic at $\xi=0$ then the sum in \Eq{conftrafo}
begins with $n=-1$ and when saturated with physical states, those
terms in the sum which contain $L_n, n\ge 1$ annihilate on these
states. The terms with $L_0$ possibly give functions of $z_i$ if the
conformal transformation depends on these, and the $L_{-1}$'s
implement the shift of the Koba-Nielsen variables, discussed below, in
the vertex. Once one takes into account the calculation of the measure
using null state decoupling then one finds that it changes so as to
compensate for the latter two effects and the scattering amplitude is
the same.

The conformal transformations $\M_i$ will also shift the Koba Nielsen
points and yield new transition functions. If the new coordinates
$\tx_i$ are given by
\[ \tx_i = \M^{-1}_i (\x_i ) , \lb{comap1}  \]
the Koba Nielsen variables are transformed according to
\[ \td z_i = (\x_i^{-1}\cc\M_i \cc\x_i )(z_i).  \lb{KN-trafo} \]
The relation between the new transition functions and
the old ones is
\[ \ttr ij = \M^{-1}_i \cc \tr ij \cc \M_j . \lb{tauijtd}     \]
Next we recall from Sect.\ \ref{VERTEXOP-1} that a conformal field of
weight $h$ transforms as follows under a conformal mapping $\M$:
\[ \Mh^{-1} \F (\x ) \Mh  =
   \left[\dd{\M^{-1} (\x)}{\x}\right]^h \F (\M^{-1} (\x )) .
\lb{Phitrafo}  \]
The covariance of the overlap equations can be inferred from
\[ \VN \F\i (\x_i ) \. \prod_{k=1}^N \Mh\k_k =
   \tV N \F\i \big( \M_i^{-1} (\x_i )\big)
   \left[\dd{\M_i^{-1}(\x_i)}{\x_i}\right]^h . \lb{Nvertex3}  \]
Namely the new vertex $\tV N$ then obeys the overlap condition
\Eq{overlap1} with $\F\i$ replaced by the transformed fields
\Eq{Phitrafo} (for each $i$) and the new transition functions \Eq{tauijtd}.

Let us determine the conformal transformations which leave a
given vertex inert. If we demand $\ttr ij=\tr ij$ for all $i,j$, then Eq.\
\Eq{tauijtd} implies
\[ \M_j=\tr ji \cc \M_i \cc \tr ij \lb{inertrafo}. \]
This shows that, given a vertex, we always have the freedom to apply
an arbitrary conformal transformation $\M_i$ to a certain leg without
changing the vertex as long as we compensate by conformal
transformations on the other legs according to the above formula; this
will turn out to be useful later. Even though the vertex remains
invariant, however, the Koba Nielsen variables will be shifted by
virtue of \Eq{KN-trafo}.  Using \Eq{inertrafo}, it is easy to see that
in this case we have $\x_i^{-1}\cc \M_i \cc \x_i = \x_j^{-1}\cc \M_j
\cc \x_j$ for all $i,j$, and therefore the conformal transformation
acts {\em in the same way} on all Koba Nielsen variables.  For global
and univalent mappings of the Riemann sphere onto itself the freedom
is consequently reduced to one M\"obius transformation, which we
identify with the transformation used at the beginning of this chapter
to gauge-fix three Koba Nielsen points to arbitrary non-coincident
points.  We stress again that, while the on-shell vertices are all
equivalent to each other under arbitrary conformal transformations,
here we have a degree of freedom for the off-shell vertices.

Finally, we note that if $\Mh$ has conformal transformation $\M$ then
$\Mh\dg$ (or $\Mh\sh$) has conformal transformation
$\G\cc\M^{-1}\cc\G$ as can be seen by taking the adjoint of Eq.\
\Eq{Phitrafo} for $\F=\vX$ (so that $h=0$) and using relation
\Eq{X-sharp}. Thus
\[ \Mh\sh=\Mh\dg=\widehat{\phantom{\quad}\G\M^{-1}\G\phantom{\quad}}.
   \lb{conf-dg} \]

\section{Vertex operators and multistring vertices} \lab{THREEVERTEX}
Having set up the multistring formalism in general terms, we now wish
to relate it to the one-string formulation of Sect.\ \ref{VERTEXOP} As
we have already mentioned, the correspondence is not one-to-one as
there are many more multistring vertices than one-string vertex
operators; of course the final answer for the computation of physical
scattering amplitude is always the same, as the vertices differ only
off shell.  We will also explain in this chapter how multistring
vertices can be constructed from three-vertices by sewing, and how
explicit expressions for them can be obtained quickly and efficiently
by means of overlap identities.

\subsection{Vertex operators from three-vertices} \lab{THREEVERTEX-1}
Given any three-string vertex $\V3(z_1,z_2,z_3)$, derived from overlap
equations, we can define a linear map
$\cV: \cF\to (\End\cF)[\![z_1,z_1^{-1},z_2,z_2^{-1},z_3,z_3^{-1}]\!]$,
which assigns to each state $\p\in\cF$ a ``vertex operator''
$\cV(\p;z_1,z_2,z_3)$ defined by its action on all states $\f\in\cF$:
\[ \cV(\p;z_1,z_2,z_3)\f
   := \R_1\left[ \V3(z_1,z_2,z_3)\kets{\f}2\kets{\p}3 \right]. \]
Note that such an operator in general is a formal series in the Koba
Nielsen variables. Out of this huge number of vertex operators,
however, only a certain class leads to physically acceptable
operators. We shall see that the requirement of mutual locality of
vertex operators is equivalent to the postulate $\tr12=\G$ which in
turn implies $z_1=\infty$ and $z_2=0$ if we put $\z\equiv\x_2$.
Imposing the physical requirement that operators associated with
physical states should be primary fields of weight one w.r.t.\ the
variable $z_3$, we will be led to the same choice for $\tr12$.
Therefore, rather than using the most general three-vertex for the
definition of vertex operators, we will in this section deal with a
class of vertices $\V3(z)\equiv\V3(\infty,0,z)$ which depend on the
third Koba Nielsen variable $z_3\equiv z$ as a parameter while $z_1$
and $z_2$ are fixed. Consequently, the fundamental identity relating
such a three-string vertex $\V3(z)$ to a one-string vertex operator
$\Vp$ is
\[ \Vp\f := \R_1\left[ \V3(z)\kets{\f}2\kets{\p}3
                       \right]. \lb{def-vertopa}\]
Remarkably, the operator $\Vp$ is automatically normal-ordered since
its action on arbitrary Fock space elements is always well-defined by the
right-hand side. Somewhat loosely we can also write\footnote{A
  specific example of this equation relating one-string vertex
  operators and three-vertices was given in \ct{West94}.}
\[ \Vp := \R_1\left[ \V3(z)\kets{\p}3 \right], \lb{def-vertop} \]
where equality is to be understood in the sense of matrix elements.
It is immediately clear that the correspondence between three-vertices
$\V3(z)$ and vertex operators $\Vp$ (for all $\p$) cannot be
one-to-one, for we can always multiply the vertex $\V3(z)$ by
conformal transformations of type \Eq{inertrafo} which leave the
vertex inert and thus do not alter the left-hand side of
\Eq{def-vertop}. We should also emphasize that these three-vertices do
not in general lead to vertex operators which satisfy all the axioms
of a vertex algebra, but nonetheless can be used to construct the Lie
algebra of physical states since on-shell the vertices are all the
same. The only difference is made by their off-shell properties. In
Sect.\ \ref{EXPLICIT} we will give two explicit examples of such
vertices, one of which does lead via the above definition to the
vertex operator we have been working with in Sect.\ \ref{VERTEXOP}. In
fact, it will be much easier to prove the axioms of a vertex algebra
than to produce the explicit expressions for the vertex operators in
the oscillator representation.

Symbolically, we have
\[ \Vp \qquad\longleftrightarrow\qquad
   \unitlength.5mm
   \begin{picture}(70,24)
    \RVpic{}{}{\kets{\p}3}z
  \end{picture} \nn \] \vspace{6mm} \\
where normal-ordering is already built by the above argument; we
will later check this explicitly for the tachyon emission vertex.

Let us now turn to the discussion of the physical requirements which
truncate the family of three-vertices.  First, we demand that for any
physical state the corresponding vertex operator defined by Eq.\
\Eq{def-vertop} is a primary field of weight one.  As already
mentioned, in this case the integrated physical vertex operator
$\p_0\equiv\Res{z}{\Vp}$ commutes with the Virasoro generators and
maps physical states into physical states. It is then straightforward
to show that under these circumstances null states decouple from the
three-vertex if all legs are saturated with physical states, viz.
\[ \Res{z}{\V3(z)\,
           L\1_{-n}\kets{\chi}1\kets{\f}2\kets{\p}3}
    &=& \Big( L_{-n}\chi \,\Big|\, \Res{z}{\Vp}\f \Big) \non
    &=& \Big( \chi \,\Big|\, \big[L_n,\Res{z}{\Vp}\big]\f \Big) +
        \Big( \chi \,\Big|\, \Res{z}{\Vp}L_n\f \Big) \non
    &=&  0. \lb{nulldecouple} \]
Cyclicity of the three-vertex (see Sect.\ \ref{E10-1} below)
guarantees that this decoupling of spurious (and in particular null
physical) states holds for any leg.

Translated into the framework of three-string vertices the
condition \Eq{Primary} for $h=1$ reads
\[ \V3(z)\left\{ -L\2_n+L\1_{-n}-(n+1)z^n
                 \right\}-z^{n+1}\dz\V3(z)=0. \lb{dzonV3} \]
We note that this condition is sufficiently strong to also fix the
overall normalization of the vertex, as the last term is clearly
sensitive to multiplication of $\V3$ by a function of $z$.  We can
relax this condition by combining it with the condition for $n=0$,
which permits us to swap the $\dz$ for $L_0$ terms. The $z^n$ term can
only come from $L\3_0$ acting on $\kets\p3$, however we may also have
$L_n$ ($n\ge1$) terms in the corresponding overlap; they annihilate on
$\p$ as it is physical.  Thus if we require
\[ \V3(z)\left\{-L\2_n+L\1_{-n}+z^n(L\2_0-L\1_0)-nz^nL\3_0
                + \mbox{ terms in } L\3_m,\ m\ge1\right\} =0,
   \lb{nodzonV3} \]
instead of \Eq{dzonV3}, we get a much larger class of three-vertices
whose normalization is no longer fixed. In order to work out the
ensuing restrictions on the transition functions we must compare Eq.\
\Eq{nodzonV3} with the overlap equation \Eq{Loverlap1} for the energy
momentum tensor. Hence, if we choose $i=1$ and
$f(\x_1)=(\x_1)^{1-n}-z^n\x_1$, then to get the above equation we
demand
\[ f(\x_1)\dd{\x_2}{\x_1} &=& -(\x_2)^{n+1}+z^n\x_2, \non
   f(\x_1)\dd{\x_3}{\x_1} &=& -nz^n\x_3+\cO\big((\x_3)^2\big). \]
The first equation implies that $\x_1=(\x_2)^{-1}$, i.e.\
\[ \tr12=\G \,; \]
from the second equation we deduce that $f$ vanishes at $\x_3=0$ which
in turn means that $\x_1(z_3)=z^{-1}$ or $\x_2(z_3)=z$. Hence if we
choose $\x_2$ as the coordinate $\z$, then $\V3(z)$ has Koba
Nielsen points $\infty$, $0$ and $z$ for legs 1, 2 and 3,
respectively. Since the parameter $z$ just the Koba Nielsen point on
leg 3, the vertex operator $\Vp$ defined by \Eq{def-vertop} can be
interpreted as describing the emission of a physical state
$\p$ from the string at the third Koba Nielsen point $z_3=z$.

The above analysis shows how an essential physical assumption
immediately fixes the transition function $\tr12$ and constrains $\tr13$
very much. On the other hand, we emphasize that the weight one
postulate is somewhat stronger than decoupling of zero norm physical
states but with this requirement decoupling is manifest and it will
lead us to a class of vertices that are sufficient for our purposes.

We also demand the following creation property (cf.\ Eq.\ \Eq{Create-prop})
of the three-vertex:
\[ \lim_{z\to0}\cV(\p,z)\ket\vo=\p . \]
This property is equivalent to
\[ \lim_{z\to0}\R_1\Vt123(z)\kets{\vo}2\va\3_{-m}
   &{\stackrel!=}& \va\1_{-m}\lim_{z\to0}\R_1\Vt123(z)\kets{\vo}2 \non
   &=& \lim_{z\to0}\R_1\left[ \Vt123(z)\kets{\vo}2 \left(\va\1_{-m}\right)\sh
                  \right], \]
i.e.\ ,
\[ \lim_{z\to0}\Vt123(z)\kets{\vo}2\left\{\va\3_{-m}+\va\1_m\right\}
    \stackrel!=0. \lb{Vz=0}   \]
Comparing with the general overlap equations for the oscillators
\Eq{oscoverlap},
\[ \Vt123(z)\kets{\vo}2
    \bigg\{ \va\3_{-m} + \sqrt{m} \sum_{n \geq 1}
            C_{mn}^{31} \frac{\va\1_n}{\sqrt{n}} +
            \big[(\G \cc \tr31)(0)\big]^m \va\1_0 \bigg\} = 0 , \]
we conclude that
\[ \lim_{z\to0}C_{mn}^{31}(z)=\d_{mn}\qquad\mbox{and}\qquad
   \lim_{z\to0}[\G\cc\tr31(z;\_)](0)=0, \]
i.e.\ (by \Eq{Cmnij}),
\[ \tr31(z;\x)=\frac1\x+\sum_{n\ge1}f_n(\x)z^n, \]
where the $f_n$'s denote some functions which are not determined by
the creation property.

We now consider how some of the axioms of the vertex algebra approach
imply restrictions for the three-vertex via Eq.\ \Eq{def-vertop}.  We
begin with the vacuum axiom \Eq{Vac} which will be certainly fulfilled
if we postulate that $\pr12:=\R_1\Vt123(z)\kets{\vo}3$ is the natural
isomorphism which identifies the Fock spaces $\cF_2$ and $\cF_1$:
\[ \pr12\, :\, \cF_2\to\cF_1,\
               \kets{\p}2\mapsto\kets{\p}1. \lb{iso-12} \]
This means that we require
\[ \R_1\Vt123(z)\kets{\vo}3\va\2_{-m}
   {\stackrel!=} \va\1_{-m}\R_1\Vt123(z)\kets{\vo}3 , \]
i.e.\ ,
\[ \Vt123(z)\kets{\vo}3\left\{\va\2_{-m}+\va\1_m\right\} \stackrel!=0. \]
Observe that this equation is valid for arbitrary $z$ unlike
\Eq{Vz=0}. Recalling the general overlap equations for the oscillators
\Eq{oscoverlap},
\[ \Vt123(z)\kets{\vo}3
    \bigg\{ \va\2_{-m} + \sqrt{m} \sum_{n \geq 1}
            C_{mn}^{21} \frac{\va\1_n}{\sqrt{n}} +
            \big[(\G \cc \tr21)(0)\big]^m \va\1_0 \bigg\} = 0 , \]
we conclude that imposing $\tr12=\G$ is equivalent to
$\cV(\ket\vo,z)=\id_{\cF}$. Happily, we are thus led to
the same choice for $\tr12$ as above although, somewhat surprisingly,
no implications for the other transition functions emerge. In fact,
below we shall see that this choice for $\tr12$ is quite powerful for
it already implies locality for the vertex operators.

As regards the injectivity axiom we recall that its information is
already encoded in the above creation property so that we do not get a
new condition. Finally, regularity is fulfilled since the vertex
contains annihilation operators only and it acts on states with a
finite occupation number. Note that the class of three-vertices we
have considered so far, do not satisfy all the axioms. We will see
that in order to recover the specific vertex operator of Sect.\
\ref{VERTEXOP} we have to fix all the transition functions.

\subsection{Sewing of three-vertices} \lab{SEWING}
Originally, $N$-vertices were computed by sewing three-vertices; the
necessary calculations used to be rather cumbersome. We will now
describe a much quicker route to their explicit determination by a
procedure, the essential steps of which were already given in
\ct{NevWes88b}. All we need to do is to figure out the transition
functions for the sewn vertex from the transition functions of the
basic three-vertex used in the sewing procedure. Once this is
accomplished we must only substitute the results into our ``master
formula'' \Eq{Nvertex1} to arrive at the final result.

Sewing of two vertices means that we turn around a leg of the second
vertex and glue it together with another leg of the first vertex
by taking the string scalar product (after identifying the two Fock
spaces, of course). This is symbolically shown in the figure below:
\[ \Vt123(z)\sew2{2'}\R_{2'}\Vt{2'}45(w)
    \qquad\longleftrightarrow\qquad
   \unitlength.5mm
   \begin{picture}(142,24)
   \put(0,0){\unitlength.5mm
             \begin{picture}(70,24)
             \Vpic123z
             \end{picture}}
   \put(70.2,15){\makebox(0,0){$\cup$}}
   \put(72,0){\unitlength.5mm
              \begin{picture}(70,24)
              \RVpic{\phantom{{}_\prime}2'}45w
              \end{picture}}
   \end{picture} \] \vspace*{10mm} \\
Note that the notation $\sew ij$ indicates that leg $i$ of the
first vertex is sewn with leg $j$ of the second vertex; of course,
this only makes sense if either of the legs has been turned around.
Under the assumption $\tr12=\G$, the above four-vertex
will turn out to be symmetric under the simultaneous interchange
of the variables $z$ and $w$ and legs 3 and 5.

As indicated above, we now must only work out the transition functions
for the new four-vertex by repeated application of the unintegrated
overlap conditions.  This means that we act with $\vX\1(\x_1)$ on leg
1 of the composite vertex and then move it through the vertex by means
of the overlap equations. For example,
\[ \lefteqn{\Big[ \Vt123 (z) \vX\1 (\x_1 ) \Big]
            \sew2{2'} \R_{2'} \Vt{2'}45 (w) } \hh{22mm} \non
&=& \Vt123 (z) \vX\2 \big(\tr21(z; \x_1)\big) \sew2{2'}
    \R_{2'} \Vt{2'}45 (w)    \non
&=& \Vt123 (z) \sew2{2'} \vX^{(2')} \big( \tr21 (z;\x_1)\big)
    \R_{2'} \Vt{2'}45 (w)    \non
&=& \Vt123 (z) \sew2{2'} \R_{2'} \Big[ \Vt{2'}45 (w)
    \vX^{(2')} \big( [\G \cc \tr21 (z)](\x_1)\big) \Big]    \non
&=& \Vt123 (z) \sew2{2'} \R_{2'} \Vt{2'}45 (w)
    \vX\5 \big( [\tr5{2'}(w) \cc \G \cc
           \tr21 (z)] (\x_1) \big), \lb{pullthrough}   \]
where the first entry in $\tr ij (z;\_)$ indicates the dependence
of the transition function on the parameter, while the second entry is
the argument of the function. In the formulas below, we will usually
omit the argument and only indicate the parameters explicitly.  Note
that in the third step we have used the relation
$\vX\i(\x)\R_iV=
 \R_i\left[V\vX^{(i)\sharp}(\x)\right]=
 \R_i\left[V\vX\i(\G(\x))\right]$
which describes how $\vX$ commutes with the reversing operator. We
conclude that
\[ \tr51(z,w)(\x) = \big[\tr5{2'}(w)\cc\G\cc\tr21(z)\big](\x). \]
The above calculation can be carried over to the determination of the
other transition functions of the sewn vertex. The final result can be
easily read off from the diagram: the general rule is to compose the
basic transition functions appropriately and to keep in mind that
every sewing operation $\cup$ is accompanied by the insertion of a
factor $\G$ into the transition function\footnote{At the risk of
  appearing overly pedantic, we emphasize once more that the insertion
  of $\G$ is due to the operation $\R$, and not to the sewing as
  such.}. Repeated application of this method yields the remaining
transition functions
\[ \tr35(z,w) &=& \tr32(z)\cc\G\cc\tr{2'}5(w), \\
   \tr14(z,w) &=& \tr12(z)\cc\G\cc\tr{2'}4(w), \\
   \tr34(z,w) &=& \tr32(z)\cc\G\cc\tr{2'}4(w). \]
where we henceforth suppress the argument $\x$. Now, as functions, we have
\[ \tr12 &=& \tr{2'}4 ,  \non
   \tr13 &=& \tr{2'}5 , \non
   \tr23 &=& \tr45 .   \]
If we furthermore take $\tr12=\G$ then the transition functions for the
composite vertex simplify to
\[ \tr14 &=& \G, \lb{tau14-gen} \\
   \tr15 &=& \tr13(w), \\
   \tr34 &=& \tr32(z), \\
   \tr35 &=& \tr31(z)\cc\tr13(w). \lb{tau35-gen} \]
We observe that the transition functions remain the same under the
transformation $3\leftrightarrow5$, $z\leftrightarrow w$. For example,
\[ \tr53(w,z)=[\tr35(w,z)]^{-1}=[\tr31(w)\cc\tr13(z)]^{-1}
                =\tr31(z)\cc\tr13(w)=\tr35(z,w). \nn \]
Hence the four-vertex is indeed symmetric under this change; at the level of
vertex operators, however, this is just locality of Theorem \ref{thm1}.
Pictorially, we have
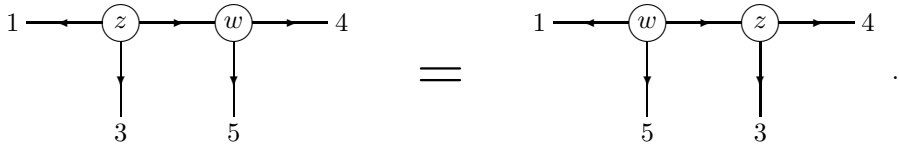
\[ \unitlength.5mm
   \begin{picture}(240,24)
   \put(0,0){\unitlength.5mm
             \begin{picture}(70,24)
             \Vpic1{}3z
             \end{picture}}
   \put(30,0){\unitlength.5mm
               \begin{picture}(70,24)
               \put(35,15){\circle{10}}
               \put(35,15){\makebox(0,0){$w$}}
               \put(52,15){\line(1,0){8}}
               \put(35,-2){\line(0,-1){8}}
               \put(40,15){\vector(1,0){12}}
               \put(35,10){\vector(0,-1){12}}
               \put(62,15){\makebox(0,0)[l]{$4$}}
               \put(35,-12){\makebox(0,0)[t]{$5$}}
               \end{picture}}
   \put(120,0){\makebox(0,0){\Huge=}}
   \put(140,0){\unitlength.5mm
               \begin{picture}(70,24)
               \Vpic1{}5w
               \end{picture}}
   \put(170,0){\unitlength.5mm
               \begin{picture}(70,24)
               \put(35,15){\circle{10}}
               \put(35,15){\makebox(0,0){$z$}}
               \put(52,15){\line(1,0){8}}
               \put(35,-2){\line(0,-1){8}}
               \put(40,15){\vector(1,0){12}}
               \put(35,10){\vector(0,-1){12}}
               \put(62,15){\makebox(0,0)[l]{$4$}}
               \put(35,-12){\makebox(0,0)[t]{$3$}}
               \end{picture}}
   \end{picture}
   . \nn \] \vspace*{10mm} \\
We stress that the overlaps are so powerful that the rather innocent
choice $\tr12=\G$ turns out to be equivalent to the principle of
locality for vertex operators. But as regards duality for vertex
operators in the sense of Theorem \ref{thm1}, we will see later that
this requires a specific choice for the other transition functions,
too.

In string theory it is often quite useful to sew vertices in different
ways. This will allow us to formulate the principle of duality and in
our context will be essential for the construction of commutators of
integrated physical vertex operators.
The  relation we would like to focus on next is
\[ \Vt123(z)\sew2{2'}\R_{2'}\Vt{2'}45(w)  =
   \Vt146(y) \sew6{6'}\R_{6'} \bar \Vt{6'}53 (z,w,y)  ,  \lb{sew1}  \]
defining some new vertex $\bar\Vt{6'}53 (z,w,y)$. The parameter $y$ on
the right-hand side is free and we can choose it as we like.  It is by
no means obvious that the left-hand side of \Eq{sew1} can always be
rewritten in the way indicated on the right-hand side; rather this is
a very nontrivial property of the underlying physical theory.
Pictorially,
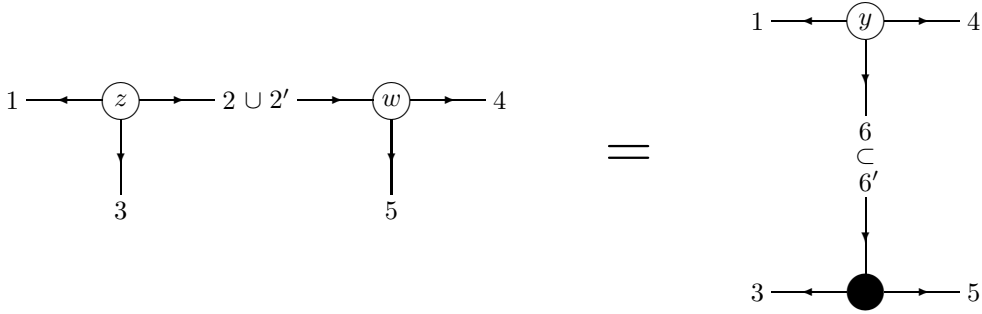
\[ \unitlength.5mm
   \begin{picture}(268,44)
   \put(0,0){\unitlength.5mm
             \begin{picture}(70,24)
             \Vpic123z
             \end{picture}}
   \put(70.2,15){\makebox(0,0){$\cup$}}
   \put(72,0){\unitlength.5mm
              \begin{picture}(70,24)
              \RVpic{\phantom{{}_\prime}2'}45w
              \end{picture}}
   \put(170,0){\makebox(0,0){\Huge=}}
   \put(198,21){\unitlength.5mm
                \begin{picture}(70,24)
                \Vpic146y
                \end{picture}}
   \put(233,0){\makebox(0,0){$\subset$}}
   \put(198,-51){\unitlength.5mm
                 \begin{picture}(70,50)
                 \put(35,15){\circle*{10}}
                 \put(18,15){\line(-1,0){8}}
                 \put(52,15){\line(1,0){8}}
                 \put(35,28){\line(0,-1){8}}
                 \put(30,15){\vector(-1,0){12}}
                 \put(40,15){\vector(1,0){12}}
                 \put(35,40){\vector(0,-1){12}}
                 \put( 8,15){\makebox(0,0)[r]{$3$}}
                 \put(62,15){\makebox(0,0)[l]{$5$}}
                 \put(35,42){\makebox(0,0)[b]{$\phantom{'}6'$}}
                 \end{picture}}
   \end{picture} \nn \] \vspace*{20mm} \\

To determine the new vertex $\bar\V3$ we use the same techniques as
above. Then the new transition functions come out to be
\[ \btr5{6'} &=& \tr31(w)\cc\tr13(y)\cc\G, \lb{tautild-1} \\
   \btr3{6'} &=& \tr31(z)\cc\tr13(y)\cc\G, \\
   \btr35 &\equiv& \tr35 = \tr31(z)\cc\tr13(w). \lb{tautild-3} \]
We observe that $\btr35$ does not depend on the free parameter $y$,
whereas for the choice $y=z$ or $y=w$ the transition functions
$\btr3{6'}$ or $\btr5{6'}$, respectively, reduce to $\G$. We shall
adopt the latter choice from now on, i.e.\ setting $y=w$, we finally obtain
\[ \btr5{6'} &=& \G, \lb{tautild-4} \\
   \btr3{6'} &=& \tr31(z)\cc\tr13(w)\cc\G. \lb{tautild-5} \]
This means that the vertex $\bar \Vt{6'}53 (z,w,w)$ is symmetric under
the simultaneous interchange of the variable $z$ and $w$ and legs $5$
and $6'$.

The above calculation together with the diagram can be regarded as the
demonstration of string duality for it expresses the same amplitude
alternatively as a sum over $s$-channel poles or $t$-channel poles.
We note, however, that the vertex $\bar\V3$ used to sew with on the
right-hand side of the above figure is in general not the same as the
vertex $\V3$ we began sewing with. We will see later that there does
exist a specific choice for the transition functions of $V(z)$ such
that $\bar\V3$ is equal to $V(z)$, too. Duality in the sense of
Theorem \ref{thm1} is a somewhat more restrictive notion which is only
satisfied by the specific vertex alluded to above. Hence the above can
be viewed as a generalisation of Theorem\ \ref{thm1}. The vertex
$\bar\V3$ will later prove useful when we analyze the commutator of
integrated physical vertex operators.

We now consider how we might construct a general class of
three-vertices $\V3 (z)$ with the above properties starting with an
arbitrary initial three-vertex $\brv\V3$ . One advantage of this way
of proceeding is that it will provide us with specific examples of
three-vertices $\V3 (z)$ and enable us to carry out the above, and
other, calculations for them. Although the above formalism is rather
elegant, to get a good feel for what is going on it is often best to
consider specific examples at least in the first instance.  It will
also allow us to see the relationship between well-known examples of
three-vertices, such as the old CSV vertex \ct{CaSchwVe69}, and the
vertices we use, and to regard the sewing of vertices above in a more
traditional manner namely the sewing of arbitrary vertices and
propagators.

To be more specific let us start with an arbitary three-vertex
$\brv\V3$, which satisfies the overlap equations for some given
transition functions $\vtr ij$; neither the vertex nor the transition
functions are assumed to depend on the parameter $z$.  Next we must
find a suitable $\V3(z)$ and hence $\bar\V3(z,w,w)$ which by the
results of the previous section is equivalent to the determination of
the conformal mappings $\M_i$ relating the vertices to one another.
{}From Eq.\ \Eq{tauijtd} we know that conformal transformations $\M_j$
exist such that
\[ \V3(z)=\brv\V3\prod_j\Mh\j_j. \]
Demanding $\tr12=\G$ implies that $\M_2=\vtr21\cc\M_1\cc\G$. We
recall that we can always choose one of the conformal transformations
at will (cf.\ Eq.\ \Eq{inertrafo}). We choose
\[ \M_1\equiv\s_z, \lb{M1-z} \]
with $\s_z(\x):=z\x$; the associated transformation $\Mh_1$ is then just
the scaling operator $\hat\s_z=z^{L_0}$. This immediately leads to
\[ \M_2=\vtr21\cc\s_z\cc\G. \lb{M2-z} \]
We also demand that $\tr23(0)=z$, since the vertex $\V3(z)$ should
have Koba Nielsen point $z_3=z$ for leg 3 in $\x_2$ coordinates. This
implies for $\M_3$
\[ \M_3(0)=\vtr31(1). \]
This condition determines only the translation part of $\M_3$, and the
remaining part is left arbitrary. For some of the specific vertices we consider
later this condition is already met and so no conformal
mapping is required on leg 3.

Having constructed $\V3(z)$, we can derive the vertex $\bar\V3(z,w,w)$
and work out the conformal mappings relating it to the initial $\brv
\V3$ as above. Let us take $\N_j$ ($j=1,2,3$) to be the conformal
transformations between $\V3(z)$ and $\bar\V3(z,w,w)$, i.e.\
\[ \bar\V3(z,w,w)=\V3(z)\prod_j\Nh\j_j. \]
{}From Eqs.\ \Eq{tauijtd}, \Eq{tautild-4} and \Eq{tautild-5} we deduce
the relations
\[ \G &=& (\N_1)^{-1}\cc\G\cc\N_2 \\
   \G\cc\btr31(w;\_)\cc\btr13(z;\_)
      &=& (\N_1)^{-1}\cc\tr13(z;\_)\cc\N_3. \]
We can choose $\N_3\equiv\id$ and then
\[ \N_1=\btr13(w;\_)\cc\G, \qquad \N_2=\G\cc\btr13(w;\_). \lb{N12-w} \]
In view of Eq.\ \Eq{conf-dg} we see that $\Nh_1\dg$ corresponds to
$(\N_2)^{-1}$.

{}From the general formula
\[ \R_1 \left[ \V3(z) \Nh\1_1 \Nh\2_2 \kets{\f}3 \right]
    = \Nh_1\dg \Vf \Nh_2    ,  \]
we obtain the associated one-string vertex operator, using
$\N\equiv\N_2$, $\Nh_1\dg\cong\N^{-1}$,
\[ \bar\cV(\f;z,w)
    &=& \R_1 \left[ \V3(z,w,w)\kets{\f}3 \right] \non
    &=& \R_1 \left[ \V3(z) \Nh\1 \Nh\2 \kets{\f}3 \right] \non
    &=& \Nh\dg \Vf \Nh \non
    &=& \left[\frac{d\N^{-1} (z)}{dz} \right]^h \cV ( \f,
           \N^{-1} (z)). \]
This relation tells us that when we put a state on their third
leg then $\V3(z)$ and $\bar\V3(z,w,w)$ lead to one-string vertex operators
that are related by the conformal transformation $\N^{-1}$. Note that it is
the transformation $\N^{-1}$ which feeds the $w$-dependence into the vertex
operator $\bar\V3(z,w,w)$.

Sewing of vertices in string theory was traditionally performed by
sewing vertices together with a propagator that involved a Feynman
like parameter $z$ (see e.g.\ \ct{AlAmBeOl71}). In contrast, we have
sewn the vertices $\V3(z)$ directly, i.e.\ with no propagator factor.
However, we can reconcile these two approaches, since the conformal
transformation between $\brv\V3$ and $\V3(z)$ can be interpreted as a
propagator and in this way of calculating we sew vertices $\brv\V3$
with propagator $\Mh_2(z)\Mh_1\sh(w)$ which, by Eqs.\ \Eq{conf-dg},
\Eq{M1-z} and \Eq{M2-z}, corresponds to the conformal transformation
$\vtr21\cc\s_{\frac z w}\cc\G$.

Finally, we can extend our calculation of the transition functions to the
$N$-vertex
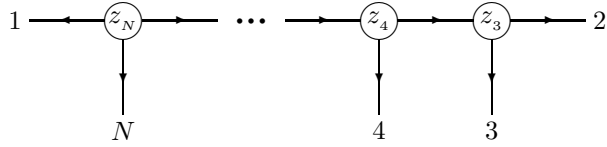
\[ \unitlength.5mm
   \begin{picture}(168,24)
   \put(0,0){\unitlength.5mm
              \begin{picture}(70,24)
              \Vpic1{}N{z_{_N}}
              \end{picture}}
   \put(66,15){\circle*{1.5}}
   \put(69,15){\circle*{1.5}}
   \put(72,15){\circle*{1.5}}
   \put(68,0){\unitlength.5mm
               \begin{picture}(70,24)
               \RVpic{}{}4{z_{_4}}
               \end{picture}}
   \put(98,0){\unitlength.5mm
               \begin{picture}(70,24)
               \put(35,15){\circle{10}}
               \put(35,15){\makebox(0,0){$z_{_3}$}}
               \put(52,15){\line(1,0){8}}
               \put(35,-2){\line(0,-1){8}}
               \put(40,15){\vector(1,0){12}}
               \put(35,10){\vector(0,-1){12}}
               \put(62,15){\makebox(0,0)[l]{$2$}}
               \put(35,-12){\makebox(0,0)[t]{$3$}}
               \end{picture}}
   \end{picture}
  . \nn \] \vspace*{10mm} \\
It is not difficult to arrive at the following results:
\[ \tr12 &=& \G, \non
   \tr1i &=& \tr13(z_i) \FOR{} 3\le i\le N, \nn \]
from which we deduce that
\[ \tr i1 &=& \tr 31(z_i) \FOR{} 3\le i\le N, \non
   \tr i2 &=& \tr 31(z_i)\cc\G \FOR{}3\le i\le N, \non
   \tr ij &=& \tr 31(z_i)\cc\tr13(z_j)
              \FOR{}3\le i,j\le N, \nn \]
expressed in terms of the transition function $\tr13$ for the basic
three-vertex.
Thus we can write down a generalization of locality for the $N$-vertex:
\[ \tr ij(z_i,z_j)= \tr ji(z_j,z_i) \FOR{}3\le i,j\le N\,, \]
i.e.\ the $N$-vertex is symmetric (in the sense of analytic
continuation of matrix elements) under simultaneous interchange of
legs $i$ and $j$ and the variables $z_i$ and $z_j$:
\[ \VN_{1\ldots N}(z_1,\ldots,z_N)
    =\VN_{12k_3\ldots k_N}(z_1,z_2,z_{k_3},\ldots,z_{k_N}) \lb{gen-loc} \]
for any permutation
$\left({3 \atop k_3}{\ldots \atop \ldots}{N \atop k_N}\right)$.

We now wish to demonstrate an analogue of duality for the $N$-string
vertex and construct the multistring analogue of
$\bar\V3(z,w,w)$. The sewing procedure is shown in the following figure:
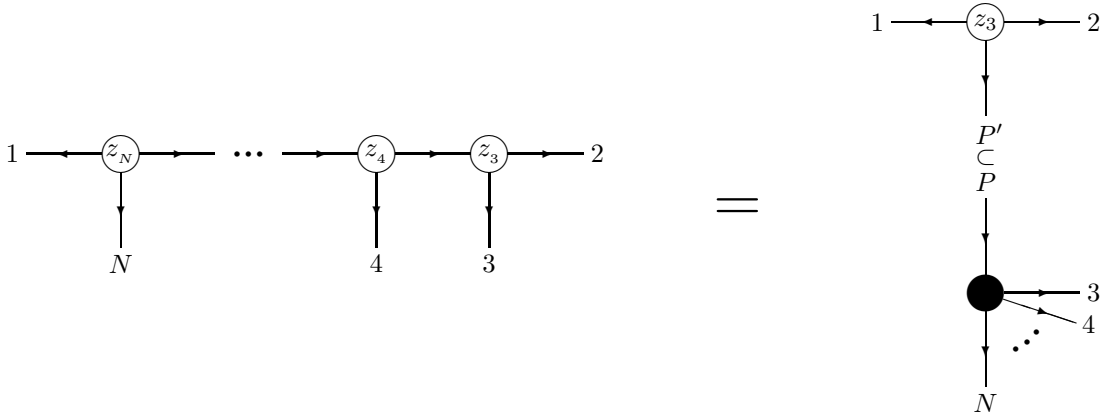
\[ \unitlength.5mm
   \begin{picture}(300,59)
   \put(0,0){\unitlength.5mm
              \begin{picture}(70,24)
              \Vpic1{}N{z_{_N}}
              \end{picture}}
   \put(66,15){\circle*{1.5}}
   \put(69,15){\circle*{1.5}}
   \put(72,15){\circle*{1.5}}
   \put(68,0){\unitlength.5mm
               \begin{picture}(70,24)
               \RVpic{}{}4{z_{_4}}
               \end{picture}}
   \put(98,0){\unitlength.5mm
               \begin{picture}(70,24)
               \put(35,15){\circle{10}}
               \put(35,15){\makebox(0,0){$z_{_3}$}}
               \put(52,15){\line(1,0){8}}
               \put(35,-2){\line(0,-1){8}}
               \put(40,15){\vector(1,0){12}}
               \put(35,10){\vector(0,-1){12}}
               \put(62,15){\makebox(0,0)[l]{$2$}}
               \put(35,-12){\makebox(0,0)[t]{$3$}}
               \end{picture}}
   \put(199,0){\makebox(0,0){\Huge=}}
   \put(230,35){\unitlength.5mm
               \begin{picture}(70,24)
               \Vpic12{\phantom{'}P'}{z_3}
               \end{picture}}
   \put(265,14){\makebox(0,0){$\subset$}}
   \put(230,-37){\unitlength.5mm
                \begin{picture}(70,50)
                \put(35,15){\circle*{10}}
                \put(51.13,9.62){\line(3,-1){8}}
                \put(52,15){\line(1,0){8}}
                \put(35,-2){\line(0,-1){8}}
                \put(35,28){\line(0,-1){8}}
                \put(39.74,13.42){\vector(3,-1){12}}
                \put(40,15){\vector(1,0){12}}
                \put(35,10){\vector(0,-1){12}}
                \put(35,40){\vector(0,-1){12}}
                \put(60.61,6.46){\makebox(0,0)[l]{$4$}}
                \put(62,15){\makebox(0,0)[l]{$3$}}
                \put(35,-12){\makebox(0,0)[t]{$N$}}
                \put(35,42){\makebox(0,0)[b]{$P$}}
                \put(43,0){\circle*{1.5}}
                \put(46,2){\circle*{1.5}}
                \put(48.5,4.67){\circle*{1.5}}
                \end{picture}}
\end{picture} \nn \] \vspace*{22.5mm} \\
The transition functions are calculated using the same techniques as
before. Consequently, they must agree with the ones above, i.e.\ the
vertex
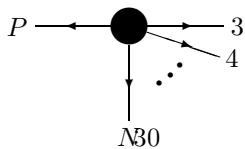
\[   \unitlength.5mm
  \begin{picture}(70,24)
   \put(35,15){\circle*{10}}
   \put(10,15){\line(1,0){8}}
   \put(51.13,9.62){\line(3,-1){8}}
   \put(52,15){\line(1,0){8}}
   \put(35,-2){\line(0,-1){8}}
   \put(39.74,13.42){\vector(3,-1){12}}
   \put(30,15){\vector(-1,0){12}}
   \put(40,15){\vector(1,0){12}}
   \put(35,10){\vector(0,-1){12}}
   \put(60.61,6.46){\makebox(0,0)[l]{$4$}}
   \put( 8,15){\makebox(0,0)[r]{$P$}}
   \put(62,15){\makebox(0,0)[l]{$3$}}
   \put(35,-12){\makebox(0,0)[t]{$N$}}
   \put(43,0){\circle*{1.5}}
   \put(46,2){\circle*{1.5}}
   \put(48.5,4.67){\circle*{1.5}}
  \end{picture} \nn \] \vspace*{10mm} \\
has transition functions
\[ \btr P3 &=& \G, \\
   \btr Pi &=& \tr13(z_i)\cc\tr31(z_3)\cc\G \FOR{} 4\le i\le N, \\
   \btr ij &=& \tr 31(z_i)\cc\tr13(z_j)
              \FOR{}3\le i,j\le N. \]

\subsection{Explicit construction of three-vertices} \lab{EXPLICIT}
We will now demonstrate how easily the formalism works by studying two
specific three-vertices, namely the CSV vertex $\CSV$ \ct{CaSchwVe69}
and the special vertex $\V3 (z)$ corresponding directly to the
one-string vertices $\Vp$ (for all choices of $\p$) used in section 2.
Studying different vertices and working out their properties is useful
even though all vertices yield the same scattering amplitudes when
sandwiched between physical states (when applying the formalism to
hyperbolic Kac Moody algebras such as $\0$, we will be concerned
exclusively with physical states). The traditional example of a
three-vertex is the $CSV$-vertex \ct{CaSchwVe69}; it is completely
characterized by the overlap functions
\[ \tr21 = \tr32 = \tr13 = \frac{1}{1-\x} \,,\qquad
   \tr31 = \tr23 = \tr12 = 1 - \frac{1}{\x} ,   \lb{CSVtauij}  \]
and therefore manifestly cyclic invariant (i.e.\ $\CSV_{123}=
\CSV_{231}= \CSV_{312}$) with Koba Nielsen points $\infty,0$ and 1.
For the matrices
$C^{ij}_{mn}$ occurring in \Eq{Nvertex2}, we obtain
\[ C^{12}_{mn} = C^{23}_{mn} = C^{31}_{mn}
     &=& (-1)^m\sqrt{\frac m n}{n \choose m}, \non
   C^{21}_{mn} = C^{32}_{mn} = C^{13}_{mn}
     &=& (-1)^n\sqrt{\frac n m}{m \choose n}, \]
and
\[ (\G\cc\tr 12)(0) = (\G\cc\tr 23)(0) = (\G\cc\tr 31)(0)
     &=& 0, \non
   (\G\cc\tr 21)(0) = (\G\cc\tr 32)(0) = (\G\cc\tr 13)(0)
     &=& 1,  \]
so that the overlaps are given by
\[ \CSV\bigg\{ \va\1_{-m}
              + \sum_{n=1}^\infty (-1)^m {n-1 \choose m-1} \va\2_n
              + \sum_{n=0}^m (-1)^n {m \choose n} \va\3_n \bigg\}
    = 0 , \lb{CSVosclap} \]
and cyclic permutations of this formula. With the above choice of
transition functions the overlaps for the Virasoro generators become
\[ \CSV\bigg\{ L\1_{-m}
              - \sum_{n=0}^\infty (-1)^{m+n} {1-m \choose n} L\2_{m+n}
              + \sum_{n=0}^{m+1} (-1)^n {m+1 \choose n} L\3_{n-1} \bigg\}
    = 0 , \]
again with their cyclically permuted analogues.

If we adopt $\CSV$ as our ``initial'' vertex, $\brv\V3$, then
the vertex $\V3(z)$ obtained by the procedure described in the
foregoing section has the transition functions
\[ \tr12 &=& \G, \non
   \tr13 &=& \s_{z^{-1}}\cc\vtr13 = \frac{\x-1}{z\x}, \non
   \tr23 &=& \G\cc\s_{z^{-1}}\cc\vtr13 = \frac{z\x}{\x-1}, \]
where we made use of Eqs.\ \Eq{tauijtd}, \Eq{M1-z} and \Eq{M2-z}.
In the above we have taken $\M_3=\id$ as the condition
$\M_3(0)=\vtr31(1)=0$ is then satisfied. We can also find the
corresponding vertex $\bar\V3(z,w,w)$ which has the transition
functions
\[ \btr12 &=& \G, \non
   \btr31 &=& \vtr31\cc\s_{\frac zw}\cc\vtr13\cc\G, \non
   \btr32 &=& \vtr31\cc\s_{\frac zw}\cc\vtr13, \]
by Eq.\ \Eq{sew1}.

As our second example we would like to identify the particular
three-vertex which makes complete contact with the
one-string vertex operators of Sect.\ \ref{VERTEXOP},
i.e.\ that three-vertex $\V3 (z)$ whose associated one-string
vertex operators $\Vp$ (defined by means of \Eq{def-vertop})
satisfy {\em all} the axioms of Sect.\ \ref{VERTEXOP} and not just
those required in Sect.\ \ref{THREEVERTEX-1}. For this special vertex
and the choice $y=w$ the vertex $\bar \Vt{6'}53$ will be the same as
the original one with the new argument $z-w$; this is just duality
(cf.\ Theorem \ref{thm2}). Let us remind the reader that although the
particular three-vertex we are about to construct has especially
simple properties and permits us to recover all the properties given
in Sect.\ \ref{VERTEXOP}, it is just one of an infinite number of
three-vertices that are equivalent once we sandwich them or their
associated $N$-vertices against physical states.

As before we need only work out the
transition functions $\tr12 , \tr23$ and $\tr31 = (\tr13 )^{-1}$,
from which the associated three-vertex follows directly as explained
in the preceding section. The task can thus be reduced to translating
the following relations (see \Eq{Tra}, \Eq{Trans-com}, \Eq{Scale-com})
\[ \frac{d}{dz} \Vf &=& \cV (L_{-1}\f, z) ,   \non
 \big[ L_0 , \Vf \big] &=& \left(z\frac{d}{dz} +h \right) \Vf  ,  \non
 \big[ L_{-1} , \Vf \big] &=& \frac{d}{dz} \Vf  ,  \lb{commLV}     \]
into the corresponding ones for the associated three-vertex $\V3$, which read
\[ \frac{d}{dz} \V3 (z) - \V3 (z) L\3_{-1} &=& 0 ,   \non
   \V3 (z) \Big\{ -L\2_0 + L\1_0 - L\3_0 - z L\3_{-1}  \Big\} &=& 0  ,  \non
   \V3 (z) \Big\{ -L\2_{-1} + L\1_1 - L\3_{-1} \Big\} &=& 0   ,
   \lb{3vLoverlap} \]
and matching them with the overlap conditions for the Virasoro
generators.  It is important that in contrast to the assumptions
previously made we here do not require the state $\f$ to be physical;
otherwise all equations would only hold up to terms involving $L\3_m$
for $m\geq 1$ which would remain undetermined. Demanding the vertices
to agree off-shell is clearly a much more stringent requirement, and
it is therefore not surprising that we will arrive at a unique answer
for $\V3 (z)$ in this case. It is also remarkable that the conditions
for $m=-1,0$ are sufficient to fix all the higher order identities.

Equation \Eq{3vLoverlap} can now be matched with \Eq{Loverlap1}
for $f(\x ) = \x $ and $f(\x ) = \x^2$, respectively. In this way,
we deduce the conditions
\[ \x_1 \frac{d\x_2}{d\x_1} &=& - \x_2 \,,\qquad \phantom{()}
   \x_1 \frac{d\x_3}{d\x_1}  =  - \x_3 - z ,        \non
   (\x_1)^2 \frac{d\x_2}{d\x_1} &=& -1 \,,\qquad
   (\x_1)^2 \frac{d\x_3}{d\x_1}  =  -1  .         \]
These differential equations are solved by $\x_i = \tr ij (\x_j)$ with
\[ \tr12 (\x ) &=& \tr21 (\x ) = \frac{1}{\x}
      \equiv \G (\x )   , \non
   \tr13 (\x ) &=& \frac{1}{\x +z} \qquad\Longleftrightarrow\qquad
   \tr31 (\x ) =  \frac{1}{\x} - z  ,   \non
   \tr23 (\x ) &=& \x + z .   \lb{3vtauij}     \]
Identifying the coordinate $\x_2$ with the global coordinate $\z$,
we get $z_2 =0$ for the second Koba Nielsen point; the other two
follow from $z_i = \tr 2{i} (0)$ and come out to be $z_1 =\infty$
and $ z_3 = z$.

As an illustration let us explicitly derive the tachyon emission
vertex operator (cf.\ \Eq{vertexop1}). We put the tachyonic state
$\ket\vv$ on the third leg. Consequently, in formula \Eq{Nvertex2} all
terms involving annihilation operators $\va\3_n$ vanish, viz.
\[ \V3(z)\kets\vv3
    = \vac \exp \bigg\{
       -\sum_{m,n \geq 1}
        \frac{\va\1_m}{\sqrt{m}} C_{mn}^{12} \frac{\va\2_n}{\sqrt{n}}
       -\sum_{i=1,2}\sum_{m\geq 1} \frac{\vv\X\va\i_m}{m}
        \big[(\G\cc\tr i3)(0)\big]^m \bigg\}
      \cN(\{\va\i_0\})\kets\vv3, \nn \]
where we have also taken into account that $(\G\cc\tr12)(0)=
(\G\cc\tr21)(0)= 0$ for our choice of $\tr12$. If we furthermore insert
$(\G\cc\tr13)(0)=z$, $(\G\cc\tr23)(0)=z^{-1}$, $C^{12}_{mn}=\d_{mn}$
and turn around the first leg we get
\[ \R_1\V3(z)\kets\vv3
    = \sum_{\vr_1,\vr_2,\vr_3\in\L \atop \vr_1+\vr_2+\vr_3=\vo}
      \hh{-1.5mm}\bras{\vr_2}2\bras{\vr_3}3
      \exp\bigg\{\sum_{m\geq1} \frac1m \va\1_{-m}\X\va\2_m
                 +i\vv\X\vX\1_<(z)
                 +i\vv\X\vX\2_>(z)\bigg\}
      z^{\vv\.\vr_2}\kets{-\vr_1}1\kets\vv3, \nn \]
also invoking formula \Eq{Nzeromode} for the zero mode term
$\cN$ (and discarding some phase factor). We can now use
$\vev{\vr_3}{\vv}=\d_{\vv,\vr_3}$ and perform the summation over
$\vr_1$ and $\vr_3$ to arrive at
\[ \R_1\V3(z)\kets\vv3
    = \sum_{\vr_2\in\L}
      \bras{\vr_2}2
      \exp\bigg\{\sum_{m\geq1} \frac1m \va\1_{-m}\X\va\2_m
                 +i\vv\X\vX\1_<(z)
                 +i\vv\X\vX\2_>(z)\bigg\}
      z^{\vv\.\vr_2}\kets{\vr_2+\vv}1. \nn \]
Finally, we consider the action of this expression on an arbitrary
state $\p$ on leg $2$. The role of the bilinear term in the
exponent is easily exhibited by noticing that it is nothing but the
natural isomorpism $\pr12$ (cf.\ \Eq{iso-12}) between the Fock spaces
$\cF_2$ and $\cF_1$:
\[ \pr12 := \R_1\V3(z)\kets\vo3 =
    \sum_{\vr_2\in\L}\bras{\vr_2}2
    \exp\bigg(\sum_{m\geq1}\frac1m\va\1_{-m}\X\va\2_m\bigg)
    \kets{\vr_2}1. \]
One can explicitly verify that
\[ \pr12 \va\2_{n_1}\cdots\va\2_{n_N}\kets{\vr}2
    = \va\1_{n_1}\cdots\va\1_{n_N}\kets{\vr}1. \]
Then, by linearity, one concludes that the operator $\pr12$ indeed
identifies $\cF_2$ with $\cF_1$.  The inverse of $\pr12$ is obviously
given by $\pr21$ and we also have $\bras{\p}1\pr12=\bras{\p}2$.  The
above expression we are interested in can now be written as
\[ \re{i\vv\.\vX\1_<(z)}
   \re{i\vv\.\vq\1}c\1_{\vv}
   z^{\vv\.\sma\1_0}
   \pr12
   \re{i\vv\.\vX\2_>(z)}, \]
where we have extracted from $\kets{\vr_2+\vv}1$ the factor
$e_{\vv}=\re{i\vv\.\vq}c_{\vv}$. Using the properties of $\pr12$ we
infer that we indeed recover the desired formula \Eq{vertexop1}. Note
that the map $\pr12$ is also responsible for the correct
normal-ordering of the final vertex operator.

The reader may be curious about the conformal transformations $\M_j$
which relate the above three-vertex to the CSV vertex, i.e.
\[ \V3 = \CSV \Mh_1 \Mh_2 \Mh_3.   \]
Applying the general prescription explained after Eq.\ \Eq{M1-z} we
arrive at
\[ \M_1(\x) = z\x \,,  \qquad
   \M_2(\x) = \frac{\x}{\x-z} \,, \qquad
   \M_3(\x) = -z^{-1}\x, \]
or, as operators,
\[ \Mh_1 = z^{L_0} \,,  \qquad
   \Mh_2 = z^{-L_0}(-1)^{L_0}\re{z^{-1}L_1} \,, \qquad
   \Mh_3 = z^{-L_0}(-1)^{L_0}. \]

Thus, sewing together $\V3(z)$ and $\V3(w)$ is equivalent to sewing
two $\CSV$'s with propagator
\[ \Mh_2(z)\Mh_1\sh(w)
    =z^{-L_0}(-1)^{L_0}\re{z^{-1}L_1}w^{L_0}
    =(-1)^{L_0}\re{L_1}\left(\frac wz\right)^{L_0}. \]
The first two factors on the right-hand side combine into what was
called twist-factor in the old days (see e.g.\ \ct{AlAmBeOl71}).

If we insert our special choice of transition functions into the
general formulas \Eq{tautild-1}--\Eq{tautild-3} for the transition
functions of the four-vertex we find
\[ \btr5{6'}(\x_{6'}) &=& \frac{1}{\x_{6'}} + y - w , \\
   \btr3{6'}(\x_{6'}) &=& \frac{1}{\x_{6'}} + y - z , \\
   \btr53 (\x_3 )     &\equiv& \tr53 (\x_3) = \x_3 + z - w . \]
Setting $y=w$, we finally obtain
\[ \btr5{6'}(\x_{6'})
   &=& \frac{1}{\x_{6'}} \qquad {\rm i.e.} \qquad
       \btr{6'}5 = \G = \tr12 ,    \non
   \btr3{6'} (\x_{6'})
   &=& \frac{1}{\x_{6'}} + w - z \qquad {\rm i.e.} \qquad
       \btr{6'}3 = \tr13 (z-w;\_) ,    \non
   \btr53 (\x_3)
   &=& \x_3 + z - w \qquad {\rm i.e.} \qquad
       \btr53 = \tr23 (z-w;\_) ,     \]
so that indeed $\bar\Vt{6'}53(z,w,w) \equiv \Vt{6'}53(z-w)$ and we have
established duality, viz.
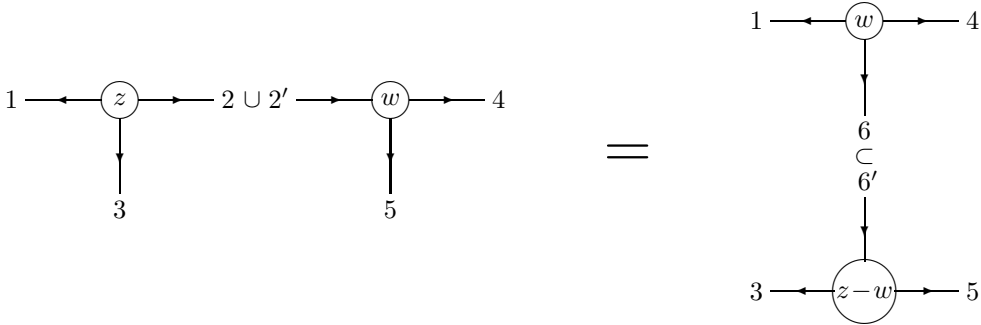
\[ \unitlength.5mm
   \begin{picture}(268,44)
   \put(0,0){\unitlength.5mm
             \begin{picture}(70,24)
             \Vpic123z
             \end{picture}}
   \put(70.2,15){\makebox(0,0){$\cup$}}
   \put(72,0){\unitlength.5mm
              \begin{picture}(70,24)
              \RVpic{\phantom{{}_\prime}2'}45w
              \end{picture}}
   \put(170,0){\makebox(0,0){\Huge=}}
   \put(198,21){\unitlength.5mm
                \begin{picture}(70,24)
                \Vpic146w
                \end{picture}}
   \put(233,0){\makebox(0,0){$\subset$}}
   \put(198,-51){\unitlength.5mm
                 \begin{picture}(70,50)
                 \put(35,15){\circle{16}}
                 \put(35,15){\makebox(0,0){$z\!-\!w$}}
                 \put(16.5,15){\line(-1,0){6.5}}
                 \put(53.5,15){\line(1,0){6.5}}
                 \put(35,29.5){\line(0,-1){6.5}}
                 \put(27,15){\vector(-1,0){10.5}}
                 \put(43,15){\vector(1,0){10.5}}
                 \put(35,40){\vector(0,-1){10.5}}
                 \put( 8,15){\makebox(0,0)[r]{$3$}}
                 \put(62,15){\makebox(0,0)[l]{$5$}}
                 \put(35,42){\makebox(0,0)[b]{$\phantom{'}6'$}}
                 \end{picture}}
   \end{picture} \nn \] \vspace*{20mm} \\
The symmetry of the new vertex under simultaneous interchanges of $z$
and $w$ and legs 3 and 5 is now manifest because $\tr53 (z-w) = \tr35
(w-z)$. The above identity is nothing but \Eq{Jac} for
vertex algebras, but now expressed in terms of three-vertices.  In
view of Theorem \ref{thm2} we have therefore established the Jacobi
identity \Eq{Jac} for the vertex operators defined by Eq.\
\Eq{def-vertop}. Note that in order to get the required off-shell
behaviour, we had to fix all transition functions.  We hope that
readers will appreciate the simplicity of our proof of this identity
which is substantially shorter, involving only simple manipulations
with overlaps and transition functions, and at the same time more
easily visualized than the one which can be found in the textbook
\ct{FLM88}.

The transition functions for the $N$-vertex are given by
\[ \tr12 &=& \G, \non
   \tr1i &=& \frac1{\x_i+z_i} \FOR{} 3\le i\le N. \nn \]
If we put formally $z_2\equiv0$ then these relations can be elegantly
summarized as
\[ \tr1i &=& \frac1{\x_i+z_i} \FOR{} 2\le i\le N, \nn \]
from which we deduce that
\[ \tr i1 &=& \frac1{\x_1} - z_i \FOR{} 2\le i\le N, \non
   \tr ij &=& \tr i1 \cc \tr1j = \x_j + z_j - z_i
    \FOR{}2\le i,j\le N. \nn \]
Recall that it was this choice of transition functions which led
to the especially simple result $\m =1$ for the measure. Identifying
the coordinate $\x_2$ with the global coordinate $\z$, we find that
$\tr 2j (0)=z_j$ for $2\le j\le N$ so that the Koba Nielsen points
indeed agree with the above formal variables $z_j$ for $2\le j \le N$;
whereas for $z_1$ we obtain the previous result $\infty$. The
symmetry of the $N$-vertex under simultaneous interchange of legs $i$
and $j$ and the variables $z_i$ and $z_j$ is now explicit: $\tr
ij(z_i-z_j)= \tr ji(z_j-z_i)$ for $2\le i,j\le N$.

We may also repeatedly apply the duality property to the $N$-vertex to
rewrite it as follows:
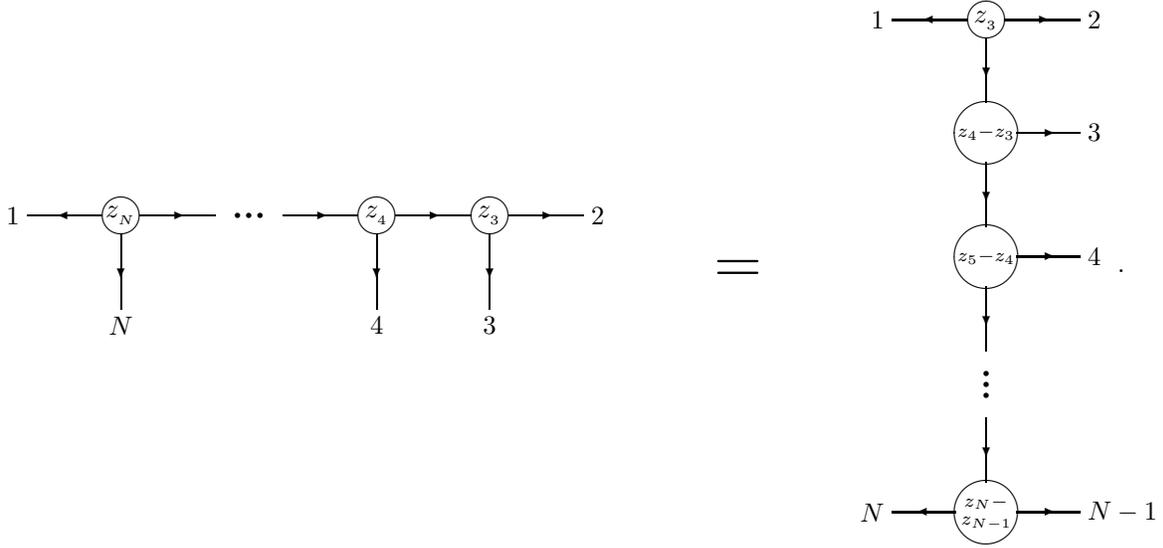
\[ \unitlength.5mm
   \begin{picture}(300,72)
   \put(0,0){\unitlength.5mm
              \begin{picture}(70,24)
              \Vpic1{}N{z_{_N}}
              \end{picture}}
   \put(66,15){\circle*{1.5}}
   \put(69,15){\circle*{1.5}}
   \put(72,15){\circle*{1.5}}
   \put(68,0){\unitlength.5mm
               \begin{picture}(70,24)
               \RVpic{}{}4{z_{_4}}
               \end{picture}}
   \put(98,0){\unitlength.5mm
               \begin{picture}(70,24)
               \put(35,15){\circle{10}}
               \put(35,15){\makebox(0,0){$z_{_3}$}}
               \put(52,15){\line(1,0){8}}
               \put(35,-2){\line(0,-1){8}}
               \put(40,15){\vector(1,0){12}}
               \put(35,10){\vector(0,-1){12}}
               \put(62,15){\makebox(0,0)[l]{$2$}}
               \put(35,-12){\makebox(0,0)[t]{$3$}}
               \end{picture}}
   \put(199,0){\makebox(0,0){\Huge=}}
   \put(230,52){\unitlength.5mm
               \begin{picture}(70,24)
               \put(35,15){\circle{10}}
               \put(35,15){\makebox(0,0){$z_{_3}$}}
               \put(52,15){\line(1,0){8}}
               \put(10,15){\line(1,0){8}}
               \put(40,15){\vector(1,0){12}}
               \put(30,15){\vector(-1,0){12}}
               \put(62,15){\makebox(0,0)[l]{$2$}}
               \put( 8,15){\makebox(0,0)[r]{$1$}}
               \end{picture}}
   \put(230,22){\unitlength.5mm
                 \begin{picture}(70,50)
                 \put(35,15){\circle{16}}
                 \put(35,15){\makebox(0,0){$\scriptstyle z_4 - z_3$}}
                 \put(53.5,15){\line(1,0){6.5}}
                 \put(35,29.5){\line(0,-1){6.5}}
                 \put(43,15){\vector(1,0){10.5}}
                 \put(35,40){\vector(0,-1){10.5}}
                 \put(62,15){\makebox(0,0)[l]{$3$}}
                 \end{picture}}
   \put(230,-11){\unitlength.5mm
                 \begin{picture}(70,50)
                 \put(35,15){\circle{16}}
                 \put(35,15){\makebox(0,0){$\scriptstyle z_5 - z_4$}}
                 \put(53.5,15){\line(1,0){6.5}}
                 \put(35,29.5){\line(0,-1){6.5}}
                 \put(43,15){\vector(1,0){10.5}}
                 \put(35,40){\vector(0,-1){10.5}}
                 \put(62,15){\makebox(0,0)[l]{$4$}}
                 \put(35,-3.5){\line(0,-1){6.5}}
                 \put(35,7){\vector(0,-1){10.5}}
                 \end{picture}}
   \put(265,-27){\circle*{1.5}}
   \put(265,-30){\circle*{1.5}}
   \put(265,-33){\circle*{1.5}}
   \put(230,-79){\unitlength.5mm
                 \begin{picture}(70,50)
                 \put(35,15){\circle{16}}
                 \put(35,15){\makebox(0,0){\shortstack{
                                           $\scriptstyle z_N -$ \\
                                           $\scriptstyle z_{N-1}$}}}
                 \put(16.5,15){\line(-1,0){6.5}}
                 \put(53.5,15){\line(1,0){6.5}}
                 \put(35,29.5){\line(0,-1){6.5}}
                 \put(27,15){\vector(-1,0){10.5}}
                 \put(43,15){\vector(1,0){10.5}}
                 \put(35,40){\vector(0,-1){10.5}}
                 \put( 8,15){\makebox(0,0)[r]{$N$}}
                 \put(62,15){\makebox(0,0)[l]{$N-1$}}
                 \end{picture}}
\end{picture}. \nn \] \vspace*{36mm} \\
Now we can write down the overlap equations for the oscillators
for this particular choice of $\VN$, as we will need them in Sect.\
\ref{E10}. We find
\[ C^{12}_{mn} = \d_{mn} ,&\qquad&
    (\G\cc\tr12)(0)=0, \non
   C^{1i}_{mn} = \sqrt{\frac n m}{m \choose n}z_i^{m-n} ,&\qquad&
    (\G\cc\tr1i)(0)=z_i\FOR{\ }3\le i\le N. \lb{C-N} \]
Thus
\[ \VN (z_3,\ldots,z_N)\bigg\{ \va\1_{-m} + \va\2_m
                + \sum_{i=3}^N\sum_{n=0}^m{m \choose n}
                   z_i^{m-n}\va\i_n \bigg\} = 0. \lb{Nosclap} \]
The Virasoro overlaps are given by
\[ \VN (z_3,\ldots,z_N)\bigg\{ L\1_{-m} - L\2_m -
    \sum_{j=3}^N \sum_{n=-1}^m{{m+1}\choose {m-n}} z_j^{m-n}
           L\j_n \bigg\} =0.           \]
We also record the generalization of \Eq{3vLoverlap} for arbitrary
$N$-vertices:
\[ \frac{\partial}{\partial z_j} \VN (z_3,\ldots,z_N)
    - \VN (z_3,\ldots,z_N) \, L\j_{-1} &=& 0
      \qquad(j\ge3),  \non
   \VN (z_3,\ldots,z_N) \,
    \bigg\{ L\1_0 - \sum_{j=2}^N
            \big( L\j_0 + z_j L\j_{-1}\big) \bigg\}  &=& 0, \non
   \VN (z_3,\ldots,z_N) \,
    \bigg\{ L\1_1 - \sum_{j=2}^N L\j_{-1}\bigg\} &=& 0,
   \lb{NvLoverlap}  \]
where, of course, $z_2 =0$. With the help of the above overlap equations
we can now explicitly verify that null states decouple when all legs of
the vertex except leg 1 are saturated with physical states
$\p_2,\ldots,\p_N$. A short calculation shows that
\[   \VN (z_3,\ldots,z_N) \kets{\p_2}2 \ldots\kets{\p_N}N \, L\1_{-m} =
  \sum_{j=3}^N \frac{\partial}{\partial z_j} \bigg( z_j^{m+1}
     \VN (z_3,\ldots,z_N ) \bigg)
      \kets{\p_2}2 \ldots \kets{\p_N}N,  \lb{NvLoverlap1}  \]
which vanishes upon integration over the variables
$z_3,\ldots,z_N$. Note that this relation is stronger than
\Eq{nulldecouple} since leg 1 is not contracted with a state.

We now come back to our claim at the beginning of this section that
the three-vertex my be regarded as an intertwining operator. To see
this let us recall the definition of such operators as given in
\ct{FrHuLe93} but with suitably modified notation.

\begin{defi} \lab{def-intertwiner} \hh{1em} \\[1.5ex]
  Let $(\cF,\cV,\one,\bw)$ be a vertex algebra and let
  $(\cF_i,\cV\i)$, $i=1,2,3$, be three modules for the vertex algebra.
  An {\bf intertwining operator of type} $\left({\cF_1 \atop \cF_3\
    \cF_2}\right)$ is a linear map $ \sI:\cF_3 \to
  \big({\rm Hom}(\cF_2,\cF_1)\big)\{z\}$, $\kets{\p}3 \mapsto
  \sI(z)\kets{\p}3$, satisfying the following properties: For every
  $\kets{\xi}1\in\cF_1$, $\kets{\f}2\in\cF_2$, $\kets{\p}3\in\cF_3$,
  \ben
\item {\bf (Regularity)}
 \[ \Res{z}{z^n\sI(z)\kets{\p}3\kets{\f}2}=0
     \FOR{ $n$ sufficiently large}; \]
\item {\bf (Translation)}
 \[ \dz\sI(z)\kets{\p}3=\sI(z)L\3_{-1}\kets{\p}3; \]
\item {\bf (Jacobi identity)}
 \[ \lefteqn{
     \Dm yzw \cV\1(\xi,z) \sI(w) \kets\p3 \kets\f2
     \Di ywz \sI(w) \kets\p3 \cV\2(\xi,z) \kets\f2}  \hh{8mm} \non
   &=&
     \Dm wzy \sI(w) \cV\3(\xi,y) \kets\p3 \kets\f2, \hh{48mm}
   \lb{intertwiner} \]
where binomial expressions have to be expanded in nonnegative
integral powers of the second variable.
 \een
\end{defi}

We are here interested in the case where $\cF_1,\cF_2,\cF_3$ are
isomorphic copies of $\cF$ and represent modules for the vertex
algebra via the adjoint action. Consequently, the intertwiner $\sI$
leads to the rather trivial fusion rules $N^1_{32}=1$. The interesting
point, however, is that the overlap equations for $\V3$ are very
reminiscent to the Jacobi identity for the intertwining operator
$\sI$. In fact, we can show that the two properties are equivalent to
each other in the same way as the Jacobi identity for vertex operators
\Eq{Jac} is related to the principles of locality and duality in
Theorem \ref{thm1}. To see this, we take matrix elements of the Jacobi
identity for interwiners, i.e.\ we pair \Eq{intertwiner} with an
arbitrary state $\bras\chi1$ from the left, and then apply
$\Res{y}{\ldots}$. Using the identity
\[ \Dm wzy = \Dm yzw \Di ywz, \]
we can mimic the proof of Theorem \ref{thm1} as presented in
\ct{FrHuLe93} and finally arrive at the following result.

\begin{theo} \lab{int-jac} \hh{1em} \\[-2.5ex]  \ben
\item
For $\chi,\p,\f,\xi\in\cF$, the formal series
$\bras\chi1\cV\1(\xi,z)\sI(w)\kets\p3\kets\f2 $ which involves only
finitely many negative powers of $w$ and only finitely many positive
powers of $z$, lies in the image of the map $\iota_{zw}$:
\[ \bras\chi1\cV\1(\xi,z)\sI(w)\kets\p3\kets\f2=\iota_{zw}f(z,w), \]
where the (uniquely determined) element $f\in\C[z,w]_S$
is of the form
\[ f(z,w)=\frac{g(z,w)}{z^rw^s(z-w)^t} \nn \]
for some polynomial $g(z,w)\in\C[z,w]$ and $r,s,t\in\Z$.
We also have
\[ \bras\chi1\sI(w)\kets\p3\cV\2(\xi,z)\kets\f2=\iota_{wz}f(z,w), \]
i.e.\ $\cV\1(\xi,z)\sI(w)\kets\p3$ agrees with
$\sI(w)\kets\p3\cV\2(\xi,z)$ as operator-valued rational functions.
\item
For $\chi,\p,\f,\xi\in\cF$, the formal series
$\bras\chi1\sI(w)\cV\3(\xi,y)\kets\p3\kets\f2$ which involves only
finitely many negative powers of $y$ and only finitely many positive
powers of $w$, lies in the image of the map $\iota_{wy}$:
\[ \bras\chi1\sI(w)\cV\3(\xi,y)\kets\p3\kets\f2=\iota_{wy}f(y+w,w), \]
with the same $f$ as above, and
\[ \bras\chi1\cV\1(\xi,y+w)\sI(w)\kets\p3\kets\f2
    =\iota_{yw}f(y+w,w), \]
i.e.\ $\cV\1(\xi,z)\sI(w)\kets\p3$ agrees with
$\sI(w)\cV\3(\xi,z-w)\kets\p3$ as operator-valued rational functions,
where the right-hand expression is to be expanded as a Laurent series
in $z-w$.
\een \end{theo} \medskip

If we now put $\sI(z)\equiv\R_1\V3(z)$ then the above theorem is just
a rigorous statement about the unintegrated overlap identities for the
three-vertex $\V3(z)$ with transition functions \Eq{3vtauij}. We note,
however, that the intertwiner Jacobi identity refers to a special
choice of transition functions whereas the unintegrated overlap
equations are valid in general. Hence it is reasonable to expect that
there is an extension of \Eq{intertwiner} which does not require a
specification of the transition functions. This would be the rigorous
(in the sense of formal calculus) analog of the overlap identities.
Besides from that point, the overlap identities are apparently much
more profound than the Jacobi identity for intertwiners since they
also hold for $N$-vertices. Finally, we emphasize that
\Eq{intertwiner} is stated as a {\em postulate} for the three-vertex
in the mathematical literature whereas it is proved here on the basis
of an explicit realization.

\section{$N$-string vertices and $E_{10}$} \lab{E10}
We now come to the heart of this paper, applying the multistring
formalism developed in the foregoing sections to the study of
hyperbolic Kac Moody algebras. For definiteness, we will only consider
$\0$ in the remainder, although some of our results are actually more
general, especially those concerning longitudinal states.  Recall that
for a given Cartan matrix $A$ the associated Kac Moody algebra $\ggA$
is recursively defined in terms of multiple commutators of the basic
Chevalley generators as explained in Sect.\ \ref{VERTEXOP} after
\Eq{Serre}. Hyperbolic Cartan matrices are indefinite and constrained
by the requirement that the deletion of any node from the Dynkin
diagram leaves a diagram which is either of finite or affine type. It
can then be shown that they possess one and only one negative
eigenvalue; the corresponding ``over-extended'' root will always be
denoted as $\vr_{-1}$ (the affine root is denoted by $\vr_0$).
Furthermore, the root lattice is Minkowskian in this case. By an
important result of \ct{Bour75} the rank of a hyperbolic Kac Moody
algebra cannot exceed $d=10$. Among the rank $d=10$ algebras the most
interesting is $\0$ with the even self-dual Lorentzian lattice $\L =
\II$ as its root lattice.  The central unsolved problem in the theory
of hyperbolic Kac Moody algebras is to find a manageable
representation for the multiple commutators of the Chevalley
generators analogous to the explicit realizations of the finite and
affine algebras; specifically, this requires to identify and discard
all those multiple commutators containing the Serre relations
\Eq{Serre} somewhere inside.

{}From Sect.\ \ref{VERTEXOP} we know already that for any given lattice
$\L$ (and in particular, any root lattice) one can construct a Lie
algebra of physical string states denoted by $\ggL$.  For indefinite
lattices, this Lie algebra, which is relatively easy to characterize,
is {\em not} the same as the Kac Moody algebra $\ggA$ generated by the
Chevalley states \Eq{Chevalley}, but contains it as a {\em proper}
subalgebra according to \Eq{inclus}. The problem then becomes one of
identifying the decoupled states which are in $\ggL$ but not in
$\ggA$, i.e.\ those physical states which cannot be reached via
multiple commutators.  The multistring formalism developed in the
foregoing sections furnishes a new method to analyze and compute
multiple commutators. This is achieved by rewriting any $N$-fold
multiple commutator in terms of an $(N+2)$-string vertex by attaching
the Lie algebra elements appearing in the commutator to the legs of
this vertex. This method permits an efficient and reasonably ``easy''
determination of the decoupled states by means of overlap identities.
Thus, instead of constructing the root space elements directly we will
search for the physical states which are in the complement of the root
space. An essential tool for this purpose is the DDF construction,
suitably adapted to the present context. For ``subcritical'' Kac Moody
algebras such as $\0$, both transversal and longitudinal DDF operators
can be shown to appear. To demonstrate the utility of the new
approach, we will show how to recover the results of Sect.\ 4.4 of
\ct{GebNic95} by explicitly determining which states in the root space
$\exam$ decouple from the three-vertex.  As for the longitudinal
states, our results go considerably beyond those of \ct{GebNic95} and
quite some way towards a more systematic understanding. The decoupling
of transversal states is even more remarkable, as it depends on the
lattice in a crucial manner and has no analog for continuous momenta
unlike the longitudinal decoupling; for the time being it remains a
mysterious phenomenon.  Nonetheless, we hope that the results
presented here convincingly support our claim that the entire
information about hyperbolic algebras such as $\0$ is encapsulated in
the $N$-vertices of the compactified string!

\subsection{The Lie algebra of physical states in terms of $N$-vertices}
\lab{E10-1}
Before turning to the hyperbolic Kac Moody algebra $\0$,
we consider the Lie algebra of physical states $\ggL$ discussed in
detail in Sect.\ \ref{VERTEXOP}. We recall that the commutator of two
physical states is given by Eq.\ \Eq{Lie-com}. Using \Eq{def-vertopa}
we can reexpress it by means of the special three-vertex $\V3(z)$
corresponding to the transition functions \Eq{3vtauij}. Specifically,
the commutator of two physical states $\p,\f$ in the Lie algebra of
physical states $\ggL$ is
\[ [\p,\f] &=& \Res{z}{\Vp\f} \non
           &=& \res_z\left\{\R_1\left[
                \V3_{123}(z)\kets{\f}2\kets{\p}3
                     \right]\right\}. \lb{def-comm} \]
We use the following pictorial representation of the commutator:
\[ [\p,\f] \qquad\longleftrightarrow\qquad
   \mbox{\Large {\rm Res}$_z$}
   \left[\hh{-3mm}
    \unitlength.5mm
    \begin{picture}(70,24)
    \RVpic{}{\kets{\f}2}{\kets{\p}3}z
    \end{picture}
    \hh{3mm}\right] \nn \]
We have thus a new and convenient way to calculate the commutator by
contracting two of the three legs of $\V3$ with the physical states
$\f$ and $\p$.  Although the state $\bra{\tilde {\bf 0}}$ is not
normalizable, the $N$-vertex does give rise to a normalizable Fock
space state upon contraction with $N-1$ states. The resulting element
of $\ggL$ can be visualized as the final state of a scattering process
of the states $\f$ and $\p$. As already pointed out before (cf.\ Eq.\
\Eq{Lie-def}), the result makes sense as a Lie algebra commutator only
on the factor space $\Pz1 / L_{-1} \Pz0$. This means that the final
state should be orthogonal to all states of the form $L_{-1}\chi$ with
$\chi \in \Pz0$; this decoupling is, however, a consequence of
\Eq{nulldecouple}. For more general $N$-vertices, it can be
established by means of Virasoro overlaps along the lines of
\Eq{NvLoverlap1}.

We next work out some simple consequences of Eq.\ \Eq{def-comm}. Let
$\p,\f,\chi\in\ggL$. Then
\[ \Res{z}{\V3(z)\kets{\chi}1\kets{\f}2\kets{\p}3}
    &=& (\chi|[\p,\f]) \non
    &=& (\f|[\chi,\p]) \non
    &=& \Res{z}{\V3(z)\kets{\f}1\kets{\p}2\kets{\chi}3}, \]
Consequently, invariance and symmetry of of the bilinear form
$(\_|\_)$ are equivalent to the on-shell cyclicity of the three-vertex
$\V3$. Note that although only the CSV vertex is manifestly cyclic,
upon contraction with physical states all three-vertices are cyclic.

The structure constants of $\ggL$ are defined by
\[ {f_{\p\f}}^\chi
    &:=& \vev{[\p,\f]}{\chi} \non
     &=& -(\th(\chi)|[\p,\f]) \non
     &=& -\Res{z}{\V3(z)\kets{\th(\chi)}1\kets{\f}2\kets{\p}3} \non
     &=& \bras\chi1\Res{z}{\R_1\V3(z)}\kets{\f}2\kets{\p}3. \lb{structure1} \]
Note that the vertex with one leg turned around gives the usual string
scalar product whereas the bare vertex leads to the invariant form for
$\ggL$. We can ``pull down'' the index $\chi$ by writing
\[ f_{\p \f \chi} := (\chi|[\p,\f]).  \lb{structure2}  \]
The corresponding metric is just the Cartan Killing metric on $\ggL$.
Since all the states used in \Eq{structure1} are physical and therefore
inert under the conformal transformations \Eq{comap1}, the structure
constants \Eq{structure1} are insensitive to the off-shell properties
of $\V3$.

Pictorially, we have
\[ {f_{\ket3 \ket2}}^{\ket1} \qquad\longleftrightarrow\qquad
   \mbox{\Large {\rm Res}$_z$}\left[
   \unitlength.5mm
   \begin{picture}(70,24)
   \RVpic{\bra1}{\ket2}{\ket3}z
   \end{picture} \right]. \nn \]
Evidently all the information about the structure constants and hence
the Lie algebra $\ggL$ is encoded into the vertex
$\Res{z}{\R_1\V3(z)}$ in this manner. Note that in the above formulas
for the structure constants we may drop the residue and put $z_3\equiv
z=1$ since we still have the freedom to fix a third (besides
$z_1=\infty$ and $z_2=0$) Koba Nielsen point due to global M\"obius
invariance.

Next we iterate \Eq{def-comm} to obtain a formula for the multiple
commutator. Evaluating the product of three-vertices amounts to
sewing them as we explained, which yields the formula
\[ [\p_N,[\p_{N-1},\ldots,[\p_3,\p_2]\ldots ]] =
   \res_{z_N}\ldots\res_{z_4}\Res{z_3}{\R_1\VN(z_3,\ldots,z_N)}
    \kets{\p_2}2\ldots\kets{\p_N}N   \lb{mult-comm}    \]
generalizing \Eq{def-comm} and \Eq{Lie-com}.
It is then natural to extend the above definition \Eq{structure1}
and to introduce structure constants for
$(N-2)$-fold commutators for $N\geq 4$. These are perhaps less familiar
objects but nonetheless very convenient.
More specifically, the generalized structure constants are associated with
$\Res{z_N\ldots z_4z_3}{\R_1\VN(\{z_i\})}$,
where $z_3,\ldots,z_N$ denote the formal variables with respect to which
the residues are to be computed:
\[ {f_{\p_N\ldots\p_2}}^{\p_1}
    &:=& \vev{[\p_N,[\p_{N-1},\ldots[\p_3,\p_2]]\ldots]}{\p_1} \non
     &=& \bras{\p_1}1
         \res_{z_N}\ldots\res_{z_4}\Res{z_3}{\R_1\VN(z_3,z_4,\ldots,z_N)}
         \kets{\p_2}2\ldots\kets{\p_N}N. \lb{struct-N} \]
Symbolically,
\[ {f_{\ket{N} \ldots \ket2 }}^{\ket1} \qquad\longleftrightarrow\qquad
   \mbox{\Large {\rm Res}$_{z_N,\ldots,z_4,z_3}$}
   \left[
   \unitlength.5mm
   \begin{picture}(168,24)
   \put(0,0){\unitlength.5mm
              \begin{picture}(70,24)
              \RVpic{\bra1}{}{\ket N}{z_{_N}}
              \end{picture}}
   \put(66,15){\circle*{1.5}}
   \put(69,15){\circle*{1.5}}
   \put(72,15){\circle*{1.5}}
   \put(68,0){\unitlength.5mm
               \begin{picture}(70,24)
               \RVpic{}{}{\ket4}{z_{_4}}
               \end{picture}}
   \put(98,0){\unitlength.5mm
               \begin{picture}(70,24)
               \put(35,15){\circle{10}}
               \put(35,15){\makebox(0,0){$z_{_3}$}}
               \put(52,15){\line(1,0){8}}
               \put(35,-2){\line(0,-1){8}}
               \put(40,15){\vector(1,0){12}}
               \put(35,10){\vector(0,-1){12}}
               \put(62,15){\makebox(0,0)[l]{$\ket2$}}
               \put(35,-12){\makebox(0,0)[t]{$\ket3$}}
               \end{picture}}
   \end{picture}
   \right]. \nn \]
We stress that the special vertex characterized by \Eq{3vtauij}
is just such that, for any $N$-vertex obtained from it by sewing, the
relevant factors in \Eq{Amplitude} combine to give unity upon fixing
the first three Koba Nielsen points to $\infty$, $0$ and $1$,
respectively, so we need no longer worry about them. In other words,
the generalized structure constants for an $(N-2)$-fold commutator of
physical string states are obtained by simply integrating the
saturated $N$-vertex over the Koba Nielsen variables $z_4,\dots,z_N$
without any extra factors.  We have already pointed out that since all
attached states are physical, we would obtain the same result with any
other $N$-vertex; however, we then would have to keep track of the
extra measure factors in \Eq{Amplitude}, which would no longer equal
unity.

For completeness and to establish the link with the more traditional
vertex operator construction based on \Eq{GO-com1}, we rephrase the
above results in terms of (ordinary) commutators of multistring
vertices. We stress that the latter are a priori different from the
prescription \Eq{Lie-com}, which maps two physical states to another
physical state in a way compatible with the properties of a Lie
bracket. As already mentioned the integrated vertex operators defined
via \Eq{def-vertop} realize the adjoint representation of the Lie
algebra of physical states. Carrying out the commutation of two such
operators corresponds to sewing $\R_1\Vt123(z)\kets{\p}3 \sew2{2'}
\R_{2'}\Vt{2'}45(w)\kets{\f}5$ and integrating their antisymmetrized
combination with respect to the attached states $\kets\p3$ and
$\kets\f5$ over $z$ and $w$.  Observe that the relative factor of
$(-1)^{\vp_3\.\vp_5}$ between these two terms, which would normally be
present, has already been taken care of by our definition
\Eq{bra-vactilde} of the bra-vacuum, which includes the cocycle
factors.  The symmetry $z\leftrightarrow w$, $3\leftrightarrow5$, of
the vertex, implies that we may apply Cauchy's theorem to find
\[ \lefteqn{
    \Oint_{|z|>|w|} \frac{dz}{2\pi i}\Oint_0\frac{dw}{2\pi i}\left\{
    \R_1\Vt123(z)\kets{\p}3 \sew2{2'} \R_{2'}\Vt{2'}45(w)\kets{\f}5
    \right\}
    - \left({z\leftrightarrow w \atop 3\leftrightarrow5}\right)}
    \hh{12mm} \non
    &=& \Oint_0\frac{dw}{2\pi i}\Oint_w\frac{dz}{2\pi i}\left\{
        \R_1\Vt123(z)\kets{\p}3 \sew2{2'} \R_{2'}\Vt{2'}45(w)\kets{\f}5
        \right\} \non
    &=& \Oint_0\frac{dw}{2\pi i}\bigg\{
        \R_1\Vt146(w) \sew6{6'} \Oint_w\frac{dz}{2\pi i}
        \R_{6'} \bar \Vt{6'}53 (z,w,w)\kets\p3\kets\f5 \bigg\},
        \lb{mult-comm1}\]
where we have used duality of Eq.\ \Eq{sew1} (with the choice $y=w$) in the
last step to reexpress the four-vertex; this derivation is the
multistring analog of the calculation leading from \Eq{GO-com1} to
\Eq{GO-com3}. We repeat, however, that \Eq{mult-comm1} is an operator on
string Fock space (with two unsaturated legs, i.e.\ a ``matrix'')
rather than a state. As such we can now identify the four-vertex on the
right with the contour integrals as the commutator of the integrated
physical vertex operators associated with $\p$ and $\f$, repectively.
Examining the transition functions $\btr ij$ for the vertex
$\bar\V3(z,w,w)$ of Eqs.\ \Eq{tautild-4} and \Eq{tautild-5}, we see
that when $z=w$ they become independent of $w$. It follows that the
residue of $\bar\V3(z,w,w)$ as $z\to w$ is independent of $w$.
Consequently, the expression $ \oint_w\frac{dz}{2\pi i} \bar \Vt{6'}53
(z,w,w)\kets\p3\kets\f5 $ is independent of $w$ and hence the $w$
integral sees only the first factor. This factor
$\oint_0\frac{dw}{2\pi i}\R_1\Vt146(w)$ is the integrated vertex
operator corresponding to the state on leg $6$.  We have in the above
calculation considered the commutator of linear operators, but we have
learnt that given the physical states $\p$ on leg $3$ and $\f$ on leg
$5$, we can assign to them the new state
\[ [\p,\f]=\Oint_w\frac{dz}{2\pi i} \R_{6'}
           \bar \Vt{6'}53 (z,w,w)\kets\p3\kets\f5, \lb{commstates} \]
which is nothing but Eq.\ \Eq{Lie-com} of Sect.\ \ref{VERTEXOP-1} by
virtue of the relation between $\bar\V3(z,w,w)$ and $\V3(z)$ of Eq.\
\Eq{N12-w} and the fact that $\V3(z)$ has conformal weight one. In
this way of viewing things, we calculate directly the commutator of
two integrated vertex operators by sewing. The resulting vertex
operator is associated with the state of Eq.\ \Eq{commstates}. So
we may equivalently think of the commutator in terms of a commutator
of states and recover the viewpoint of Sect.\ \ref{VERTEXOP}.
We conclude that \Eq{mult-comm1} can be reexpressed as
\[    \Oint_{|z|>|w|} \frac{dz}{2\pi i}\Oint_0\frac{dw}{2\pi i}\left\{
    \R_1\Vt123(z)\kets{\p}3 \sew2{2'} \R_{2'}\Vt{2'}45(w)\kets{\f}5
    \right\}
    - \left({z\leftrightarrow w \atop 3\leftrightarrow5}\right)
     = \Oint_0\frac{dw}{2\pi i}
        \R_1\Vt146(w) \kets{[\p,\f]}6,  \lb{mult-comm2}   \]
which corresponds to the last line of \Eq{GO-com3}. We repeat that
these arguments are valid only on-shell. Since the reader may be
puzzled by the fact that we employed the three-vertex in \Eq{def-comm}
to define the commutator of states whereas in the discussion above we
used four-vertices to introduce the commutator, we will now clarify
this point. Suppose we put a (physical) state $\x$ on leg 2 in
\Eq{mult-comm2}. Then we find the state $[[\p,\f],\x]$ on leg 1. This
seems to contradict our definition \Eq{mult-comm} which would give the
state $[\p,[\f,\x]]$ on leg 1 as the result of the scattering process.
This superficial inconsistency is resolved by noticing that the
left-hand side of \Eq{mult-comm2} contains two terms both of which are
of the form \Eq{mult-comm}. Hence the identity \Eq{mult-comm2} would
lead to $[\p,[\f,\x]]- [\f,[\p,\x]]= [[\p,\f],\x]$ which is just the
Jacobi identity! This means that applying the commutator of two
integrated vertex operators to a state is the same as acting on the
state with the integrated vertex operator associated with the
commutator of states. Hence we are again (cf.\ the discussion after
\Eq{Lie-jac}) led to conclude that the integrated vertex operators
realize the adjoint representation of the Lie algebra of states.

Just as before we can now iterate the above calculation to obtain a
formula for the multiple commutator. We wish to evaluate
\[ \Oint_0\frac{dz_3}{2\pi i}
   \Oint_{|z_4|>|z_3|}\frac{dz_4}{2\pi i}\ \ldots\
   \Oint_{|z_{N+1}|>|z_{N}|}\frac{dz_{N+1}}{2\pi i}
   \R_1 \V{N+1}(z_3,z_4,\ldots,z_{N+1})
   \kets{\p_3}3\kets{\p_4}4\ldots\kets{\p_{N+1}}{N+1} \pm \ldots\, ,
    \lb{multicom1} \]
where the dots stand for the remaining terms appearing in the multiple
commutator with appropriate signs and reorderings of the contours. In
analogy with the arguments leading from \Eq{GO-com1} to \Eq{GO-com2}
we employ the generalization of locality \Eq{gen-loc} to arrive at the
nested contour integral
\[ \Oint_0\frac{dz_3}{2\pi i}
   \Oint_{z_3}\frac{dz_4}{2\pi i}\ \ldots\
   \Oint_{z_N}\frac{dz_{N+1}}{2\pi i}
   \R_1 \V{N+1}(z_3,z_4,\ldots,z_{N+1})
   \kets{\p_3}3\kets{\p_4}4\ldots\kets{\p_{N+1}}{N+1}, \]
which again has two unsaturated legs. As before, we interpret the
$(N+1)$-vertex sandwiched with $\kets{\p_3}3 \kets{\p_4}4 \ldots
\linebreak[0] \kets{\p_{N+1}}{N+1}$ as the multiple commutator of the
associated integrated physical vertex operators. Employing duality as
discussed at the end of Sect.\ \ref{SEWING} we thus obtain
\[ \Eq{multicom1}
   &=&
   \Oint_0\frac{dz_3}{2\pi i} \R_1\Vt12{P'}(z_3) \sew{P'}{P}
   \Oint_{z_3}\frac{dz_4}{2\pi i}\ \ldots\
   \Oint_{z_N}\frac{dz_{N+1}}{2\pi i}
   \R_P\bar\V{N}_{P34\ldots N+1}(z_3,z_4,\ldots,z_{N+1})
   \kets{\p_3}3\kets{\p_4}4\ldots\kets{\p_{N+1}}{N+1} \non
   &=&
   \Oint_0 \frac{dz_3}{2\pi i} \R_1 \Vt123(z_3)\,
   \kets{[\p_{N+1},[\p_N,\ldots,[\p_4,\p_3]]\ldots]}3, \]
where we have defined
\[ \lefteqn{
   [\p_{N+1},[\p_N,\ldots,[\p_4,\p_3]]\ldots]} \hh{8mm} \non
   &:=&   \Oint_{z_3}\frac{dz_4}{2\pi i}\ \ldots\
          \Oint_{z_N}\frac{dz_{N+1}}{2\pi i}
          \R_P\bar\V{N}_{P34\ldots N+1}(z_3,z_4,\ldots,z_{N+1})
          \kets{\p_3}3\kets{\p_4}4\ldots\kets{\p_{N+1}}{N+1}.
   \lb{multicom2}  \]
This shows that the multiple commutator of integrated vertex operators
yields a vertex operator which is associated with the above state on
leg 3. Hence we may think of the multiple commutator of the states
$\p_3, \ldots, \p_{N+1}$ as given by this state and so recover the
viewpoint of Sect.\ \ref{VERTEXOP}. Note that the two definitions
\Eq{mult-comm} and \Eq{multicom2} are completely equivalent. To see
this we recall that $V$ and $\bar V$ agree on-shell so that we may insert
for $\bar\V{N}_{P34\ldots N+1}(z_3,z_4,\ldots,z_{N+1})$ the
expression $\VN_{P34\ldots N+1}(z_4-z_3,\ldots,z_{N+1}-z_N)$ (cf.\
the diagram before \Eq{C-N}). Upon performing the shift of variables
$z_i\to z_i'= z_i-z_{i-1}$ (for $4\le i\le N+1$) the integrals become
residues around zero and thus we indeed recover \Eq{mult-comm}.

\subsection{$\0$: a brief review} \lab{E10-2}
Most of the information about the hyperbolic Kac Moody algebra $\0$
available in the mathematical literature can be gleaned from
\ct{KaMoWa88}; for more general information about hyperbolic (and
indefinite) Kac Moody algebras, readers are advised to consult
\ct{Kac90} as well as
\ct{Mood79,FeiFre83,Fren85,KaMoWa88,Sacl89,FeFrRi93,MarSci93}. We will
here briefly summarize \ct{GebNic95,GebNic94}, whose notation and
conventions we will adhere to, and where readers can find more details
about the specific results needed here.

The simple roots of $\0$ are given by the lattice vectors
$$ \begin{array}{l@{\quad=\quad(}r@,r@,r@,r@,r@,r@,r@,r@,r@{\|}l}
   \vr_{-1} & 0 & 0 & 0 & 0 & 0 & 0 & 0 & 1 & -1 & 0),  \non
   \vr_{ 0} & 0 & 0 & 0 & 0 & 0 & 0 & 1 & -1 & 0 & 0),  \non
   \vr_{ 1} & 0 & 0 & 0 & 0 & 0 & 1 & -1 & 0 & 0 & 0),  \non
   \vr_{ 2} & 0 & 0 & 0 & 0 & 1 & -1 & 0 & 0 & 0 & 0),  \non
   \vr_{ 3} & 0 & 0 & 0 & 1 & -1 & 0 & 0 & 0 & 0 & 0),  \non
   \vr_{ 4} & 0 & 0 & 1 & -1 & 0 & 0 & 0 & 0 & 0 & 0),  \non
   \vr_{ 5} & 0 & 1 & -1 & 0 & 0 & 0 & 0 & 0 & 0 & 0),  \non
   \vr_{ 6} & -1 & -1 & 0 & 0 & 0 & 0 & 0 & 0 & 0 & 0),  \non
   \vr_{ 7} & \frc12 & \frc12 & \frc12 & \frc12 &
   \frc12 & \frc12 & \frc12 & \frc12 & \frc12 & \frc12),  \non
   \vr_{ 8} & 1 & -1 & 0 & 0 & 0 & 0 & 0 & 0 & 0 & 0)    ,   \nn
\end{array} $$
and correspond to the Dynkin diagram
\[ \unitlength1mm
   \begin{picture}(66,10)
   \multiput(1,0)(8,0){9}{\circle*{2}}
   \put(49,8){\circle*{2}}
   \multiput(2,0)(8,0){8}{\line(1,0){6}}
   \put(49,1){\line(0,1){6}}
  \end{picture}   \]
The Cartan matrix is $A_{ij} = \vr_i \X \vr_j$ ($-1\leq i,j \leq 8$),
where the scalar product is to be computed with the Minkowski metric
$(+...+|-)$ on the lattice.  The root lattice of $\0$ generated by
these simple roots is the unique even self-dual Lorentzian lattice
$\II$ in ten dimensions (see \ct{Conw83} for details).  In the
following, the $\9$ null root $\vd$ will play an important role; it is
\[  \vd=\sum_{i=0}^8n_i\vr_i
       = (0,\ 0,\ 0,\ 0,\ 0,\ 0,\ 0,\ 0,\ 1\|1), \lb{nullroot} \]
where the coefficients $n_i$ (called {\bf marks} of $\9$)
can be read off from
\[       \left[\begin{array}{*{9}{c}}
                &   &   &   &    &    &  3 &   &   \\
               0& 1 & 2 & 3 &  4 &  5 &  6 & 4 & 2
               \end{array} \right]. \]
The fundamental Weyl chamber $\eC$ of $\0$ is the convex cone
generated by the {\bf fundamental weights} $\vL_i$ which obey
$\vL_i \X \vr_j = - \d_{ij}$ and are explicitly given by
\[ \vL_i = - \sum_{j=-1}^8 (A^{-1})_{ij} \vr_j
           \FOR{$i=-1,0,1,\ldots 8$} \]
in terms of the inverse Cartan matrix. Thus,
\[ \vL \in \eC    \qquad\Longleftrightarrow\qquad
   \vL = \sum_{i=-1}^8 k_i \vL_i \]
for $ k_i \in \Z_+$.  A special feature of $\0$ is that its root
and weight lattices coincide as the lattice is self-dual.  Since Weyl
transformations preserve multiplicities and since every root can be
brought into $\eC$ by means of a Weyl transformation, it is sufficient
to consider only roots belonging to $\eC$.  Note also that the null
root is just the first fundamental weight $\vL_{-1} = \vd$ and that
all other fundamental roots are timelike.  Hence the fundamental Weyl
chamber lies entirely within the lightcone, touching it at precisely
one edge. This is a very special feature of $\0$, and also important
for the DDF construction of \ct{GebNic95}, as otherwise one would have
to deal with two or even more null directions at the same time.

A useful notion in the theory of hyperbolic Kac Moody algebras is the
{\bf level} $\ell$ of a given root $\vL$, which is defined as the
number of times the over-extended root $\vr_{-1}$ occurs in it
(counted negatively for negative roots); equivalently, it is the
number of $e_{-1}$ generators (for positive roots) or minus the number
of $f_{-1}$ generators (for negative roots) in the multiple commutator
corresponding to the root $\vL$. A convenient formula for $\ell$ is
\[   \ell \equiv \ell (\vL) := -\vd \X \vL     \lb{level}  \]
where $\vd$ is given in \Eq{nullroot}. Obviously, $\ell$ is not
invariant under the full Weyl group $\eW (\0)$ but only under its
affine Weyl subgroup $\eW(\9)$.  The level derives its importance from
the fact that it grades the hyperbolic algebra with respect to its
affine subalgebra.  This is the content of the following important
theorem \ct{FeiFre83,FeFrRi93}.

\begin{theo} \lab{thm4} \hh{1em} \\[1ex]
Suppose that $x$ is an element of $\0$ at level $\cl$.
Then it can be represented as a linear combination of $(\cl -1)$-fold
commutators of level-one elements, viz.
\[ x = [x_1,[x_2,\ldots[x_{\cl-1},x_{\cl}]\ldots]]    \]
where each $x_i$ contains exactly one generator $e_{-1}$
in the right-most position, i.e.
\[ x_1 = [e_{i_1},[e_{i_2},\ldots[e_{i_k},e_{-1}]\ldots]]     \]
with $i_\nu \in \{0,1,\ldots,8 \}$, and similarly for the other $x_i$.
\end{theo} \medskip

Specializing to $\0$, we can thus write
\[ \0 = \bigoplus_{\cl=-\infty}^\infty E_{10}^{[\cl]}  \lb{graded1}      \]
with
\[ E_{10}^{[\cl]}
    = \underbrace{\big[ E_{10}^{[1]},\big[ E_{10}^{[1]},\ldots,
                  \big[ E_{10}^{[1]},}_{(\cl-1) {\rm\ times}}
                   E_{10}^{[1]} \big]\big]\ldots\big]  \lb{graded2}   \]
for positive $\cl$ where $E_{10}^{[\cl]}$ is the level-$\cl$ sector of
the full algebra (for negative $\cl$, one must take the contragredient
representation generated by multiply commuting $E_{10}^{[-1]}$; this
amounts to a simple replacement of all $e_i$'s by $f_i$'s).  The
level-zero sector $E_{10}^{[0]}= \9$ is just the affine subalgebra,
and the level-one elements $E_{10}^{[1]}$ are known to constitute the
so-called basic representation \ct{Kac90,KaMoWa88}. In terms of the
graded decomposition \Eq{graded1} one can now in principle study the
hyperbolic algebra level by level. All higher level sectors form (in
general reducible) representations of the affine subalgebra, and since
they can be generated through multiple commutators of level-one
elements, it should be possible to understand them in terms of the
irreducible representations obtained by taking multiple products of
level-one representations.  This idea has been successfully applied so
far only to the level-two states, where it was possible to derive a
general multiplicity formula \ct{KaMoWa88,FeiFre83}, but (to the best
of our knowledge) not to level three and beyond. The problem here is
that not all the representations appearing in the product occur in the
Kac Moody algebra, because some of them drop out on account of the
Serre relation $[e_{-1},[e_{-1},e_0]]=0$. To identify the ones which
are not present would require complete knowledge of the theory of
irreducible representations of Kac Moody algebras of arbitrary level.

In \ct{GebNic95} an attempt was made to tackle the problem from a
different and at the same time more ``physical'' point of view by
exploiting the vertex operator construction of affine Kac Moody
algebras \ct{GodOli85,FreKac80}, which associates the elements of the
affine algebra with the physical string vertex operators for the
emission of tachyons and photons. The modern formalism of vertex
algebras \ct{FLM88,Borc86} relates the elements of a given indefinite
(or generalized) Kac Moody algebra to the higher (massive) excitations
of a string. A major advantage of this construction is that it not
only affords a concrete realization of the abstract algebra but also
that the $\Lo$ physical state condition \Eq{Phys-cond1} immediately
implies the Serre relations, so the corresponding elements of the
(free) Lie algebra are eliminated from the outset. However, this does
not mean that there is an ``easy'' realization of the hyperbolic
algebra as we must now identify the states which are elements of the
Lie algebra of physical states $\ggI$ associated with the $\0$ root
lattice, but {\em not} in $\0$. Although the problem of isolating
decoupled states bears some resemblance to the problem of discarding
multiple commutators involving the Serre relations, one should realize
that the decoupled states have nothing to do with the Lie algebra
elements that must be discarded because of the Serre relations, as the
latter never appear in the vertex algebra construction.

As explained in \ct{GebNic95} all states in $\ggL$ can be represented
in terms of DDF operators.  The new, although not completely
unexpected, feature is the relevance of longitudinal DDF for
subcritical (i.e.\ rank $d<26$) Kac Moody algebras. The longitudinal
states decouple only for $d=26$, in which case one is led to the
``fake monster'' $\gfake$ of \ct{Borc92} (taking into account extra
lightlike simple roots). In the DDF framework, the level-zero and
level-one states are still relatively easy to understand: the affine
subalgebra corresponds to the tachyon and photon states, and the
level-one states can be reinterpreted as purely transversal states
built on the tachyon state $\ket{\vr_{-1}}$ (remember that $\vr_{-1}$
is the over-extended root) and its orbit under the action of the Weyl
group $\eW (\9)$. Beyond level one it was shown in \ct{GebNic95} by
explicit computation that longitudinal states appear while at the same
time certain transversal states decouple. The root spaces can then be
understood rather explicitly in terms of polarization states for the
corresponding root by analyzing which physical states cannot be
reached via multiple commutators.

Let us now recall some basic features of the DDF construction. As is
well known (see e.g.\ \ct{Scher75}), the DDF operators require for
their definition a tachyon momentum $\vv$ (thus $\vv^2 =2$) and a
lightlike vector $\vk = \vk (\vv)$ (thus $\vk^2 =0$) obeying $\vk\X\vv
=1$.  For continuous momenta, such vectors can always be found and
rotated into a convenient frame, but on the lattice this is in general
not possible. For this reason we must extend the root lattice by
rational points in order to apply the DDF construction in the present
setting. To see this let us define the {\bf DDF decomposition} of a
given level $\ell$ root $\vL$ by \ct{GebNic95}
\[ \vL = \vv - n\vk    \lb{DDFmomenta} \]
where $\vk \equiv \vk (\vv):= -\frc{1}{\ell} \vd$, and $n:= 1 - \frc12
\vL^2$ is the number of steps required to reach the root by starting
from $\vv$ and decreasing the momentum by $\vk$ at each step.
Consequently, we have a factor $\ell^{-1}$ in the definition of $\vk$
for level $\ell$, so clearly neither $\vv$ nor $\vk$ will in general
be elements of lattice any more.  To analyze the full algebra at {\em
  arbitrary} level we have to make use of the rational extension of
the lattice (which is again self-dual, unlike the ``intermediate''
lattices for finite $\ell$).  In addition, we need a set of
transversal polarization vectors $\vxc = \vxc (\vv , \vk )$ subject to
$\vxc \X \vv = \vxc \X \vk = 0$, which for convenience we assume to be
orthonormalized. We here use letters $c,...=1,...,8$ from the
beginning of the alphabet for the transversal indices as we reserve
the letters $i,j,...$ for the labeling of the one-string Fock spaces.
Given these data, we can define the {\bf transversal DDF operators} by
(cf.\ \ct{DeDiFu72})
\[ \Av{c}{m} := \Res{z}
   {\vxc \X \vP (z) \re{im\vk(\vv) \. \vX (z) } }. \lb{Atransv}\]
These operators describe the emission of a photon with momentum $m\vk$
and polarization $\vxc$ for the string.  Note that no normal-ordering
is required here. It is, however, required in the definition of {\bf
  longitudinal DDF operators}, which we will also need and which are
given by (cf.\ \ct{Brow72})
\[ A^-_m (\vv) := \sL_m(\vv) -\frc12 \sum_{a=1}^8 \sum_{n\in \Z}
  \Ord \Av{a}{n} \Av{a}{m-n} \Ord      \lb{Along1}  \]
with
\[ \sL_m (\vv) := \Res{z}{ \ord \Big( - \vv\X\vP (z) + \frc12 m
   \frac{d}{dz} \log \big(\vk\X\vP(z)\big)
   \Big) \re{im\vk(\vv)\cdot \vX (z)}\ord }  \lb{Along2}   \]
and
\[ \Ord\Av{a}{m}\Av{b}{n}\Ord
    := \cases{ \Av{a}{m}\Av{b}{n} & if $m\le n$, \cr
               \Av{b}{n}\Av{a}{m} & if $m>n$, } \]
where we used $\vv\X\vk =1$.
Unlike in \ct{GebNic95} we will work with the operators $A^-_m$
instead of $\sL_m$, because they commute with the transversal DDF
operators. In the sequel we will be careful to indicate the dependence
of the DDF operators on their associated DDF momenta as we did above.
In string theory these labels are usually omitted as one considers
only a single set of DDF operators for a fixed pair $(\vv,\vk)$ of DDF
momenta, but here we will be required to simultaneously use {\em
  different} sets of DDF operators. Strictly speaking, we should even
include the lightlike moment and write $A^a_m(\vv, \vk)$. Here we can
suppress the second label because for $\0$ all $\vk (\vv)$ in the
fundamental Weyl chamber are proportional to the null root $\vd$, with
the proportionality factor being unambiguously determined by $\vv$.

In the remainder of this subsection, we shall provide some more
details about the specific example studied in \ct{GebNic95}.  The
particular root space analyzed there is associated with the level-two
root $\vL = \vL_7$ which is dual to the simple root $\vr_7$, and this
will also be our principal example here. Explicitly, $\vL_7$ is given
by
\[ \vL_7=\left[\begin{array}{*{9}{c}}
                &   &   &   &    &    &  7 &   &   \\
               2& 4 & 6 & 8 & 10 & 12 & 14 & 9 & 4
               \end{array} \right]
        =(0,\ 0,\ 0,\ 0,\ 0,\ 0,\ 0,\ 0,\ 0\|2), \]
so $\vL_7^2 = -4$. Its decomposition into two level-one
tachyonic roots is $\vL_7=\vr+\vs+2\vd$, where
\[ \vr:=\vr_{-1}
      &=&\left[\begin{array}{*{9}{c}}
                &   &   &   &    &    &  0 &   &   \\
               1& 0 & 0 & 0 &  0 &  0 &  0 & 0 & 0
               \end{array} \right]
        =(0,\ 0,\ 0,\ 0,\ 0,\ 0,\ 0,\ 1, -1\|0), \non
   \vs&:=&
         \left[\begin{array}{*{9}{c}}
                &   &   &   &    &    &  1 &   &   \\
               1& 2 & 2 & 2 &  2 &  2 &  2 & 1 & 0
               \end{array} \right]
       = (0,\ 0,\ 0,\ 0,\ 0,\ 0,\ 0, -1, -1\|0). \nn \lb{vrvs}  \]
Since $n=1-\frc12\vL_7^2=3$ we have $\vL_7=\vv-3\vk (\vv)$ with
$\vk (\vv):=-\frc12 \vd$ and
\[ \vv:=\vr+\vs-\vk
       =(0,\ 0,\ 0,\ 0,\ 0,\ 0,\ 0,\ 0, -\frc32\|\frc12). \nn  \lb{vv} \]
Observe the extra factor of $\frc12$ in $\vk (\vv)$ as appropriate for
level two, whereas we have $\vk(\vr) = \vk (\vs)= -\vd$.  As
anticipated neither $\vk$ nor $\vv$ are elements of $\II$.
Nevertheless, since $\vv\X\vk=1$, the action of the DDF operators on
the tachyonic ground-state $\ket{\vv}$ is well defined.  The three
sets of transversal polarization vectors associated with the tachyon
states $\ket{\vr}$, $\ket{\vs}$ and $\ket{\vv}$ will be denoted by
$\vxc\equiv\vxc(\vv) , \vza\equiv\vza(\vr)$ and $\vyb\equiv\vyb(\vs)$,
respectively. Explicitly,
\[  \vxi^1 = \vze^1 = \vet^1
    &:=&(1,\ 0,\ 0,\ 0,\ 0,\ 0,\ 0,\ 0,\ 0\|0), \non
         &\vdots& \non
   \vxi^7 = \vze^7 = \vet^7
    &:=&(0,\ 0,\ 0,\ 0,\ 0,\ 0,\ 1,\ 0,\ 0\|0); \nn \\[1em]
   \vze^8 &:=&(0,\ 0,\ 0,\ 0,\ 0,\ 0,\ 0,\ 1,\ 1\|1), \non
   \vet^8 &:=&(0,\ 0,\ 0,\ 0,\ 0,\ 0,\ 0, -1,\ 1\|1), \non
   \vxi^8 &:=&(0,\ 0,\ 0,\ 0,\ 0,\ 0,\ 0,\ 1,\ 0\|0). \nn \lb{polarizations} \]
(the notation here is slightly different from \ct{GebNic95}).
In accordance with definitions \Eq{Atransv} and \Eq{Along2}, the
associated DDF operators are designated by $\Av{c}{m}, \Ar{a}{m}$ and
$\As{b}{m}$, and $A^-_m(\vv), A^-_m(\vr)$ and $A^-_m(\vs)$,
respectively.

Beyond $\ell =2$ longitudinal states appear, while certain transversal
states decouple. All elements of $\exam$ can be reached via the
following commutators of level-one states and their $\eW (\9 )$
rotated versions \ct{GebNic95}:
\[ && [\Ar{a}{-1} \ket{\vr} \, , \, \As{b}{-1} \ket{\vs} ], \non {}
   && [\Ar{a}{-1} \Ar{b}{-1} \ket{\vr} \, ,\, \ket{\vs} ] ,  \non  {}
   && [\Ar{a}{-2} \ket{\vr} \, ,\, \ket{\vs}] .  \lb{commlev2}  \]
The resulting states can be expressed in terms of DDF operators
appropriate to the new groundstate $\ket{\vv}$ as follows
(no summation on repeated indices):
\[ \Av{a}{-2} \Av{b}{-1} \ket{\vv}&&
     \FOR{$a,b$ arbitrary}   ;  \non
   \Av{a}{-1}\Av{b}{-1}\Av{c}{-1}\ket{\vv}&&
     \FOR{$a \neq b,c \, ; \, b\neq c$} ; \non
   \Big( \Av{a}{-3} - \Av{a}{-1}\Av{b}{-1}\Av{b}{-1} \Big) \ket{\vv}&&
     \FOR{$a \neq b$}  ;  \non
   \Big( 5\Av{a}{-3}^i + \Av{a}{-1}\Av{a}{-1}\Av{a}{-1}\Big) \ket{\vv}&&
     \FOR{$a$ arbitrary}  ;    \non
   \Av{a}{-1} A^-_{-2}(\vv) \ket{\vv}&&
     \FOR{$a$ arbitrary} ;  \lb{E10-lev2}  \]
Altogether, we get $64 + 2 \X 56 + 2\X 8 = 192 $ states in agreement
with the formula in \ct{KaMoWa88}. Despite the fact that this number
coincides with the number of transversal states, we explicitly see the
appearance of longitudinal as well as the disappearance of some
transversal states. The orthogonal complement of $\exam$ in
$\ggI^{(\smvL_7)}$ w.r.t.\ the usual string scalar product
\Eq{contraform} (for which $(A^a_{-m})\dg=A^a_m$) is spanned by
the nine missing states
\[  &&  A^-_{-3} (\vv) \ket{\vv}  ,    \non
    &&  \bigg( 2 \Av{a}{-3} - 8 \Av{a}{-1} \Av{a}{-1} \Av{a}{-1}
               + 3 \Av{a}{-1} \sum_{b=1}^8 \Av{b}{-1} \Av{b}{-1} \bigg)
        \ket{\vv}.  \lb{E10-notlev2} \]
These states {\em cannot} be reached by multiple commutation of the
$\0$ Chevalley generators. Observe that neither \Eq{E10-lev2}
nor \Eq{E10-notlev2} are $SO(8)$ covariant.

Our main point in the remaining sections will be to demonstrate that
these decoupled states can be directly and rather efficiently
determined by means of the multistring vertices introduced in the
foregoing sections. Thus, we here take the opposite approach to the
one followed in \ct{GebNic95}: instead of computing the root space
elements \Eq{E10-lev2} directly, we seek to identify the states
\Eq{E10-notlev2} which are {\em not} in the root space by showing that
they decouple from the string vertex. As we will see the overlap
identities discussed before furnish the requisite tools for this task.
In accordance with the representation \Eq{graded2}, we represent the
general level-$\ell$ element as an $(\ell -1)$-fold commutator of
level-one elements. By formula \Eq{struct-N} above, the corresponding
string state can be alternatively viewed as the outcome of a string
scattering process with $\ell$ incoming level-one states on the
external legs of the string vertex, i.e.  as the ``out-state''
emerging on the unsaturated leg (so this process involves a vertex
with $N= \ell +1$ external legs). The scattering process is depicted
in the diagram below.
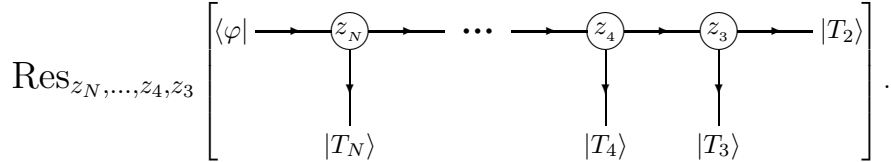
\[ \mbox{\Large {\rm Res}$_{z_N,\ldots,z_4,z_3}$}
   \left[
   \unitlength.5mm
   \begin{picture}(168,24)
   \put(0,0){\unitlength.5mm
              \begin{picture}(70,24)
              \RVpic{\bra\f}{}{\ket{T_{N}}}{z_{_{N}}}
              \end{picture}}
   \put(66,15){\circle*{1.5}}
   \put(69,15){\circle*{1.5}}
   \put(72,15){\circle*{1.5}}
   \put(68,0){\unitlength.5mm
               \begin{picture}(70,24)
               \RVpic{}{}{\ket{T_4}}{z_{_4}}
               \end{picture}}
   \put(98,0){\unitlength.5mm
               \begin{picture}(70,24)
               \put(35,15){\circle{10}}
               \put(35,15){\makebox(0,0){$z_{_3}$}}
               \put(52,15){\line(1,0){8}}
               \put(35,-2){\line(0,-1){8}}
               \put(40,15){\vector(1,0){12}}
               \put(35,10){\vector(0,-1){12}}
               \put(62,15){\makebox(0,0)[l]{$\ket{T_2}$}}
               \put(35,-12){\makebox(0,0)[t]{$\ket{T_3}$}}
               \end{picture}}
   \end{picture} \hh{1.5mm}
   \right]. \nn \]
The crux of the matter is now that not every physical state at
level $\ell$ can be produced by scattering level-one states (remember
that these are just the transversal states associated with the
tachyonic groundstate $\ket{\vr_{-1}}$). Whether a given level-$\ell$
state $\f$ decouples or not can be tested by attaching it to the above
vertex: if $\f$ decouples, this contraction must vanish for all
possible choices of level-one states on the other legs.  That is, $\f$
must then obey the condition
\[ \V{\ell +1} \kets{\f}{1}
               \kets{T_2}2\kets{T_3}3 \cdots \kets{T_{\ell+1}}{\ell+1}
     = 0,  \lb{decouple}     \]
where we have schematically denoted the transversal level-one states
by $\kets{T_i}{i}$, so $T_i \in E_{10}^{[1]}$ for all $i$.  Although
the relation \Eq{decouple} must be verified {\em for all} such states
$T_i$, the task is substantially facilitated by momentum conservation
which reduces the calculation to a finite number of checks for any
given state $\f$ at level $\ell$. Note also that the total ``level
number'' is conserved by the vertex; this is a consequence of momentum
conservation along the axis defined by the null root $\vd$.

In the remaining section we will exhibit the basic decoupling
mechanism in specific examples. For this purpose we distinguish
between longitudinal decoupling (for states $\f$ involving
longitudinal DDF operators) and transversal decoupling (for states
$\f$ involving only transversal DDF operators). As it turns out, the
decoupling of longitudinal states is still relatively straightforward
to analyze, and in particular does {\em not} depend on special
properties of the lattice, though on its dimension (or rank) $d$; in
fact, decoupling remains valid in the continuum (with possible
consequences for Liouville theory for $1<d<25$.) Transversal
decoupling, on the other hand, is far more subtle, depending on the
properties of the lattice in a crucial way, and having no continuum
analog.

\subsection{Decoupling of longitudinal states} \lab{E10-3}
Recall that in the course of the DDF construction we switched from the
usual oscillator basis $\{\amm\}$ to the DDF basis
$\{A^a_m(\vv),A^-_m(\vv),L_m\}$ which nicely separates the Fock space
into physical and unphysical states. For the discussion of decoupling
of longitudinal states it turns out, however, that it is more useful
to go over to the ``non-diagonal'' basis $\{A^a_m(\vv),K_m(\vv),L_m\}$
used in the proof of the no-ghost theorem (see \ct{GodTho72}). Given DDF
momenta $\vv$ and $\vk(\vv)$ we take the usual set of transversal DDF
oscillators, $A^a_m(\vv)$, and define
\[ K_m(\vv):=\vk\X\va_m\equiv k_\m\amm, \]
so that we have the commutation relations
\[ [K_m(\vv),K_n(\vv)]=0, \qquad
   [A^a_m(\vv),K_n(\vv)]=0, \qquad
   [L_m,K_n(\vv)]=-nK_{m+n}(\vv). \]
Thus both sets $\{A^-_m(\vv),L_m\}$ and $\{K_m(\vv),L_m\}$ commute
with the transversal DDF operators, and the only difference is that
the second set is not diagonal. Consequently, the $K_n$ operators do
not generate physical states. The general form of a longitudinal DDF
state in terms of the $LK$ basis is
\[ A^-_{-n_1}(\vv)\ldots A^-_{-n_N}(\vv)\ket\vv
    &=& \Bigg[\sum_{i_1\ge\ldots\ge i_I;\ j_1\ge\ldots\ge j_J
                    \atop i_1+\ldots+j_J=n}
              y_{i_1,\ldots,i_I;j_1,\ldots,j_J}
              L_{-i_1}\ldots L_{-i_I}
              K_{-j_1}(\vv)\ldots K_{-j_J}(\vv) \non
    & & \phantom{\Bigg[}
              +\sum_{j_1\ge\ldots\ge j_J \atop j_1+\ldots+j_J=n}
               x_{j_1,\ldots,j_J}K_{-j_1}(\vv)\ldots K_{-j_J}(\vv)\Bigg]
     \ket{\vv-n\vk}, \lb{lDDF-LK} \]
where $n:=n_1+\ldots+n_N$. The coefficients $y$ and $x$ are uniquely
determined by the physical state conditions, i.e.\ by the requirement
that the state is annihilated by the positive $L_n$'s. Note that each
term involving only $K_n$'s (i.e.\ the second sum) is expected to
vanish for $d=26$ since we know from the no-ghost theorem that in the
critical dimension the longitudinal states become null and hence no
terms without Virasoro generators can occur.

Expressed in terms of the $LK$ basis we find for the lowest
longitudinal DDF states:
\[ A^-_{-1}\ket\vv &=& -L_{-1}\ket{\vv-\vk}, \non[1ex]
   A^-_{-2}\ket\vv &=& \left\{-L_{-2}+3L_{-1}K_{-1}
                        +\frc{26-d}4[-K_{-2}+K_{-1}^2]\right\}
                       \ket{\vv-2\vk}, \non[1ex]
   A^-_{-3}\ket\vv
    &=& \big\{-L_{-3}+4L_{-2}K_{-1}+\frc52L_{-1}K_{-2}
              -\frc{17}2L_{-1}K_{-1}^2 \non
    & & \phantom{\big\{}
              +\frc{2(26-d)}3[-K_{-3}+3K_{-2}K_{-1}
               -2K_{-1}^3]\big\}\ket{\vv-3\vk}, \non[1ex]
   A^-_{-4}\ket\vv
    &=& \big\{-L_{-4}+5L_{-3}K_{-1}+3L_{-2}K_{-2}-13L_{-2}K_{-1}^2
              +\frc73L_{-1}K_{-3}
              -16L_{-1}K_{-2}K_{-1}+\frc{71}3L_{-1}K_{-1}^3 \non
    & & \phantom{\big\{}
              +\frc{26-d}8[-10K_{-4}
              +40K_{-3}K_{-1}+13K_{-2}^2-86K_{-2}K_{-1}^2
              +43K_{-1}^4]\big\}\ket{\vv-4\vk}, \non[1ex]
   A^-_{-2}A^-_{-2}\ket\vv
    &=& \big\{L_{-2}^2-3L_{-3}K_{-1}-6L_{-2}L_{-1}K_{-1}
              +\frc{34-d}2L_{-2}K_{-2}
              +\frc{46-d}2L_{-2}K_{-1}^2 \non
    & & \phantom{\big\{}
              +9L_{-1}^2K_{-1}^2
              +5L_{-1}K_{-3}+\frc{3(44-d)}2L_{-1}K_{-2}K_{-1}
              +\frc{146-3d}2L_{-1}K_{-1}^3 \non
    & & \phantom{\big\{}
              +\frc{26-d}{16}[-12K_{-4}+48K_{-3}K_{-1}
              +(46-d)K_{-2}^2-2(82-d)K_{-2}K_{-1}^2
              +(82-d)K_{-1}^4]\big\}\ket{\vv-4\vk}, \nn \]
where we suppressed the $\vv$ dependence for notational convenience.
We notice that $A^-_{-1}$ leads to physical null states and should not
be used for building physical states.  Rewriting the $A^-_m$'s in
terms of the $LK$ basis quickly becomes rather cumbersome for higher
excitations but we shall see below that we can circumvent most of this
calculation because we need only the information about the $x$
coefficients for decoupling.

We now consider the three-vertex with a longitudinal DDF state of the
form \Eq{lDDF-LK} on leg 1 and arbitrary physical states on legs 2 and
3, respectively:
\[ \unitlength.5mm
   \begin{picture}(70,24)
     \put(35,15){\circle{10}}
     \put(35,15){\makebox(0,0){$1$}}
     \put(10,15){\line(1,0){8}}
     \put(52,15){\line(1,0){8}}
     \put(35,-2){\line(0,-1){8}}
     \put(30,15){\vector(-1,0){12}}
     \put(40,15){\vector(1,0){12}}
     \put(35,10){\vector(0,-1){12}}
     \put( 8,15){\makebox(0,0)[r]{
      $\th\left(A^{(1)-}_{-n_1}(\vv)\ldots
                A^{(1)-}_{-n_N}(\vv)\kets{\vv}1\right)$}}
     \put(62,15){\makebox(0,0)[l]{$\kets{\rm phys}2$}}
     \put(35,-12){\makebox(0,0)[t]{$\kets{\rm phys'}3$}}
   \end{picture}. \] \vspace*{10mm} \\
Note that in accordance with Eq.\ \Eq{structure1} we have not turned
around leg 1 but have instead put in the Chevalley involution $\th$.
Our aim is to derive by means of the overlap equations some simple
criteria for the longitudinal DDF state to decouple. Since the
$A^-_{-m}$'s are integrated physical vertex operators we could invoke
Eq.\ \Eq{tDDFoverlap} to feed them straight through the vertex:
\[ \VN\sum_{j=1}^NA^{(j)-}_{-m}(\vv)=0. \lb{lDDFoverlap} \]
This identity, however, gives no insight into the decoupling
mechanism. On the other hand, if we rewrite the longitudinal DDF in
terms of the $LK$ basis and move those through the vertex step by step
then we do get powerful statements about the decoupling. First, we
observe that all terms of the form $L_{-i_1}\ldots
L_{-i_k}K_{-i_{k+1}}\ldots K_{-i_I}\ket{\vv-n\vk}$ vanish when fed
through the vertex with physical states on the other legs. This
follows from the discussion in Sect.\ \ref{E10-1} and the fact that
the Chevalley involution commutes with the Virasoro generators. Thus
only the $K$ terms contribute on-shell.  As regards these terms we
will next exploit the DDF construction and assume that the physical
states on the other legs are transversal DDF states with lightlike
vectors $\vk',\vk''$ which are proportional to the vector $\vk$ needed
for $A^-_m(\vv,\vk)$:
\[ \unitlength.5mm
   \begin{picture}(70,24)
     \put(35,15){\circle{10}}
     \put(35,15){\makebox(0,0){$1$}}
     \put(10,15){\line(1,0){8}}
     \put(52,15){\line(1,0){8}}
     \put(35,-2){\line(0,-1){8}}
     \put(30,15){\vector(-1,0){12}}
     \put(40,15){\vector(1,0){12}}
     \put(35,10){\vector(0,-1){12}}
     \put( 8,15){\makebox(0,0)[r]{
      $\th\left(A^{(1)-}_{-n_1}(\vv,\vk)\ldots
                A^{(1)-}_{-n_N}(\vv,\vk)\kets{\vv}1\right)$}}
     \put(62,15){\makebox(0,0)[l]{
      $A^{(2)a_1}_{-l_1}(\vr,\vk'')\ldots
       A^{(2)a_L}_{-l_L}(\vr,\vk'')\kets{\vr}2$}}
     \put(35,-12){\makebox(0,0)[t]{
      $A^{(3)b_1}_{-m_1}(\vs,\vk')\ldots
       A^{(3)b_M}_{-m_M}(\vs,\vk')\kets{\vs}3$}}
   \end{picture}. \] \vspace*{10mm} \\
Before applying the overlap equation for the $K_{-j}$'s we note that
\[ K_n(\vv)A^{a_1}_{-l_1}(\vr,\vk'')\ldots
            A^{a_L}_{-l_L}(\vr,\vk'')\ket{\vr} &=& 0, \non
   K_n(\vv)A^{b_1}_{-m_1}(\vs,\vk')\ldots
            A^{b_M}_{-m_M}(\vs,\vk')\ket{\vs} &=& 0, \]
for $n>0$ because of $\vk\X\vk''=\vk\X\vze^{a_i}=0$ and
$\vk\X\vk'=\vk\X\vet^{b_j}=0$, respectively\footnote{This is a point
  where the construction could go wrong if the fundamental Weyl
  chamber contained more than one null direction, because the products
  of the null vectors and the polarization vectors would then no
  longer vanish in general.}. In view of the overlap Eq.\
\Eq{CSVosclap} we therefore conclude that any $K\1_{-j}$, when fed
through the vertex, becomes $K\3_0(\vv)\equiv\vk\X\va\3_0$ acting on
$\kets\vs3$ (note that we pick up one minus sign from the Chevalley
involution and one from the overlaps giving the stated result). Thus
it is just a {\it number} $\K:=\vk\X\vs$ (or $\vk'=\K\vk$) and the net
result of feeding a product of longitudinal DDF operators through the
vertex is a {\it polynomial} $P(\K)$ which we will refer to as {\bf
  (longitudinal) decoupling polynomial}.  We will write
$P^{[3]}_{n_1,\ldots,n_N}(\K)$ for the polynomial encoding the
decoupling of the longitudinal DDF state $A^-_{-n_1}\ldots
A^-_{-n_N}\ket\vv$ from the three-vertex.  Using the decomposition in
Eq.\ \Eq{lDDF-LK} it has the general form
\[ P^{[3]}_{n_1,\ldots,n_N}(\K)
    :=\sum_{j_1\ge\ldots\ge j_J \atop j_1+\ldots+j_J=n}
      x_{j_1,\ldots,j_J}\K^J. \lb{decpoly-1} \]
This allows us to rewrite the result of the above diagram as
\[ P^{[3]}_{n_1,\ldots,n_N}(\K)\X\left[ \hh{14mm}
   \unitlength.5mm
   \begin{picture}(162,24)
     \put(35,15){\circle{10}}
     \put(35,15){\makebox(0,0){$1$}}
     \put(10,15){\line(1,0){8}}
     \put(52,15){\line(1,0){8}}
     \put(35,-2){\line(0,-1){8}}
     \put(30,15){\vector(-1,0){12}}
     \put(40,15){\vector(1,0){12}}
     \put(35,10){\vector(0,-1){12}}
     \put( 8,15){\makebox(0,0)[r]{$\kets{-\vv+n\vk}1$}}
     \put(62,15){\makebox(0,0)[l]{
      $A^{(2)a_1}_{-l_1}(\vr,\vk'')\ldots
       A^{(2)a_L}_{-l_L}(\vr,\vk'')\kets{\vr}2$}}
     \put(35,-12){\makebox(0,0)[t]{
      $A^{(3)b_1}_{-m_1}(\vs,\vk')\ldots
       A^{(3)b_M}_{-m_M}(\vs,\vk')\kets{\vs}3$}}
   \end{picture}\right]. \lb{dia-decpoly} \]
After some algebra, we arrive at the following list of
decoupling polynomials for $\sum_i n_i \leq 7$:
\[ P^{[3]}_2(\K) &=& \frc{26-d}4\K(\K-1), \\[1ex]
   P^{[3]}_3(\K) &=& -\frc{2(26-d)}3\K(\K-1)(2\K-1), \\[1ex]
   P^{[3]}_4(\K)
    &=& \frc{26-d}8\K(\K-1)(43\K^2-43\K+10), \\[1ex]
   P^{[3]}_{2,2}(\K)
    &=& \frc{26-d}{16}\K(\K-1)[(82-d)\K^2-(82-d)\K+12], \\[1ex]
   P^{[3]}_5(\K)
    &=& -\frc{26-d}3\K(\K-1)(2\K-1)(29\K^2-29\K+6), \\[1ex]
   P^{[3]}_{3,2}(\K)
    &=& -\frc{26-d}{30}\K(\K-1)(2\K-1)
         [(558-5d)\K^2-(558-5d)\K+84], \\[1ex]
   P^{[3]}_6(\K)
    &=& \frc{26-d}{24}\K(\K-1)
         [1568\K^4-3136\K^3+2237\K^2-669\K+70], \\[1ex]
   P^{[3]}_{4,2}(\K)
    &=& \frc{26-d}{288}\K(\K-1)
         [(54686-387d)\K^4-(109372-774d)\K^3 \non
    & & \phantom{\frc{26-d}{288}\K(\K-1)[}
          +(75754-477d)\K^2-(21068-90d)\K+1920], \\[1ex]
   P^{[3]}_{3,3}(\K)
    &=& \frc{26-d}{72}\K(\K-1)
         [(8773-128d)\K^4-(17546-256d)\K^3 \non
    & & \phantom{\frc{26-d}{72}\K(\K-1)[}
          +(11819-160d)\K^2-(3046-32d)\K+216], \\[1ex]
   P^{[3]}_{2,2,2}(\K)
    &=& -\frc{26-d}{192}\K(\K-1)
         [(34460-660d+3d^2)\K^4-(68920-1320d+6d^2)\K^3 \non
    & & \phantom{-\frc{26-d}{192}\K(\K-1)[}
          +(45076-768d+3d^2)\K^2-(10616-108d)\K+768], \\[1ex]
   P^{[3]}_7(\K)
    &=& -\frc{26-d}{60}\K(\K-1)(2\K-1)
         [6367\K^4-12734\K^3+8873\K^2-2506\K+240], \\[1ex]
   P^{[3]}_{5,2}(\K)
    &=& -\frc{26-d}{420}\K(\K-1)(2\K-1)
         [(173602-1015d)\K^4-(347204-2030d)\K^3 \non
    & & \phantom{-\frc{26-d}{420}}
         +(236278-1225d)\K^2-(62676-210d)\K+5400], \\[1ex]
   P^{[3]}_{4,3}(\K)
    &=& -\frc{26-d}{420}\K(\K-1)(2\K-1)
         [(125851-1505d)\K^4-(251702-3010d)\K^3 \non
    & & \phantom{-\frc{26-d}{420}}
         +(167909-1855d)\K^2-(42058-350d)\K+3000], \\[1ex]
   P^{[3]}_{3,2,2}(\K)
    &=& -\frc{26-d}{120}\K(\K-1)(2\K-1)
         [(94916-1396d+5d^2)\K^4
         -(189832-2792d+10d^2)\K^3 \non
    & & \phantom{-\frc{26-d}{120}}
         +(124364-1624d+5d^2)\K^2-(29448-228d)\K+2160]. \]
We make the following important observations.

All decoupling polynomials vanish identically in the critical
dimension $d=26$ reflecting the decoupling of all longitudinal
states there, in accord with the no-ghost theorem.

In general, decoupling takes only place at certain values of $\K$,
namely at those zeros of the polynomials which are of the form
$\K=\frac1\cl$ for $\cl\in\Nat$. In view of diagram \Eq{dia-decpoly}
one might argue that this only constitutes a sufficient criterion for
decoupling and in general the remaining diagram could also account for
decoupling. But this is not the case since the oscillator part of the
longitudinal state, which is the relevant piece of information for
decoupling, has been completely absorbed into the decoupling
polynomial and this is {\it independent} of the choice of transversal
level-one states on the other legs. Of course, the remaining diagram
could vanish for a special choice of transversal states, but for
decoupling it had to vanish for arbitrary transversal level-one states
on legs 2 and 3.

Once a ``purely'' longitudinal DDF state (i.e.\ involving only
$A^-_{-m}$'s) decouples then also the whole transversal Heisenberg
module built on it will also decouple because the transversal DDF
oscillators commute with the $LK$ basis. This shows that the question
of missing longitudinal states can be completely solved in terms of
the decoupling polynomials.

By construction, the decoupling polynomials have vanishing constant
term, i.e.\ have a zero at $\K=0$. But $\vk\X\vs=0$ means that the
state on leg 3 is an element of the $\9$ subalgebra. Since, by level
conservation, we then must have a level-one element on leg 2 we can
interpret the zero at $\K=0$ as the fact that the transversal
level-one DDF states form a representation of $\9$.  In view of the
antisymmetry of the Lie bracket, interchanging legs 2 and 3 should
make no difference for the decoupling, so that we also expect a zero
at $\K=1$. Indeed, the above examples show this feature and we will
give a general proof below.

Inspection of the list of examples tempts us to conjecture
a generic zero at $\K=\frc12$ for odd total mode number, i.e.
\[ P^{[3]}_{n_1,\ldots,n_N}(\frc12)=0\FOR{ $n_1+\ldots+n_N$ odd}. \]
In other words, no (purely) longitudinal DDF states with total odd
mode number can occur as level-two elements of $\0$. This is in
agreement with our example $\vL_7$ described in the last section
where we found $A^-_{-3}(\vv)\ket{\vv}$ not to be an element of $\0$.
Below, we will indicate how to prove the conjecture in general.

So far, we have always referred to $\0$ in our discussion of
decoupling. But in the derivation of the decoupling polynomials we
actually did not use the fact that the momenta lie on a lattice but
merely exploited features of the old DDF construction which are of
course also valid for the continuum case. Thus the pattern of
longitudinal decoupling as presented above also applies to {\it any}
bosonic string model. In particular, we can now analyze other
hyperbolic Kac Moody algebras if we take the corresponding root
lattice as momentum lattice for the string.

It is clear that the decoupling polynomial in \Eq{decpoly-1} has
degree $n$ at most. Taking further into account the universal factor
$\K(\K-1)$, there can be at most $n-1$ linearly independent decoupling
polynomials. For fixed $n$, on the other hand, we know that there are
$p_1(n)-p_1(n-1)$ physical longitudinal DDF states and hence the same
number of decoupling polynomials. In other words, whereas the number
of longitudinal DDF states grows exponentially the number of
independent decoupling polynomials increases linearly. This means that
there is a wealth of decoupling of longitudinal states. In fact, for
large $n$ (which corresponds to highly excitated massive states)
almost all longitudinal states decouple and only a small fraction
leads to $\0$ elements. This feature explains, for example, why the
level-two multiplicity formula for $\0$ found in \ct{KaMoWa88} stays
close to the number of transversal states with slowly increasing
deviation. But this argument will also apply in a more general context
to all hyperbolic Kac Moody algebras. There, the common feature of all
explicitly known root multiplicities is that they agree with or lie
very near to the number of transversal states (see e.g.\ \ct{Kac90}
for a table of root multiplicities of the hyperbolic extension of
$A_1^{(1)}$). Longitudinal decoupling thus gives a qualitative
explanation of this phenomenon.

It is possible to extract some information about the general structure
of decoupling polynomials from the physical state conditions. More
specifically, let us consider the implications of the condition that a
longitudinal DDF state of the form \Eq{lDDF-LK} is annihilated by
$L_1$:
\[ 0 & {\stackrel!=} & \Bigg\{
    \sum_{i_1,\ldots,i_I \atop i_1+\ldots+i_I=n}y_{i_1,\ldots,i_I}
    \Big[
     (1+i_1)L_{1-i_1}L_{-i_2}\ldots L_{-i_k} K_{-i_{k+1}}\ldots K_{-i_I}
     +\ldots \non
    & & \phantom{\sum_{i_1+\ldots+i_I=n}y_{i_1,\ldots,i_I}\Big\{}
     +(1+i_k)L_{-i_1}\ldots L_{-i_{k-1}} L_{1-i_k} K_{-i_{k+1}}\ldots
      K_{-i_I} \non
    & & \phantom{\sum_{i_1+\ldots+i_I=n}y_{i_1,\ldots,i_I}\Big\{}
     +i_{k+1}L_{-i_1}\ldots L_{-i_k} K_{1-i_{k+1}}K_{-i_{k+2}}\ldots K_{-i_I}
     +\ldots \non
    & & \phantom{\sum_{i_1+\ldots+i_I=n}y_{i_1,\ldots,i_I}\Big\{}
     +i_I L_{-i_1}\ldots L_{-i_k} K_{-i_{k+1}}\ldots K_{-i_{I-1}} K_{1-i_I}
    \Big] \non
    & &{}
     +\sum_{j_1\ge\ldots\ge j_J \atop j_1+\ldots+j_J=n}
       x_{j_1,\ldots,j_J}\Big[
        j_1 K_{1-j_1}K_{-j_2}\ldots K_{-j_J}
        +\ldots+
        j_J K_{-j_1}\ldots K_{-j_{J-1}}K_{1-j_J}
       \Big] \Bigg\}\ket{\vv-n\vk} . \nn \]
In order to fulfill this condition the coefficients of all linearly
independent combinations of $LK$ basis elements have to vanish. Let us
concentrate on terms which involve only $K$'s. We claim that they can
only come from the second sum. Indeed, terms of the form
$L_{-1}K_{-i_2}\ldots K_{-i_I}\ket{\vv-n\vk}$ with $i_2\ge\ldots\ge
i_I$ and $i_2+\ldots+i_I=n-1$ do not give rise to pure $K$ monomials
since $[L_1,L_{-1}]=2L_0$ and $L_0K_{-i_2}\ldots
K_{-i_I}\ket{\vv-n\vk}=0$. Thus we arrive at the following necessary
condition:
\[ 0 & {\stackrel!=} &
      \sum_{j_1\ge\ldots\ge j_J \atop j_1+\ldots+j_J=n}
       x_{j_1,\ldots,j_J}\Big[
        j_1 K_{1-j_1}K_{-j_2}\ldots K_{-j_J}
        +\ldots+
        j_J K_{-j_1}\ldots K_{-j_{J-1}}K_{1-j_J}
       \Big], \lb{K-conds} \]
where we put $K_0\equiv1$ due to $\vk\X(\vv-n\vk)=1$. This already
severely constrains the $L$ independent part of a longitudinal
physical state for it yields $p_1(n-1)$ linear equations for the
$p_1(n)$ coefficients $x_{j_1,\ldots,j_J}$. Consequently it also
determines a major part of the decoupling polynomials. For example, if
we add up all linear equations we obtain
\[ 0 =
       \sum_{j_1\ge\ldots\ge j_J \atop j_1+\ldots+j_J=n}
       x_{j_1,\ldots,j_J}[j_1+\ldots+j_J] \non
     =
       n\sum_{j_1\ge\ldots\ge j_J \atop j_1+\ldots+j_J=n}
       x_{j_1,\ldots,j_J}, \]
but, in view of Eq.\ \Eq{decpoly-1}, this is nothing but the statement
that the decoupling polynomial vanishes at $\K=1$, i.e.\ we can indeed
always pull out the factor $(\K-1)$. We have also checked for the
above list of decoupling polynomials that, for odd $n$, the zeros at
$\K=\frac12$ also follow from the information encoded in Eq.\
\Eq{K-conds}, but we have not yet succeeded in finding a general
proof. Note that the condition \Eq{K-conds} is ``universal'' in the
sense that (apart from the factor $(26-d)$ inside) it is independent
of $d$ and thus the same for all subcritical strings.

If we want to probe higher-level decoupling we encounter a new
subtlety: the decoupling polynomials will depend on the Koba Nielsen
variables as parameters. The reason for this is that we can always
only fix three Koba Nielsen points whereas there are $N$ of them for
the $N$-vertex. Using the overlap equations \Eq{Nosclap} we conclude
that any $K\1_{-m}$, when fed through the vertex, becomes
$\sum_{i=3}^N z_i^m K\i_0$ so that the decoupling polynomial for the
process
\[ \unitlength.5mm
  \begin{picture}(70,24)
   \put(35,15){\circle{10}}
   \put(35,15){\makebox(0,0){$z_i$}}
   \put(10,15){\line(1,0){8}}
   \put(51.13,9.62){\line(3,-1){8}}
   \put(52,15){\line(1,0){8}}
   \put(35,-2){\line(0,-1){8}}
   \put(30,15){\vector(-1,0){12}}
   \put(39.74,13.42){\vector(3,-1){12}}
   \put(40,15){\vector(1,0){12}}
   \put(35,10){\vector(0,-1){12}}
   \put( 8,15){\makebox(0,0)[r]{
      $\th\big(A^{(1)-}_{-n_1}(\vv,\vk)\ldots
                A^{(1)-}_{-n_N}(\vv,\vk)\kets{\vv}1\big)$}}
   \put(60.61,6.46){\makebox(0,0)[l]{$\kets{T_3}3$}}
   \put(62,15){\makebox(0,0)[l]{$\kets{T_2}2$}}
   \put(35,-12){\makebox(0,0)[t]{$\kets{T_N}N$}}
   \put(43,0){\circle*{1.5}}
   \put(46,2){\circle*{1.5}}
   \put(48.5,4.67){\circle*{1.5}}
  \end{picture} \] \vspace*{10mm} \\
  is given by
\[ P^{[N]}_{n_1,\ldots,n_N}(\K)
    :=\sum_{m_1\ge\ldots\ge m_J \atop m_1+\ldots+m_J=n}
      \sum_{i_1,\ldots,i_J=3}^N
      x_{m_1,\ldots,m_J}
      (z_{i_1})^{m_1}\cdots(z_{i_J})^{m_J} \K^J, \lb{decpoly-N} \]
with $n:=n_1+\ldots+n_N$.

\subsection{On the decoupling of transversal states} \lab{E10-4}
We now turn to the decoupling of transversal states, which is our most
striking result. It is here that the special properties of the lattice
enter in a crucial and still mysterious way, as there is no analogous
decoupling in the continuum, unlike for the longitudinal states
discussed in the preceding section.  Due to our lack of general
understanding we will in this section limit ourselves to the
discussion of the specific example corresponding to the root space
$\exam$. For the explicit calculations we will mostly work with the
vertex $\CSV$ whose cyclic symmetry implies the cyclic symmetry of the
associated overlap equations, which were given in Sect.\
\ref{EXPLICIT}. We will also need the expressions of the DDF states
entering \Eq{commlev2} in terms of oscillators. Beside the obvious
formulas
\[ \Ar{a}{-1}\ket{\vr} = (\vza[-1]) \ket{\vr + \vd}  , \qquad
   \As{b}{-1}\ket{\vs} = (\vyb[-1]) \ket{\vs + \vd}  , \lb{osc1} \]
we shall need
\[ \Ar{a}{-2}\ket{\vr} = \Big\{ (\vza[-2]) + 2(\vza[-1])
           (\vda{-1}) \Big\} \ket{\vr + 2\vd}   \lb{osc2}    \]
as well as
\[ \Av{c}{-3} \ket{\vv}
    &=& \Big\{ (\vxc[-3]) + \frc32(\vxc[-2])(\vda{-1})
               + \frc34 (\vxc[-1])(\vda{-2}) \non
    & & \phantom{\Big\{}
        + \frc98 (\vxc[-1])(\vda{-1})(\vda{-1})
        \Big\} \ket{\vv + \frc32\vd}  \lb{osc3} \]
and
\[ \Av{c}{-1} \Av{d}{-1} \Av{e}{-1} \ket{\vv}
    &=& \Big\{ (\vxc[-1])(\vxd[-1])(\vxe[-1]) \non
    & & \phantom{\Big\{}
        + \frc32 \d^{(cd}(\vxi^{e)}\X\va_{-1})
          \big[ \frc14(\vda{-1})(\vda{-1}) +
                \frc12 (\vda{-2}) \big]
        \Big\} \ket{\vv + \frc32\vd},    \lb{osc4}   \]
where $(cde)$ means symmetrization in the indices $c,d,e$ with
strength one. These formulas can be read off from the appendix of
\ct{GebNic95}; the only point to remember is that because $\vk (\vr) =
\vk (\vs)=-\vd$ and $\vk (\vv)= -\frc12 \vd$, we must replace $\vd$ by
$\frc12 \vd$ everywhere in the appropriate formulas of \ct{GebNic95}
to get \Eq{osc3} and \Eq{osc4}. Furthermore, we will need to make use
of the Chevalley involuted states obtained by replacing $\va^\mu_m$
and $\ket{\vv}$ by $-\va^\mu_m$ and $\ket{-\vv}$, respectively, in the
above expressions, cf.\ Eq.\ \Eq{Chevinv}.

To prove that a given state $\p$ decouples, we will have to show that
expressions like
\[ && \V3
   \Big\{ \kets{\p}{1}
          \XO \Ar[2]{a}{-1} \kets{\vr}{2}
          \XO \As[3]{b}{-1} \kets{\vs}{3} \Big\}, \lb{decouple1a} \\[.5ex]
   && \V3
    \Big\{ \kets{\p}{1}
           \XO \Ar[2]{a}{-2} \kets{\vr}{2}
           \XO \kets{\vs}{3} \Big\}, \lb{decouple1b} \\[.5ex]
   && \V3
    \Big\{  \kets{\p}{1}
            \XO \Ar[2]{a}{-1} \Ar[2]{b}{-1} \kets{\vr}{2}
            \XO \kets{\vs}{3} \Big\},  \lb{decouple1c} \]
vanish for all possible choices of the transversal indices\footnote{We
  here write out the tensor product symbol $\XO$ unlike in previous
  formulas in order to render the expressions more transparent.}.
Since we will only attach physical states to the legs, any
three-vertex could be used. However, for all practical calculations,
and in particular those involving oscillator overlaps, we will make
use of the CSV vertex $\CSV$. Before proceeding with the calculation,
however, we note an important simplification due to the following
identity which is a special case of \Eq{tDDFoverlap} and the
transversal analog of \Eq{lDDFoverlap}:
\[ \V3 \sum_{j=1}^3 \, \Ar[j]{a}{m} = 0. \]
It is crucial here that the DDF operators refer to the same DDF
momenta although they act on states in different Fock spaces which in
general are constructed on other tachyonic grounstates with different
DDF momenta, and we therefore must exercise some care to indicate
precisely which DDF operators are meant.  As an example let us apply
the above identity to the two-string state $\Ar[2]{a}{-1}\kets{\vr}{2}
\XO \As[3]{b}{-1}\kets{\vs}{3}$, so the first leg is not saturated. We
get
\[ \V3
    \Big\{ \Ar[2]{a}{-1}\kets{\vr}{2}
           \XO \As[3]{b}{-1} \kets{\vs}{3} \Big\} =
   \V3
    \Big\{ \kets{\vr}{2}
           \XO \Ar[3]{a}{-1} \As[3]{b}{-1} \kets{\vs}{3}
          +\kets{\vr}{2}
           \XO \As[3]{b}{-1} \kets{\vs}{3} \, \Ar[1]{a}{-1}
    \Big\}.   \lb{feedDDF1}  \]
The first term on the right-hand side tells us that the expressions
\Eq{decouple1a} and \Eq{decouple1c} are related. The last term on the
right-hand side of \Eq{feedDDF1} contains a DDF operator acting on the
first leg; when this operator acts on the Chevalley involuted state
$\ket{-\vv}$, we obtain
\[  \Ar{a}{-m} \ket{- \vv}
     = \Res{z}{\vza\X\vP(z) \re{i2m\vk(-\vv)\.\vX(z)}}\ket{-\vv}
     \equiv \vza_\mu A^\mu_{2m}(-\vv) \ket{-\vv},  \]
because $\vk(-\vv) = +\frc12\vd = -\frc12\vk(\vr)$ (note the
contraction with the polarization vector $\vza_\mu$ on the right-hand
side whereas the operator $A^a_{2m}(-\vv)$ is defined with
$\bx^a_\mu$, which differs from $\vza_\mu$ for $a=8$). Thus the
creation operator is converted into an annihilation operator whose
index is multiplied by two (for higher levels, the index will be
similarly multiplied by $\ell$).  Below we will apply this identity to
a state built with odd-moded DDF operators on the tachyonic
groundstate, and so $\Ar[1]{a}{-m}$ will commute through to annihilate
the state.  Consequently, the second term on the right-hand side of
\Eq{feedDDF1} vanishes in this case, and the expressions
\Eq{decouple1a} and \Eq{decouple1c} are, in fact, equivalent.

To evaluate \Eq{decouple1a} and \Eq{decouple1b} in practice, we first
express the states on legs 2 and 3 in terms of oscillators according
to \Eq{osc1} and \Eq{osc2}, and then feed through the creation
operators $\va_{-n}$ (with $n>0$) to the other legs where they become
annihilation or momentum operators acting on the state $\p$ by the
oscillator overlap equations \Eq{CSVosclap}.  Next we commute these
$\va_n$'s (with $n\ge0$) through the creation operators defining the
state $\p$ until they hit the oscillator vacuum.  If any creation
operators are still left on any of the legs, we repeat this process
until all creation operators have been eliminated.  The result of the
calculation is some polynomial involving products of the polarization
vectors and DDF momenta in various combinations multiplying
\[ \V3 \, \kets{\vt_1}{1} \XO \kets{\vt_2}{2} \XO \kets{\vt_3}{3}  =
    \d_{\vt_1 +\vt_2 +\vt_3 , 0}, \lb{decouple2}      \]
where $\vt_1, \vt_2 ,\vt_3$ are the momenta of the three states
attached to the three legs (in the example to be discussed, we have
$\vt_1 =-\vr - \vd , \vt_2 =-\vs -\vd$ and $\vt_3 =\vv + \frc32 \vd$).
Clearly, this result just expresses momentum conservation. It also
shows why we have to apply the Chevalley involution to some of the
states, since the scalar product evidently vanishes if all three
momenta correspond to positive roots. In other words we have to
arrange for the momenta (i.e.\ the roots) to respect momentum
conservation, which leaves only a finite number of states to be
checked. To establish the decoupling it then remains to show that the
prefactor of \Eq{decouple2} vanishes for all possible choices of
polarizations.

We will now show how this works for the above example with the
vertex $\CSV$, using the Chevalley involuted version of
\Eq{osc1} on legs 2 and 3. Leaving the state $\p$ on the first
leg arbitrary for the moment (it will be specified below), we get
\[ \lefteqn{
   \CSV \Big\{ \kets{\p}{1}
               \XO (\vza[-1]\2) \kets{-\vr -\vd}{2}
               \XO (\vyb[-1]\3) \kets{-\vs -\vd}{3}
        \Big\} } \hh{8mm} \non
   &=&
   \CSV \Big\{ (\vza\X\vyb) \kets{\p}{1}
               \XO \kets{-\vr - \vd}{2}
               \XO \kets{-\vs - \vd}{3} \non
   & & \phantom{\CSV\Big\{}
                +\big(\vza[1]\1-\vza[0]\1\big) \kets{\p}{1}
               \XO \kets{-\vr - \vd}{2}
               \XO (\vyb[-1]\3) \kets{-\vs -\vd}{3} \Big\} \non
   &=&
   \CSV \bigg\{ \bigg[ (\vza\X\vyb)
                      +\bigg( \vyb\X\vr
                             +\sum_{n=1}^\infty \vyb[n]\1 \bigg)
                       \big(\vza[1]\1-\vza[0]\1\big) \bigg]
                \kets{\p}{1}
               \XO \kets{-\vr - \vd}{2}
               \XO \kets{-\vs - \vd}{3}
        \bigg\}, \hh{10mm} \lb{trans1} \]
where we have used the overlaps \Eq{CSVosclap} for
$\va\2_{-1}$ and $\va\3_{-1}$, respectively.
We now take $\p$ to be either $\Av{c}{-3}\ket{\vv}$ or
$\Av{c}{-1}\Av{d}{-1}\Av{e}{-1}\ket{\vv}$ whose form in terms of
oscillators is given in \Eq{osc3} and \Eq{osc4}, respectively. After
some oscillator algebra we obtain
\[ \lefteqn{
   \bigg[ (\vza\X\vyb)
         +\bigg( \vyb\X\vr
                +\sum_{n=1}^\infty \vyb[n] \bigg)
          \big(\vza[1]-\vza[0]\big) \bigg]
          \Av{c}{-3}\ket{\vv} } \hh{8mm} \non
    &=& \Big\{ \big[(\vza\X\vyb)-(\vyb\X\vr)(\vza\X\vv)\big]
               \big[(\vxc[-3]) + \frc32(\vxc[-2])(\vda{-1})
                    + \frc34 (\vxc[-1])(\vda{-2}) \non
    & & \phantom{\Big\{ \big[(\vza\X\vyb)-(\vyb\X\vr)(\vza\X\vv)\big]\big[}
        + \frc98 (\vxc[-1])(\vda{-1})(\vda{-1})\big] \non
    & & \phantom{\Big\{}
        +\big[(\vza\X\vxc)(\vyb\X\vr)-(\vza\X\vv)(\vyb\X\vxc)\big]
               \big[ \frc34(\vda{-2})
                    +\frc98(\vda{-1})(\vda{-1}) \big] \non
    & & \phantom{\Big\{}
        -3(\vza\X\vv)(\vyb\X\vxc)\big[ (\vda{-1})+1 \big]
        \Big\} \ket{\vv + \frc32\vd} \]
for \Eq{osc3} and
\[ \lefteqn{
   \bigg[ (\vza\X\vyb)
         +\bigg( \vyb\X\vr
                +\sum_{n=1}^\infty \vyb[n] \bigg)
          \big(\vza[1]-\vza[0]\big) \bigg]
          \Av{c}{-1}\Av{d}{-1}\Av{e}{-1}\ket{\vv} } \non
    &=& \Big\{ \big[(\vza\X\vyb)-(\vyb\X\vr)(\vza\X\vv)\big]
               \big[(\vxc[-1])(\vxd[-1])(\vxe[-1]) \non
    & & \phantom{\Big\{ \big[(\vza\X\vyb)-(\vyb\X\vr)(\vza\X\vv)\big]\big[}
        + \frc32 \d^{(cd}(\vxi^{e)}\X\va_{-1})
          [ \frc14(\vda{-1})(\vda{-1}) + \frc12 (\vda{-2}) ] \big] \non
    & & \phantom{\Big\{}
        +3\big[ (\vyb\X\vr)(\vza\X\vxi^{(c})
              -(\vza\X\vv)(\vyb\X\vxi^{(c}) \big]
         (\vxd[-1])(\vxi^{e)}\X\va_{-1}) \non
    & & \phantom{\Big\{}
        + \frc32 \d^{(cd}\big[ (\vxi^{e)}\X\vza)(\vyb\X\vr)
                              -(\vxi^{e)}\X\vyb)(\vza\X\vv) \big]
          \big[ \frc14(\vda{-1})(\vda{-1}) + \frc12 (\vda{-2}) \big] \non
    & & \phantom{\Big\{}
        +6(\vza\X\vxi^{(c})(\vet^{|b|}\X\vxd)(\vxi^{e)}\X\va_{-1})
        \Big\} \ket{\vv + \frc32\vd} \]
for \Eq{osc4}. When we insert the above two expressions into
formula \Eq{trans1} the remaining oscillators on leg 1 can be removed
by feeding them back again onto the other two legs. A glimpse at the
overlaps shows that the result is very simple because we only have
groundstates on the other two legs. This means that $\va\1_{-m}$ just
becomes $-\va\3_0$. Consequently, we can replace $\vda{-m}\1$ by $-1$
and $\vxc[-m]$ by $\vxc\X\vs$, respectively. Let us now consider the
following linear combination of the above states:
\[ \p:= \bigg( F \Av{c}{-3} + G \Av{c}{-1} \Av{c}{-1} \Av{c}{-1}
               + H \Av{c}{-1} \sum_{d=1}^8 \Av{d}{-1} \Av{d}{-1} \bigg)
        \ket{\vv}. \lb{lincomb} \]
If we put this on leg 1 in \Eq{trans1} and take into account the
preceding calculations we finally get
\[ \lefteqn{
   \CSV \Big\{ \kets{\p}{1}
               \XO (\vza[-1]\2) \kets{-\vr -\vd}{2}
               \XO (\vyb[-1]\3) \kets{-\vs -\vd}{3}
        \Big\} } \hh{8mm} \non
   &=& (\vxc\X\vs)\big[(\vza\X\vyb)-(\vyb\X\vr)(\vza\X\vv)\big]
       \bigg\{-\frc18F+G\big[(\vxc\X\vs)^2-\frc38\big]
              +H\bigg(\sum_{d=1}^8(\vxd\X\vs)^2-\frc54\bigg)\bigg\}
              \non
   & &{} +\big[(\vza\X\vxc)(\vyb\X\vr)-(\vza\X\vv)(\vyb\X\vxc)\big]
       \bigg\{\frc38F+3G\big[(\vxc\X\vs)^2-\frc18\big]
              +H\bigg(\sum_{d=1}^8(\vxd\X\vs)^2
                      -\frc54\bigg)\bigg\} \non
   & &{} +6G(\vza\X\vxc)(\vyb\X\vxc)(\vxc\X\vs)
         +2H(\vxc\X\vs)\sum_{d=1}^8(\vxd\X\vs)
           \big[(\vza\X\vxd)(\vyb\X\vr)-(\vza\X\vv)(\vyb\X\vxd)\big]
           \non
   & &{} +2H\sum_{d=1}^8\big[ (\vza\X\vxc)(\vyb\X\vxd)(\vxd\X\vs)
                             +(\vza\X\vxd)(\vyb\X\vxc)(\vxd\X\vs)
                             +(\vza\X\vxd)(\vyb\X\vxd)(\vxc\X\vs)
                            \big]. \]
Note that the momenta were just right for \Eq{decouple2} to give 1 on
the right-hand side. For decoupling the above expression has to vanish
for any choice of $\vza$ and $\vyb$. This analysis has to be performed
case by case.
\ben
\item[(1)] Let $\vxc\X\vs=0$, i.e.\ $c=1,\ldots,7$. Then the
  right-hand side reduces to
\[ \big[(\vza\X\vxc)(\vyb\X\vr)-(\vza\X\vv)(\vyb\X\vxc)\big]
       \big[\frc38F-\frc38G-\frc14H\big]
   +2H\sum_{d=1}^8 (\vxd\X\vs)
       \big[ (\vza\X\vxc)(\vyb\X\vxd)
            +(\vza\X\vxd)(\vyb\X\vxc) \big]. \nn \]
 \ben
 \item[(i)] If $a=b=8$ or both $a\neq8$ and $b\neq8$ then this expression
   identically vanishes.
 \item[(ii)] If either $a=8$ or $b=8$ then we are left with the relation
   \[ F-G-\frc{10}3H=0. \lb{transeq1} \]
 \een
\item[(2)] Let $\vxc\X\vs=-1$, i.e.\ $c=8$. Then the right-hand side
  simplifies to
\[ & &    \big[(\vza\X\vyb)-(\vyb\X\vr)(\vza\X\vv)\big]
          \big[\frc18F-\frc58G+\frc14H\big] \non
   & &{} +\big[(\vza\X\vxc)(\vyb\X\vr)-(\vza\X\vv)(\vyb\X\vxc)\big]
          \big[\frc38F+\frc{21}8G-\frc14H\big] \non
   & &{} +2H\sum_{d=1}^8 (\vxd\X\vs)
           \big[-(\vza\X\vxd)(\vyb\X\vr)
                +(\vza\X\vv)(\vyb\X\vxd)
                +(\vza\X\vxc)(\vyb\X\vxd)
                +(\vza\X\vxd)(\vyb\X\vxc) \big] \non
   & &{} -6G(\vza\X\vxc)(\vyb\X\vxc)
         -2H\sum_{d=1}^8(\vza\X\vxd)(\vyb\X\vxd), \]
where we have also used $\sum_{d=1}^8(\vxd\X\vs)=-1$ and
$\sum_{d=1}^8(\vxd\X\vs)^2=1$.
 \ben
 \item[(i)] If both $a\neq8$ and $b\neq8$ then only the first term and
   the last term contribute and we get the condition
 \[ F-5G-14H=0. \lb{transeq2} \]
 \item[(ii)] If either $a=8$ or $b=8$ then the expression identically
   vanishes.
 \item[(iii)] If $a=b=8$ then all terms are nonzero and we find after
   insertion of the various scalar products
\[ 17F+11G+18H=0. \]
 \een
\een
Note that the last condition is just
$24\times$\Eq{transeq1}$-7\times$\Eq{transeq2}, i.e.\ the three
equations are linearly dependent and only two of three parameters
$F,G,H$ are determined. If we put $F=2$ then it is easy to check that
we get $G=-8$ and $H=3$.  Hence we have verified that the linear
combinations appearing in \Eq{E10-notlev2} indeed decouple in the
expression \Eq{decouple1a}.

It remains to show that this linear combination also decouples in
\Eq{decouple1b}. The calculation is completely analogous to the one above.
Putting the Chevalley involuted version of \Eq{osc2} on leg 2 and
using the oscillator overlaps we get
\[ \lefteqn{
   \CSV \Big\{ \kets{\p}{1}
               \XO \big[-(\vza[-2]\2)
                        +2(\vza[-1]\2)(\vda{-1}\2)\big]
                   \kets{-\vr -2\vd}{2}
               \XO \kets{-\vs}{3}
        \Big\} } \hh{8mm} \non
   &=&
   \CSV \Big\{ \big[ \vza[0]\1-2\vza[1]\1+\vza[2]\1 \non
   & &\phantom{\CSV \Big\{\big[}
                   +2\big(\vza[1]\1-\vza[0]\1\big)
                     \big(\vda{1}\1-\vda{0}\1\big) \big]
                \kets{\p}{1}
               \XO \kets{-\vr - 2\vd}{2}
               \XO \kets{-\vs}{3}
        \Big\}. \lb{trans2} \]
Explicitly, we obtain
\[ \lefteqn{
              \big[ \vza[0]\1-2\vza[1]\1+\vza[2]\1
                   +2\big(\vza[1]\1-\vza[0]\1\big)
                     \big(\vda{1}\1-\vda{0}\1\big) \big]
          \Av{c}{-3}\ket{\vv} } \hh{8mm} \non
    &=&        \big[ -3\vza[0]\1+2\vza[1]\1+\vza[2]\1 \big]
          \Av{c}{-3}\ket{\vv} \non
    &=& \Big\{ -3(\vza\X\vv)
                \big[(\vxc[-3]) + \frc32(\vxc[-2])(\vda{-1})
                     +(\vxc[-1])[ \frc34(\vda{-2})
                                 +\frc98(\vda{-1})(\vda{-1})] \big] \non
    & & \phantom{\Big\{}
        +(\vza\X\vxc) \big[ \frc32(\vda{-2})
                            +\frc94(\vda{-1})(\vda{-1})
                            +3(\vda{-1}) \big]
        \Big\} \ket{\vv + \frc32\vd} \]
for \Eq{osc3} and
\[ \lefteqn{
              \big[ \vza[0]\1-2\vza[1]\1+\vza[2]\1
                   +2\big(\vza[1]\1-\vza[0]\1\big)
                     \big(\vda{1}\1-\vda{0}\1\big) \big]
          \Av{c}{-1}\Av{d}{-1}\Av{e}{-1}\ket{\vv} } \hh{8mm} \non
    &=&        \big[ -3\vza[0]\1+2\vza[1]\1+\vza[2]\1 \big]
          \Av{c}{-1}\Av{d}{-1}\Av{e}{-1}\ket{\vv} \non
    &=& \Big\{ -3(\vza\X\vv)
               \big[ (\vxc[-1])(\vxd[-1])(\vxe[-1])
                    +\frc32\d^{(cd}(\vxi^{e)}\X\va_{-1})
          [ \frc14(\vda{-1})(\vda{-1}) + \frc12 (\vda{-2})] \big] \non
    & & \phantom{\Big\{}
        +6(\vza\X\vxi^{(c})(\vxd[-1])(\vxi^{e)}\X\va_{-1})
        +3\d^{(cd}(\vxi^{e)}\X\vza)
          \big[ \frc14(\vda{-1})(\vda{-1}) + \frc12 (\vda{-2}) \big]
        \Big\} \ket{\vv + \frc32\vd} \hh{8mm} \]
for \Eq{osc4}. When we insert the above two expressions into
formula \Eq{trans2} the remaining oscillators on leg 1 can be removed
by feeding them back again onto the other two legs. This amounts to
replacing $\vda{-m}$ by $-1$ and $\vxc[-m]$ by $\vxc\X\vs$,
respectively. If we put the linear combination \Eq{lincomb} on leg 1
in \Eq{trans2} and take into account the preceding calculations we
finally get
\[ \lefteqn{
   \CSV \Big\{ \kets{\p}{1}
               \XO \big[-(\vza[-2]\2)
                        +2(\vza[-1]\2)(\vda{-1}\2)\big]
                   \kets{-\vr -2\vd}{2}
               \XO \kets{-\vs}{3}
        \Big\} } \hh{8mm} \non
   &=& -3(\vxc\X\vs)(\vza\X\vv)
       \bigg\{-\frc18F+G\big[(\vxc\X\vs)^2-\frc38\big]
              +H\bigg(\sum_{d=1}^8(\vxd\X\vs)^2-\frc54\bigg)\bigg\}
              \non
   & &{} +2(\vza\X\vxc)
       \bigg\{-\frc98F+3G\big[(\vxc\X\vs)^2-\frc18\big]
              +H\bigg(\sum_{d=1}^8(\vxd\X\vs)^2
                      -\frc54\bigg)\bigg\} \non
   & &{} +4H(\vxc\X\vs)\sum_{d=1}^8(\vxd\X\vs)(\vza\X\vxd). \]

This decoupling analysis again has to be performed case by case.
\ben
\item[(1)] Let $\vxc\X\vs=0$, i.e.\ $c=1,\ldots,7$. Then the
  right-hand side reduces to
\[ 2(\vza\X\vxc)\big[-\frc98F-\frc38G-\frc14H\big]. \nn \]
 \ben
 \item[(i)] If $a=8$ then this expression identically vanishes.
 \item[(ii)] If $a\neq8$ then we are left with the relation
   \[ 3F+G+\frc23H=0. \lb{transeq3} \]
 \een
\item[(2)] Let $\vxc\X\vs=-1$, i.e.\ $c=8$. Then the right-hand side
  simplifies to
\[ 3(\vza\X\vv)\big[-\frc18F+\frc58G-\frc14H\big]
   +2(\vza\X\vxc)\big[-\frc98F+\frc{21}8G-\frc14H\big]
   -4H\sum_{d=1}^8(\vxd\X\vs)(\vza\X\vxd). \]
 \ben
 \item[(i)] If $a=8$ we get the condition
 \[ F-G-\frc{10}3H=0. \lb{transeq4} \]
 \item[(ii)] If $a\neq8$ then the expression identically vanishes.
 \een
\een
We observe that Eq.\ \Eq{transeq1} and Eq.\ \Eq{transeq4} are
identical and that Eq.\ \Eq{transeq3} can be written as
$4\times$\Eq{transeq1}$-$\Eq{transeq2}. This means that the last two
equations are compatible with the previous ones and hence we have
verified that the linear combinations appearing in \Eq{E10-notlev2}
are indeed not elements of the root space $\exam$.

This result is remarkable in several ways. Firstly, it exemplifies an
as yet ill understood mechanism by which certain transversal physical
states decouple from other transversal ones. Secondly, the calculation
crucially depends on properties of the root lattice; this means that
the result will be completely different when we perform the analysis for
another lattice. Finally, we should stress again that, by this method,
we have found eight states orthogonal to the root space $\exam$
without explicitly computing a single commutator! All this makes
transversal decoupling even more miraculous a phenomenon
than longitudinal decoupling.

\begin{appendix}

\section{The measure} \lab{MEASURE}
We consider an infinitesimal shift of a fixed single Koba Nielsen
point, $z_i$, say:
\[ \td z_j := z_j+\d_{ij}\e\qquad\forall j. \]
If this shift is implemented by conformal transformations $\M_j$
acting on the $N$-vertex it then follows that
\[ -\frac{\partial}{\partial z_i}\VN(\{z_j\})
   =\lim_{\e\to0}\frac1\e\big[\VN(\{z_j\})-\tV{N}(\{\td z_j\})\big]
   =\VN(\{z_j\})\lim_{\e\to0}\frac1\e\bigg[\one-\prod_{j=1}^N\Mh\j_j\bigg]. \]
For the scattering amplitudes Eq.\ \Eq{Amplitude} this implies
\[ \lefteqn{
    \Oint\frac{dz_1}{2\pi i} \ldots \Oint\frac{dz_N}{2\pi i}
    \Delta(z_1,z_2,z_3)\,
        \frac{\partial}{\partial z_i}\bar\m(\{z_j\}) \,
        \VN(\{z_j\})
        \kets{\p_1}1\ldots\kets{\p_N}N } \hh{8mm} \non
   &=&{}
   -\Oint\frac{dz_1}{2\pi i} \ldots \Oint\frac{dz_N}{2\pi i}
    \Delta(z_1,z_2,z_3)\,
        \bar\m(\{z_j\}) \,
        \frac{\partial}{\partial z_i}\VN(\{z_j\})
        \kets{\p_1}1\ldots\kets{\p_N}N \non
   &=&{}
    \Oint\frac{dz_1}{2\pi i} \ldots \Oint\frac{dz_N}{2\pi i}
    \Delta(z_1,z_2,z_3)\,
        \bar\m(\{z_j\}) \,
        \VN(\{z_j\})
        \lim_{\e\to0}\frac1\e\bigg[\one-\prod_{j=1}^N\Mh\j_j\bigg]
        \kets{\p_1}1\ldots\kets{\p_N}N, \lb{measure1} \]
where
\[ \Delta(z_1,z_2,z_3)
    :=\frac{(z_1 - z_2)(z_2 - z_3)(z_1 - z_3)}
             {\big(z_1 - z_1^{(0)}\big)
              \big(z_2 - z_2^{(0)}\big)
              \big(z_3 - z_3^{(0)}\big)}. \]
This means that, once we know the operators $\Mh_j$,
we can determine the function $\bar\m$ by solving the differential
equation
\[ \frac{\partial}{\partial z_i}\bar\m(\{z_j\})
    = \bar\m(\{z_j\})
      \lim_{\e\to0}\frac1\e\bigg[\one-\prod_{j=1}^N\Mh\j_j\bigg], \]
which is to be understood as a relation between operators acting on
arbitrary states.

Let us now determine the conformal transformations implementing the
above shift of the Koba Nielsen point $z_i$. We have
\[ \tx_j(\z)=[\M_j^{-1}\cc\x_j](\z), \]
i.e.,
\[ \M_j=\x_j\tx_j^{-1}. \]
Since we are only interested in infinitesimal changes of the Koba
Nielsen points we consider the expansion of $\x_j$ and $\tx_j$ around
$z_j$ and $\td z_j$, respectively:
\[ \x_j(\z)  &=& \x_j'(\z-z_j)+\cO\big[(\z-z_j)^2\big], \\
   \tx_j(\z) &=& \tx_j'(\z-\td z_j)+\cO\big[(\z-\td z_j)^2\big], \]
where
\[ \x_j' \equiv\left.\dd{\x_j}{\z}\right|_{\z=z_j},\qquad
   \tx_j'\equiv\left.\dd{\tx_j}{\z}\right|_{\z=\td z_j}. \]
Inverting the above expansion for $\tx_j(\z)$ we get
\[ \tx_j^{-1}(\z)\approx\td z_j+\frac1{\tx_j'}\z. \]
Hence the conformal transformations associated with the infinitesimal
shift of the Koba Nielsen point $z_i$ have the form
\[ \M_j \approx \x_j'\left(\td z_j-z_j+\frac1{\tx_j'}\z\right)
        = \e\d_{ij}\x_j'+\frac{\x_j'}{\tx_j'}\z, \]
or, as operators,
\[ \Mh_j &\approx& \re{\e\d_{ij}\x_j'L_{-1}}
                   \left(\frac{\x_j'}{\tx_j'}\right)^{L_0} \non
         &\approx& \left[\one+\e\d_{ij}\x_j'L_{-1}\right]
                   \left(\frac{\x_j'}{\tx_j'}\right)^{L_0}. \]
Suppose now that the states $\p_j$ for $j\neq i$ occurring in
\Eq{measure1} are all physical and that $\p_i\equiv\Omega\in\Pz0$,
i.e.\ $L_n\Omega=0$ $\forall n\ge0$. It then follows that
\[ \lefteqn{
    \prod_{j=1}^N\Mh\j_j
    \kets{\p_1}1\ldots\kets{\p_{i-1}}{i-1}\kets{\Omega}i
    \kets{\p_{i+1}}{i+1}\ldots\kets{\p_N}N } \hh{8mm} \non
   &=&
    \prod_{j=1 \atop j\neq i}^N
    \left(\frac{\x_j'}{\tx_j'}\right)
    \left[\one+\e\x_i'L\i_{-1}\right]
    \kets{\p_1}1\ldots\kets{\p_{i-1}}{i-1}\kets{\Omega}i
    \kets{\p_{i+1}}{i+1}\ldots\kets{\p_N}N. \]
Obviously $L\i_{-1}\kets{\Omega}i$ is a null physical state and will
not contribute to a physical string scattering amplitude (cf.\ Def.\
\ref{def3}):
\[  \Oint\frac{dz_1}{2\pi i} \ldots \Oint\frac{dz_N}{2\pi i}
     \Delta(z_1,z_2,z_3)\,
        \m(\{z_j\}) \,
        \VN(\{z_j\})
        L\i_{-1}\kets{\Omega}i
        \prod_{j=1 \atop j\neq i}^N\kets{\p_j}j=0, \]
where $\m\equiv\x_i'\bar\m$ denotes the true measure determined by
decoupling. Consequently, we are left with the following differential
equation for the measure:
\[ \frac{\partial}{\partial z_i}
   \left(\frac{\m}{\x_i'}\right)
   &=& \frac{\m}{\x_i'}
       \lim_{\e\to0}\frac1\e\bigg[\one-\prod_{j=1 \atop j\neq i}^N
                                \frac{\x_j'}{\tx_j'}\bigg] \non
   &=& \frac{\m}{\x_i'}
       \frac{\partial}{\partial z_i}
       \log\bigg[\prod_{j=1 \atop j\neq i}^N\x_j'\bigg], \]
which has the solution (up to multiplication by a constant)
\[ \m=\prod_{j=1}^N\x_j'
     =\prod_{j=1}^N\left.\frac{\partial\x_j}{\partial\z}\right|_{\z=z_j}. \]
If we choose $\x_k$, say, as the coordinate $\z$, then the measure
takes the form
\[ \m=\prod_{j=1}^N\left.\frac{\partial\tr jk}{\partial\x_k}
                         \right|_{\x_j=0}. \]
Let us work out some examples.
\ben
\item[(i)] For the choice \ct{Love70}
\[ \x_j:=\frac{(\z-z_j)(z_{j+1}-z_{j-1})}
              {(\z-z_{j-1})(z_{j+1}-z_j)}, \]
or,
\[ \tr ji(\x_i):=
    \frac{\big[ (z_{i+1}-z_i)(z_{i-1}-z_j)\x_i
               +(z_{i+1}-z_{i-1})(z_j-z_i) \big] (z_{j+1}-z_{j-1}) }
         {\big[ (z_{i+1}-z_i)(z_{i-1}-z_{j-1})\x_i
               +(z_{i+1}-z_{i-1})(z_{j-1}-z_i) \big] (z_{j+1}-z_j) },
             \]
we find the measure to be
\[ \m=\prod_{j=1}^N\frac{z_{j+1}-z_{j-1}}
                        {(z_{j+1}-z_j)(z_j-z_{j-1})}, \]
while the zero mode term \Eq{Nzeromode} after some algebra is given by
\[ \cN=\prod_{i<j}
       \left[-\frac{(z_{j+1}-z_j)(z_j-z_{j-1})
                    (z_{i+1}-z_i)(z_i-z_{i-1})}
                   {(z_{j+1}-z_{j-1})(z_{i+1}-z_{i-1})
                    (z_j-z_i)^2}\right]^{-\frac12\vp_i\.\vp_j}. \]
When we rearrange this product and use momentum conservation we arrive
at the formula
\[ \cN=\prod_{j=1}^N
       \left[\frac{(z_{j+1}-z_j)(z_j-z_{j-1})}
                  {z_{j+1}-z_{j-1}}\right]^{\frac12\vp_j^2}\
       \prod_{i<j}(z_i-z_j)^{\vp_i\.\vp_j}. \]
Note that for tachyon scattering, i.e.\ $\vp_j^2=2$ $\forall j$, the
measure is cancelled by the first product in the zero mode term.
\item[(ii)] For the simple choice:
\[ \x_j(\z):=\z-z_j, \]
or,
\[ \tr ji(\x_i):=\x_i+z_i-z_j, \]
the measure comes out to be a constant,
\[ \m=1, \]
while the zero mode term \Eq{Nzeromode} is given by
\[ \cN=\prod_{i<j}(z_i-z_j)^{\vp_i\.\vp_j}. \]
\item[(iii)] For the ``vertex operator choice'':
\[ \tr12=\G,\qquad
   \tr1i=\frac1{\x_i+z_i},\qquad
   \tr i1=\frac1{\x_1}-z_i,\qquad
   \tr ij=\x_j+z_j-z_i, \]
we may take $\x_2\equiv\z$ so that $z_2=0$, $z_1=\infty$ and the
measure becomes
\[ \m=-\frac1{z_1^2}, \]
which is cancelled by the Faddeev Popov determinant
(for $z_2=0$, $z_1 \to \infty$)
\[ (z_1-z_2)(z_2-z_3)(z_1-z_3)=(-z_1^2)z_3, \]
to give a finite result as $z_1 \to \infty$.
Evaluation of the zero mode term finally yields
\[ \cN=\prod_{2\leq i<j\leq N}(z_i-z_j)^{\vp_i\.\vp_j}. \]
\een
For tachyon scattering, i.e.\ $\vp_j^2=2$ $\forall j$, we
observe that the product $\m\cN$ of the measure and the zero mode term
always reproduces the well-known Koba Nielsen formula.
\end{appendix}


\end{document}